\documentclass{aa}
\usepackage[varg]{txfonts}

\usepackage{natbib}
\bibpunct{(}{)}{;}{a}{}{,}

\usepackage{hyperref}
\usepackage{multirow}
\usepackage{amsmath}
\usepackage{subfig}

\usepackage{graphicx}
\graphicspath{{./images/}}

\def\Sch{Schwarzschild}
\def\SdS{Schwarzschild\discretionary{--}{--}{--}de~Sitter}

\providecommand{\dd}{\mathrm{d}}
\providecommand{\oder}[2]{\frac{\dd #1}{\dd #2}}
\providecommand{\Msun}{\ensuremath{\mathrm{M_{\odot}}}}
\providecommand{\euler}{\ensuremath{\mathrm{e}}}
\providecommand{\eto}[1]{\ensuremath{\euler^{#1}}}
\providecommand{\rhocent}{\ensuremath{\rho_{\mathrm{c}}}}
\providecommand{\pcent}{\ensuremath{p_{\mathrm{c}}}}
\providecommand{\epow}[1]{e^{#1}}
\providecommand{\oder}[2]{\frac{\partial #1}{\partial #2}}

\newcommand{\circledd}[1]{\raisebox {.4 pt}{\textcircled {\raisebox {-.7 pt}{\small #1}}}}

\begin{document}

\title{Dark matter halos modeled by polytropic spheres influenced by the relict cosmological constant and trapping polytropes forming supermassive black holes}
	
\author{Zden\v{e}k Stuchl\'{\i}k \thanks{\email{zdenek.stuchlik@physics.slu.cz}}
		\and
		Jan Novotn{\'y} \thanks{\email{jan.novotny@physics.slu.cz}}
		\and
		Jan Hlad\'{\i}k \thanks{\email{jan.hladik@physics.slu.cz}}
	   }

\institute{Research Centre for Theoretical Physics and Astrophysics, Institute of Physics, Silesian University in Opava,\\ Bezru\v{c}ovo n\'am.~13, 746\,01 Opava, Czech Republic}\label{inst1}
	
\date{Received: date / Accepted: date}
	
\abstract%
% context
%{Dark matter and repulsive dark energy govern evolution of the Universe, but they also influence structure of galaxies and their clusters.}%
{}
% aims
{We study dark matter halos modeled by general relativistic polytropic spheres in spacetimes with the repulsive cosmological constant representing vacuum energy density, governed by a polytropic index $n$ and a relativistic (cosmological) parameter $\sigma$ ($\lambda$) determining the ratio of central pressure (vacuum energy density) and central energy density of the fluid.}%
% methods
{To give mapping of the polytrope parameters for matching extension and mass of large dark matter halos, we study properties of the polytropic spheres and introduce an effective potential of the geodesic motion in their internal spacetime. Circular geodesics enable us to find the limits of the trapping polytropes with central regions containing trapped null geodesics; supermassive black holes can be formed due to the instability of the central region against gravitational perturbations. Stability of the polytropic spheres relative to radial perturbations is determined. We match extension and mass of the polytropes to those of dark matter halos related to large galaxies or galaxy clusters, with extension $100 < \ell/\mathrm{kpc} < 5000$ and gravitational mass $10^{12} < M/\Msun < 5 \times 10^{15}$. The velocity radial profiles of circular geodesics in the polytrope spacetimes are numerically compared to the observed velocity profiles.}%
% results
{The observed velocity profiles simulated by the phenomenological dark matter halo density profiles can be well matched also by the velocity profiles of the exact polytrope spacetimes. The matching is possible by the non-relativistic polytropes for each value of $n$, with relativistic parameter $\sigma \leq 10^{-4}$ and very low central energy density. Surprisingly, the matching works for ``spread'' relativistic polytropes with $n > 3.3$ and $\sigma \geq 0.1$ when the central density can be much larger. The trapping polytropes forming supermassive black holes must have $n > 3.8$ and $\sigma > 0.667$. We thus explain the mass and structure of large galaxies and galaxy clusters, their extension limited by the cosmic repulsion, and the existence of black holes with mass $M > 10^{10}\,\Msun$ in very large galaxies; we suggest black holes with $M \sim 10^{12}\,\Msun$ in large galaxy clusters.}%
% conclusions
{}
	
    
\titlerunning{Relativistic polytropes as models of dark matter halos}
\authorrunning{Stuchl\'{\i}k Z. et al.}

\maketitle

%%%%%%%%%%%%%%%%%%%%%%%%%%%%%%%%%%%%%%%%%%%%%%%%%%%%%%%%%%%%%%%%%%%%%%%%%%%%%%%%%%%%%%%%%%%%%%%%%%%%%%%%%%%%%%%%%%%%%%%%%%%%%%%%%%%%%%%%%%%%	

\section{Introduction}\label{intro}
The inflationary paradigm~\citep{Guth:1981:PRD:,Lin:1990:InfCos:,Guth:2014:PhLB} and a wide range of cosmological observations indicate the existence of dark energy demonstrating a repulsive gravitational effect; specifically, it can be the vacuum energy, i.e., a very small relict repulsive cosmological constant $\Lambda > 0$~\citep{Kra-Tur:1995:GENRG2:,Ost-Ste:1995:NATURE:,Kra:1998:ASTRJ2:,Bah-etal:1999:SCIEN:,ArP-Muk-Ste:2000:PHYRL:,Wan-etal:2000:ASTRJ2:,Carroll:2001:LRR}. Observations of distant Ia-type supernova explosions indicate that starting at the cosmological redshift $z \approx 1$ expansion of the Universe is accelerated~\citep{Rie-etal:2004:ASTRJ2:}. In accord with the inflationary paradigm, the total energy density of the Universe is very close to the critical energy density $\rho_{\mathrm{crit}}$, corresponding thus to an almost flat universe~\citep{Spe-etal:2007:ASTJS:3yrWMAP,Tristram:2024:AaA}. The cosmological tests demonstrate that the dark energy represents about $70\%$ of the energy content of the observable universe ($\rho_{\mathrm{vac}} \sim 0.7\rho_{\mathrm{crit}}$) \citep{Huterer:2018:RPPh}; then about $25\%$ of the Universe energy content is in the form of dark matter, and the rest $5\%$ corresponds to the baryonic matter and the other forms of matter and fields \citep{Brout:2022:ApJ}. These results are confirmed by measurements of cosmic microwave background anisotropies by the space satellite observatory PLANCK~\citep{PlanckB:2020:AaA}. The dark energy equation of state is very close to those corresponding to the vacuum energy, i.e., to the repulsive cosmological constant. The value of the relict cosmological constant is estimated to be $\Lambda \approx 1.1\times 10^{-56}\,\mathrm{cm^{-2}}$, and the vacuum mass density $\rho_{\mathrm{vac}} \sim 10^{-29}\,\mathrm{g\,cm^{-3}}$ \citep{PlanckB:2020:AaA}. 

The inflationary universe stands at the base of the $\Lambda$CDM model that combines the role of dark energy represented by the relict cosmological constant $\Lambda$ and cold dark matter, assumed to be relevant in the recent era of the Universe evolution \cite{Mukhopadhyay:2008:IJMPD}. The $\Lambda$CDM model can be considered the standard cosmological model of the recent era of cosmological investigations, besides some difficulties with the interpretation of some recent results \citep{Blanchard:2024:OJAp}. The exact form of the inflationary universe remains an open problem under extensive debate, but the cosmic observations are introducing strong constraints on the considered inflationary models \citep{Planck:2020:AaA}. 

The cosmological and astrophysical consequences of the effect of the relict cosmological constant estimated due to the cosmological tests have been studied extensively not only for the cosmological models~\citep{Peebles:2003:RvMP}, but also for the Einstein--Strauss vacuola models \citep{Stu:1983:BULAI:,Stu:1984:BULAI:,Uza-Ell-Lar:2011:GENRG2:2MassExp:,Gre-Lak:2010:PHYSR4:,Fle-Dup-Uza:2013:PHYSR4:,Arra:2014:PHYSR4:,Far-Lap-Pra:2015:JCAP:,Far:2016:PDU:}, or in the framework of the McVittie model \citep{McV:1933:MONRAS:} of mass concentrations immersed in the expanding universe \citep{Nol:1998:PHYSR4:,Nol:1999:CLAQG:,Nan-Las-Hob:2012:MONRAS:,Kal-Kle-Mar:2010:PHYSR4:,Lak-Abd:2011:PHYSR4:,Sil-Fon-Gua:2013:PHYSR4:,Nol:2014:CLAQG:}.

Very important role of the repulsive cosmological constant has been demonstrated also for astrophysical processes (accretion disks, jets) related to active galactic nuclei~\citep{Stu-Char-Sche:2018:EPJC,Stu-etal:2020:Universe} and their central supermassive black holes~\citep{Stu-Cal:1991:GENRG2:,Lak:2002:PHYSR4:BendLiCC:,Stu-Hle:2002:ACTPS2:,Stu-Sla:2004:PHYSR4:,Kra:2004:CLAQG:,Kra:2005:DARK:CCPerPrec,Kra:2007:CLAQG:Periapsis,Cru-Oli-Vil:2005:CLAQG:GeoSdSBH,Stu-Sla-Hle:2000:ASTRA:,Sla-Stu:2005:CLAQG:,Stu:2005:MODPLA:,Ser:2008:PHYSR4:CCLens,Mul:2008:GENRG2:FallSchBH,Sch-Zai:2008:0801.3776:CCTimeDelay,Vil-etal:2013:ASTSS1:PhMoChgAdS:,Rez-Zan-Fon:2003:ASTRA:,Kag-Kun-Lam:2006:PHYLB:SolarSdS,Ali:2007:PHYSR4:EMPropKadS,Che:2008:CHINPB:DkEnGeoMorSchw,Ior:2009:NEWASTR:CCDGPGrav,Hac-etal:2010:PHYSR4:KerrBHCoStr:,Kol-Stu:2010:PHYSR4:CurCarStrLoops,Hen-Mom:2011:EPJC:,Hen-Pan-Mou:2012:GENRG:,Gu-Cheng:2007:GENRG2:CircLoopKdS,Wan-Che:2012:PHYLB:CirLoopPerTens,Stu-Sla-Kov:2009:CLAQG:,Pug-Stu:2016:ApJS:,Pug-Stu:2024:EPJC:}. On the other hand, the Kerr superspinars representing an alternative to the supermassive black holes in active galactic nuclei, based on the String theory~\citep{Gim-Hor:2004:hep-th0405019:GodHolo,Boy-etal:2003:PHYSR4:HoloProtChron,Gim-Hor:2009:PHYLB:AstVioSignStr,Stu-Sch:2012:CLAQG:} and exhibiting a variety of unusual physical phenomena~\citep{deFel:1974:ASTRA:,deFel:1978:NATURE:InstabNS,Stu:1980:BULAI:,Hio-Mae:2009:PHYSR4:KerrSpinMeas,Stu-Sch:2013:CLAQG:UHEKerrGeo}, could also be relevant \citep{Stu-Hle-Tru:2011:CLAQG:}. The pseudo-Newtonian potential related to the spherically symmetric spacetimes with the repulsive cosmological constant~\citep{Stu-Kov:2008:INTJMD:PsNewtSdS} serves well in studies of the motion of interacting galaxies~\citep{Stu-Sch:2011:JCAP:CCMagOnCloud}.

The general relativistic polytropic spheres in spacetimes with the repulsive cosmological constant have been extensively studied in \citep{Stu-Hle-Nov:2016:PHYSR4:}, generalizing thus the case of spheres with uniform distribution of energy density (but radii dependent distribution of pressure) corresponding formally to the polytropes with index $n = 0$ that can serve as a test bed for properties of the polytropes because their structure equations can be solved in terms of elementary functions~\citep{Stu:2000:ACTPS2:,Boh:2004:GENRG2:,Nil-Ugg:2000:ANNPH1:GRStarPoEqSt,Boe-Fod:2008:PHYSR4:}. Two important results were obtained: first, the polytropic spheres cannot exceed the static radius where their gravitational attraction is just balanced by cosmic repulsion \citep{Stu:1983:BULAI:,Stu-Hle:1999:PHYSR4:}, giving thus a natural limit on gravitationally bound systems in the accelerated Universe; second, extension and mass of the polytropic spheres can be comparable even with extension and mass of dark matter halos of large galaxies and galaxy clusters \citep{Stu-Hle-Nov:2016:PHYSR4:}. 

We have shown that the polytropic spheres considered in the non-relativistic regime can be satisfactorily applied to describe and qualify the dwarf galaxies demonstrating the so-called cusp-core problem~\citep{Nov-Stu-Hla:AA:2021}. In the present paper, we extend our previous works and, taking fully into account the role of the relict cosmological constant, we realize a detailed mapping of the parameters of polytropic spheres enabling matching to the extension and mass of large dark matter halos connected with large galaxies and their clusters, with inclusion of a detailed analysis of the trapping polytropes allowing for the creation of supermassive black holes in their central region \citep{Stu-etal:2017:JCAP:}.

The general relativistic polytropic spheres enable a very useful, physically relevant, idealization of fluid configurations under various conditions, giving a simple and coherent picture of all the potentially relevant relativistic phenomena influencing matter configurations across different distance scales. Well known is the application of the polytropic equations of state for modelling of neutron (or quark) stars. In the basic approximation, the degenerated Fermi gas can be represented by the equation of state with polytropic index $n = 3/2$ in the non-relativistic limit, and $n = 3$ in the ultrarelativistic limit \citep{Sha-Teu:1983:CompStar:}. Polytropic state equations with various values of the polytropic index $n$ are used to give precise approximation of relativistic equations of state governing the interior of neutron stars \citep{Oze-Psa:2009:PRD:, Lat-Pra:2001:ASTRJ2:NS}; one can even use several polytropic state equations to cover the neutron star core \citep{Alv-Bla-Typ:2017:AN}.

On the other hand, extremely extended general relativistic polytropic spheres can serve as models of halos of dark matter in galaxies or even galaxy clusters; the preliminary results \citep{Stu-Hle-Nov:2016:PHYSR4:} demonstrate that non-relativistic polytropes can represent halos made of cold dark matter, while the relativistic polytropes can represent halos made of hot dark matter, or could be applied in more complex situations, mixing the cold and hot dark matter, or other influences (as those of standard baryonic matter). Very interesting are galaxies existing in the early stages of expansion of the Universe, observed at the cosmological redshift $z > 6$. Of particular interest are the active nuclei of galaxies containing supermassive black holes that in some cases have masses exceeding $10^{10}M_{\odot}$ \citep{Zio:2005:NUOC2:GalCollObj}, as the standard explanation of successive growing of black holes having initially mass of the stellar order, $M \leq 100 M_{\odot}$, requires specially ordered conditions during growing in such small time scales to such extreme values of the black hole mass. 

Despite enormous efforts in both theoretical and experimental particle physics, composition of dark matter remains unknown; a large variety of possible (but controversial) candidates of both cold and hot (warm) dark matter can be considered as acceptable. Due to the lack of a clear dark matter candidate coming from particle physics, we are free to choose any parameters of the polytrope equation of state to test observationally relevant predictions of the general relativistic polytropic spheres related to their extension, mass, and the corresponding velocity curves. Here, in matching the extension and mass of halos related to large galaxies and galaxy clusters, we test the simple relativistic or non-relativistic polytropic spheres represented by a single ensemble of the polytrope parameters.

The cold dark matter (CDM) halos are recently considered as the most natural explanation of hidden structure of galaxies, enabling correct treatment of the motion in their external parts ~\citep{Bos:1981:ASTRJ1:21cmSpiGal,Rub:1982:HIContNorGal:SaSbScGal}, or as sources of gravitational wells of galaxy clusters explaining their binding \citep{Bar:etal:2015:MNRAS:GalLensing:,Sar-etal:2014:APJ:DMtoEOS:}.

The CDM halos are usually treated in the Newtonian approximation~\citep{Bin-Tre:1988:GalacDynam:,Ior:2010:MONNR:GalOrbMoDarkMat,Nav-Fre-Whi:1997:ASTRJ2:UniDeProHiCl,Stu-Sch:2011:JCAP:CCMagOnCloud,Cre-Stu:2013:IJMPD:}, or in the pseudo-Newtonian approximation \citep{Stu-Sch:2011:JCAP:CCMagOnCloud}. We shall consider here as the halo model the fully general relativistic spherically symmetric static configurations of perfect fluid with a polytropic equation of state~\citep{Too:1964:ASTRJ2:} and modify them by introducing the vacuum energy represented by the repulsive cosmological constant restricted by the recent cosmological observations. Details of the physical processes inside the polytropic spheres are not considered; the power law relating the total pressure to the total energy density of matter is assumed. The polytropic approximation seems to be applicable in the dark matter models assuming weakly interacting particles (see, e.g.,~\cite{Bor:1993:EarlyUniv:,Kol-Tur:1990:EarUni:,Cre-Stu:2013:IJMPD:}).

The rough estimates realized in \cite{Stu-Hle-Nov:2016:PHYSR4:} indicate that the non-relativistic polytropic spheres could represent the CDM halos. However, there is a strong indication of the failure of the CDM halo model at small scales in recent galaxies \citep{Gar-etal:2017:ARx,Mur-etal:2017:Arxiv}. On the other hand, applicability of the warm dark matter (WDM) model \citep{Bod-Jer-Vik:2009:Apj,Bau-etal:2016:JCAP:WDM}, especially in the case of primeval galaxies, is seriously considered \citep{Lap-Dan:2015:JCAP:}. There is a variety of possible non-cold dark matter halo candidates, starting at the standard possibilities of WDM represented by sterile neutrinos \citep{Adhi-etal:2017:JCAP} or axions \citep{Mar-Sil:2014:MONNR:, Hui-etal:2017:PRD:Ultra}. Anderhalden et al. offer mixed C\&{}WDM matter \citep{And-etal:2012:JCAP:}; also, self-interacting DM is seriously taken into account \citep{Rac-etal:PRD:2016:}.
	
For description of most of the non-cold DM halo models, the relativistic polytropic spheres have to be relevant, and they thus deserve attention in our study. Therefore, we present here a detailed mapping on the possibility of matching the extension and mass of dark matter halos of large galaxies or galaxy clusters by extremely extended polytropic spheres that could occur for sufficiently large polytropic indexes $3.3 < n < 5$ when such polytropes can exist, if the relativistic parameter is close to the critical values $\sigma_\mathrm{crit}(n)$ \citep{Nil-Ugg:2000:ANNPH1:GRStarPoEqSt,Stu-Hle-Nov:2016:PHYSR4:}. Recall that for $\sigma = \sigma_\mathrm{crit}(n)$ and vanishing vacuum energy, the polytropic spheres have unlimited extension, but the relict cosmological constant introduces a relevant restriction on their extension, which will be demonstrated in the present paper for selected representative values of the polytropic index $n$.

Moreover, we have demonstrated a very interesting property of highly relativistic polytropes that contain a region of trapped null geodesics \citep{Nov-Hla-Stu:2017:PRD:trapping}. We have shown that in the trapping zone an efficient gravitational instability causes gravitational collapse of the matter and conversion of the zone of trapping to a central black hole containing nearly $10^{-3}$ of the halo mass \citep{Stu-etal:2017:JCAP:}. If the polytropic parameters are properly tuned, the trapping polytrope can be extremely extended, representing thus a dark matter halo of a large galaxy or a galaxy cluster, having mass $M_\mathrm{halo-galaxy} > 10^{12}\,\Msun$, and the central supermassive black hole created in the trapping zone can have mass $M_\mathrm{BH} > 10^9\,\Msun$ \citep{Stu-etal:2017:JCAP:}. The black hole mass can even exceed $M\sim{}10^{12}\,\Msun$ if centered in a large galaxy cluster with $M_\mathrm{halo-cluster} > 10^{15}\,\Msun$. The gravitational instability of the central "trapping" region of extremely extended trapping polytropes can thus serve as an alternative model of formation of supermassive black holes in the high-redshift ($z > 6$) large galaxies or their clusters. As the original study concerning the gravitational instability of extended trapping polytropes has been done under a simplified assumption on vanishing cosmological constant, while its influence on the extension of such polytropic spheres can be crucial, we devote in the present paper special attention to the study of the role of the cosmological constant to constrain the mass and extension of the extremely extended trapping polytropes. \footnote{On the other hand, the study of the geodesic structure of the internal polytrope spacetime can be applied in a complementary study of trapping of weakly-interacting particles (neutrinos or other types that could represent the dark matter) inside the neutron stars or quark stars.}

In the present paper, we give the detailed description of the general relativistic polytropes in spacetimes containing the relict cosmological constant, including discussion of their stability relative to radial perturbations. As a first step in modeling the DM halos using the polytropic spheres, we present a detailed study of the possibility to match extension and mass of both the non-relativistic and relativistic polytropic spheres to those corresponding to DM halos related to large galaxies or their clusters. For selected typical polytropic halos, we give the density and metric coefficient radial profiles. We also present a detailed study of the circular geodesics in the interior of the polytropic spheres, as the velocity radial profiles of the circular geodesics of the polytropic internal spacetimes should correspond to the velocity profiles of stars at the outer regions of large galaxies. A more detailed study of the velocity radial profiles, combining the polytropic regions representing the outer regions of large galaxies with the galaxy bulge and galaxy disc regions dominating the inner regions of large galaxies, is planned for a future paper. 

%%%%%%%%%%%%%%%%%%%%%%%%%%%%%%%%%%%%%%%%%%%%%%%%%%%%%%%%%%%%%%%%%%%%%%%%%%%%%%%%%%%%%%%%%%%%%%%%%%%%%%%%%%%%%%%%%%%%%%%%%%%%%%%%%%%%%%%%%%%%

\section{Polytropic spheres with cosmological constant}\label{eost}
The line element of a spherically symmetric, static spacetime, expressed in terms of the standard \Sch{} coordinates, reads
	\begin{equation}
	   \dd s^{2} = - \eto{2\Phi}\, c^{2} \dd t^{2} + \eto{2\Psi}\,\dd r^{2} + r^{2} (\dd\theta^{2} + \sin^{2} \theta\,\dd\phi^{2})\, .  \label{tpa007e1}
	\end{equation}
The metric contains just two unknown functions of the radial coordinate, $\Phi(r)$ and $\Psi(r)$. Matter of the static configuration is assumed to be a perfect fluid with the stress-energy tensor
    \begin{equation}
	   T^\mu_{\hphantom{\mu}\nu} = (p + \rho c^{2}) U^{\mu} U_{\nu} + p\,\delta^{\mu}_{\nu}\, ,  \label{e3}
	\end{equation}
where $U^{\mu}$ denotes the four-velocity of the fluid. In the rest-frame of the fluid $\rho = \rho (r)$ represents the mass-energy density and $p = p(r)$ represents the isotropic pressure.

We restrict our attention to the simplest direct relation between the mass-energy density and pressure given by the polytropic equation of state
	\begin{equation}
	   p = K \rho^{1 + \frac{1}{n}}\, ,  \label{grp6}
	\end{equation}
where $n$ denotes the `polytropic index' assumed to be a given constant. $K$ is a constant determined by the thermal characteristics of a given fluid spherical configuration by specifying the density $\rhocent$ and pressure $p_{\mathrm{c}}$ at the center of the polytropic sphere --- it is determined by the total mass and radius of the configuration, and the so-called relativistic parameter \citep{Too:1964:ASTRJ2:}
	\begin{equation}
	   \sigma  \equiv \frac{\pcent}{\rhocent c^{2}}\  = \frac{K}{c^{2}}\rhocent^{1/n}\, .  \label{grp17-18}
	\end{equation}
For a given pressure, the density is a function of temperature, thus the constant $K$ contains the temperature implicitly.

Note that the polytropic equation represents a limiting form of the parametric equations of state for a completely degenerate gas at zero temperature, relevant, e.g., for neutron stars. Then both $n$ and $K$ are universal physical constants \citep{Too:1964:ASTRJ2:}. In fact, the simple polytropic law assumption enables one to obtain basic properties of the fluid configurations governed by the relativistic laws. For example, the equation of state of the ultrarelativistic degenerate Fermi gas is determined by the polytropic equation with the adiabatic index $\Gamma = 4/3$ corresponding to the polytropic index $n = 3$, while the non-relativistic degenerate Fermi gas is determined by the polytropic equation of state with $\Gamma = 5/3$, and $n = 3/2$ \citep{Sha-Teu:1983:CompStar:}.

The structure of the general relativistic polytropic spheres is determined by the Einstein field equations
	\begin{equation}
	   R_{\mu\nu} - \frac{1}{2}Rg_{\mu\nu} + \Lambda g_{\mu\nu} = \frac{8\pi G}{c^4} T_{\mu\nu}
	\end{equation}
containing the dark vacuum energy that is represented by the cosmological constant term, and by the local energy-momentum conservation law
	\begin{equation}
	   T^{\mu\nu}_{\hphantom{\mu\nu};\nu} = 0\, .
	\end{equation}

\subsection{Structure equations}
For the structure of the polytropic spheres, only the $(t)(t)$ and $(r)(r)$ components of the Einstein field equations are relevant (the $(\theta)(\theta)$ and $(\phi)(\phi)$ components give dependent equations) \citep{Too:1964:ApJ:}. The structure is governed by the two structure functions. The first one $\theta(r)$ is related to the mass-energy density radial profile $\rho(r)$ and the central density $\rhocent$ \citep{Too:1964:ASTRJ2:}
	\begin{equation}
	   \rho = \rhocent\theta^{n}\, ,  \label{grp14}
	\end{equation}
with the boundary condition $\theta(r = 0) = 1$. The second one is the mass function of the polytropic configuration given by the relation
	\begin{equation}
	  m(r) = \int^{r}_{0} {4 \pi r'^{2} \rho \, \dd r'}\, ,  \label{grp11}
	\end{equation}
with the integration constant chosen to be $m(0) = 0$, to guarantee the smooth spacetime geometry at the origin (see Ref. \citep{Mis-Tho-Whe:1973:Gra:}). At the edge of the configuration at $r = R$, there is $\rho(R) = p(R) = 0$, and the total mass of the polytropic configuration $M = m(R)$. Outside the polytropic configuration, the spacetime is described by the vacuum Schwarzschild-de Sitter metric.

Projection of $T^{\mu\nu}_{\hphantom{\mu\nu};\nu} = 0$ orthogonal to $U^{\mu}$ implies the equation of hydrostatic equilibrium describing the balance between the gravitational force and pressure gradient that can be put into the Tolman--Oppenheimer--Volkoff (TOV) form modified by the presence of a non-zero cosmological constant \citep{Stu:2000:ACTPS2:}:
	\begin{equation}
	   \frac{\dd p}{\dd r} = - (\rho c^{2} + p) \frac{G m(r)/c^{2} - \Lambda r^{3}/3 + 4\pi G p r^{3}/{c^4}}{r \left[r-2G m(r)/c^{2} - \Lambda  r^{3}/3 \right]}\, . \label{tpa007e16}
	\end{equation}

Then the structure equations of the polytropic spheres related to the two structure functions, $\theta(r)$ and $m(r)$, and the three parameters $n$, $\sigma$, $\Lambda$, can be put into the form \citep{Stu-Hle-Nov:2016:PHYSR4:}
	\begin{align}
	   \frac{\sigma(n + 1)}{1 + \sigma\theta}\,r\,\oder{\theta}{r}\left(1-\frac{2Gm(r)}{c^{2}r}-\frac{1}{3}\Lambda r^{2}\right) + \nonumber\\
	   + \frac{Gm(r)}{c^{2}r} - \frac{1}{3}\Lambda r^{2} & = -\frac{G}{c^{2}}\sigma\theta\oder{m}{r}\, , \label{grp24}\\
	   \oder{m}{r} & = 4\pi r^{2} \rhocent\theta^{n}\, .  \label{grp25}
	\end{align}

Introducing the length factor $\mathcal{L}$ giving characteristic length scale of the polytropic sphere
	\begin{equation}
	   \mathcal{L} = \left[\frac{(n + 1)K\rhocent^{1/n}}{4\pi G\rhocent}\right]^{1/2} = \left[\frac{\sigma(n + 1)c^{2}}{4\pi G\rhocent}\right]^{1/2}\, ,  \label{grp26}
	\end{equation}
and the mass factor $\mathcal{M}$ giving characteristic mass scale of the polytropic sphere
	\begin{equation}
	   \mathcal{M} = 4\pi \mathcal{L}^3 \rho_{\rm c} = \frac{c^2}{G}\sigma (n + 1)\mathcal{L}\, ,  \label{grp26a}
	\end{equation}
the structure equations (\ref{grp24}) and (\ref{grp25}) can be transformed into dimensionless form by introducing a dimensionless radial coordinate
	\begin{equation}
	   \xi = \frac{r}{\mathcal{L}}\, , \label{grp27}
	\end{equation}
and dimensionless quantities
	\begin{eqnarray}
	   \nu(\xi) & = & \frac{m(r)}{4\pi \mathcal{L}^{3}\rhocent} \, ,  \label{grp28}\\
	   \lambda & = & \frac{\rho_{\mathrm{vac}}}{\rhocent}\, , \label{grp29}
	\end{eqnarray}
where $\nu(\xi)$ represents the dimensionless mass function and $\lambda$ represents the dimensionless cosmological constant related to the polytropic sphere. The vacuum energy density $\rho_\mathrm{vac} c^2$ and the cosmological constant $\Lambda$ are related by
	\begin{equation}
	   \rho_{\mathrm{vac}}c^{2} = \frac{\Lambda c^4}{8\pi G} = \lambda \rho_\mathrm{c} c^2\, . \label{grp30}
	\end{equation}

The dimensionless structure equations (\ref{grp24}) and (\ref{grp25}) take the form
	\begin{align}
	   \xi^{2}\oder{\theta}{\xi}\frac{1-2\sigma(n + 1)\left(\nu/\xi + \lambda\xi^{2}/3\right)}{1 + \sigma\theta} + \nonumber \\
	   + \nu(\xi) - \frac{2}{3}\lambda\xi^{3} & =  - \sigma\xi\theta\oder{\nu}{\xi}\, ,  \label{grp31}\\
	   \oder{\nu}{\xi} & =  \xi^{2}\theta^{n}\, . \label{grp32}
	\end{align}
For fixed parameters $n$, $\sigma$, $\lambda$, equations (\ref{grp31}) and (\ref{grp32}) have to be simultaneously solved under the boundary conditions
	\begin{equation}
	   \theta(0) = 1\, , \quad \nu(0) = 0\, .  \label{grp33}
	\end{equation}
It follows from (\ref{grp32}) and (\ref{grp33}) that $\nu(\xi) \sim \xi^{3}$ for $\xi \to 0$ and, according to Eq.\,(\ref{grp31}), there is
	\begin{equation}
	   \lim_{\xi\to 0_{+}} \oder{\theta}{\xi} = 0\, . \label{grp34}
	\end{equation}
The boundary of the polytropic sphere, $r = R$, is represented by the first zero point of $\theta(\xi)$, denoted as $\xi_{1}$:
	\begin{equation}
	   \theta(\xi_{1}) = 0\, . \label{grp35}
	\end{equation}
The solution $\xi_{1}$ determines the surface radius of the polytropic sphere, and the solution $\nu(\xi_{1})$ determines its gravitational mass. In Fig.~\ref{Fig_tlak} we illustrate possible types of the behaviour of the function $\theta(\xi)$, including the limiting case governed by the value of the cosmological parameter. 

The radius of the polytropic sphere reads
    \begin{equation}
	   R = \mathcal{L}\xi_{1}\, ,    \label{grp47}
	\end{equation}
while the mass of the sphere is given by
	\begin{equation}
	   M = \mathcal{M} \nu(\xi_1) = \frac{c^{2}}{G}\mathcal{L}\sigma(n + 1) \nu(\xi_{1})\, .  \label{grp48}
	\end{equation}

\subsection{Characteristics of the polytropic spheres}\label{properties}
The general relativistic polytropic spheres are determined by the functions $\theta(\xi)$ and $\nu(\xi)$ of the dimensionless coordinate $\xi$ and by the length and mass scales, $\mathcal{L}$, $\mathcal{M}$.

The energy density, pressure, and mass-distribution radial profiles are given by the relations
	\begin{align}
	   \rho(\xi) & = \rhocent\theta^{n}(\xi)\, ,  \label{grp49}\\
	   p(\xi) & = \sigma\rhocent\theta^{n + 1}(\xi)\, ,  \label{grp50}\\
	   M(\xi) & = M\frac{\nu(\xi)}{\nu(\xi_{1})} \equiv \mathcal{M}\nu(\xi)\, . \label{grp51}
	\end{align}
The temporal metric coefficient of the internal spacetime of the polytrope can be expressed in the form
	\begin{equation}
	   \eto{2\Phi} = (1 + \sigma\theta)^{-2(n + 1)}\left\{1-2\sigma(n + 1) \left[\frac{\nu(\xi_{1})}{\xi_{1}} + \frac{1}{3}\lambda\xi_{1}^{2}\right] \right\}\, , \label{grp52}
	\end{equation}
while the radial metric coefficient of the internal spacetime reads
	\begin{equation}
	   \eto{-2\Psi} = 1 - 2\sigma(n + 1)\left[\frac{\nu(\xi)}{\xi} + \frac{1}{3}\lambda\xi^{2}\right]\, . \label{grp43}
	\end{equation}
In the exterior of the polytropic sphere, at $\xi > \xi_1$, the metric coefficients take the form
	\begin{equation}
	   \eto{2\Phi} = \eto{-2\Psi} = 1 - 2\sigma(n + 1)\left[\frac{\nu(\xi_1)}{\xi} + \frac{1}{3}\lambda\xi^{2}\right]\, . \label{grp43a}
	\end{equation}

The compactness, determining the effectiveness of the gravitational binding of the polytropic sphere, is given by the relation
	\begin{equation}
	   \mathcal{C} \equiv \frac{GM}{c^{2}R} = \frac{1}{2}\frac{r_{\mathrm{g}}}{R} = \frac{\sigma(n + 1)\nu(\xi_{1})}{\xi_{1}} \label{grp42}
	\end{equation}
where we have introduced the standard gravitational radius of the polytropic sphere that reflects its gravitational mass in length units, $r_{\mathrm{g}} = 2GM/c^{2}$. The compactness $\mathcal{C}$ of the polytropic sphere can be represented by the gravitational redshift of radiation emitted from the surface of the polytropic sphere \citep{Hla-Stu:2011:JCAP:}. It is clear that the polytropic spheres representing the DM halos must be of extremely low compactness, as the gravitational radius has to be located deeply in the central region of such spheres.

The expressions for the gravitational energy and binding energy of the polytropic spheres can be found in~\cite{Stu-Hle-Nov:2016:PHYSR4:}, where also the analytical expression in terms of elementary functions is given for the special class of uniform energy density spheres, i.e., polytropes with index $n = 0$.\footnote{For an alternative way of expressing the $n = 0$ polytropes see~\cite{Stu:2000:ACTPS2:}.}

Notice that for vanishing cosmological constant, the polytrope structure equations are fully determined only by the parameters $n$ and $\sigma$, but the extension and mass scales are governed also by the central energy density $\rho_c$. The cosmological constant breaks this degeneracy of the structure equations, as the central energy density enters them due to its mixing with the vacuum energy density in the parameter $\lambda$.

%%%%%%%%%%%%%%%%%%%%%%%%%%%%%%%%%%%%%%%%%%%%%%%%%%%%%%%%%%%%%%%%%%%%%%%%%%%%%%%%%%%%%%%%%%%%%%%%%%%%%%%%%%%%%%%%%%%%%%%%%%%%%%%%%%%%%%%%%%%%

\section{External spacetime and length scales}
The external vacuum \SdS{} spacetime is represented by the same gravitational mass parameter $M$ and the same cosmological constant $\Lambda$ as those characterizing the internal spacetime of the polytropic sphere and is given by the metric coefficients
    \begin{equation}
	   \eto{2\Phi} = \eto{-2\Psi} = 1 - \frac{2GM}{c^{2} r} -\frac{1}{3}\Lambda r^{2}\, .
	\end{equation}
There are two event horizons related to the external vacuum spacetime --- the inner black hole horizon, and the outer cosmological horizon \citep{Stu:1983:BULAI:,Stu-Hle:1999:PHYSR4:,Stu-Sla-Hle:2000:ASTRA:}. In astrophysically plausible situations, even for the most massive black holes in the central part of giant galaxies, e.g., in the quasar TON~618 with the mass $M \sim 6.6\times{}10^{10}\,\Msun$ \citep{Zio:2008:CHIAA:MassBHU}, or for masses related to the whole giant galaxies containing an extended DM halo, and for clusters of galaxies having mass up to $M \sim 10^{15}\,\Msun$, the related black hole horizon radius (hidden deeply inside the galaxy or the cluster) and the cosmological horizon radius are given with very high precision by the simple formulae \citep{Stu-Sla-Kov:2009:CLAQG:}
	\begin{equation}
	   r_{\mathrm{h}} = r_{\mathrm{g}} = \frac{2GM}{c^{2}}\, ,\quad r_{\mathrm{c}} = \left(\frac{\Lambda}{3}\right)^{-1/2}\, .
	\end{equation}

The event horizons thus give two characteristic length scales of the SdS spacetimes. Of course, the black hole horizon is located inside the polytropic spheres being thus physically irrelevant, while the cosmological horizon is located outside the polytropic sphere, at extremely large distance from the polytropic sphere for the observationally given value of the relict cosmological constant $\Lambda \sim 10^{-56}\,\mathrm{cm}^{-2}$. These two characteristic length scales can be combined to give a dimensionless parameter characterizing the \SdS{} geometry~\citep{Stu-Hle:1999:PHYSR4:}
	\begin{equation}
	   y = \frac{\Lambda}{12}\,r_{\mathrm{g}}^{2} = \frac{1}{4}\left(\frac{r_{\rm g}}{r_{\rm c}}\right)^2\, .
	\end{equation}
For the observationally estimated repulsive cosmological constant $\Lambda \sim 1.1\times 10^{-56}\,\mathrm{cm^{-2}}$, the cosmological parameter $y$ takes extremely small values, if we consider the stellar mass black holes and galactic center black holes; very small values are obtained even for the largest compact objects of the universe, i.e., the extremal central supermassive black holes in the active galactic nuclei and the giant galaxies or their clusters~\citep{Stu-Sla-Hle:2000:ASTRA:,Stu:2005:MODPLA:}.

A third length scale characterizing the SdS vacuum spacetimes determines the boundary of the gravitationally bound system behind which the cosmic repulsive effects start to be effective. The third scale is given by the so-called static radius (sometimes also called turn-around radius)~\citep{Stu-Hle:1999:PHYSR4:,Stu-Sla:2004:PHYSR4:,Stu:2005:MODPLA:,Gre-Lak:2010:PHYSR4:,Stu-Sch:2011:JCAP:CCMagOnCloud,Ara:2013:arXiv:,Ara:2013:MODPLA:,Ara:2014:arXiv:,Far-Mor-Pra:2014:PHYSR4:,Far-Lap-Pra:2015:JCAP:} defined by
\begin{equation}
r_{\mathrm{s}} = \frac{r_{\mathrm{g}}}{2y^{1/3}}\,,
\end{equation}
where the gravitational attraction of the central mass source is just balanced by the cosmic repulsion. All test particle (or fluid) bound orbits, e.g., the circular orbits, have to be located inside the static radius sphere \citep{Stu-Sla:2004:PHYSR4:,Stu-Sla-Kov:2009:CLAQG:}.

Detailed discussion of the properties of the general relativistic polytropic spheres in spacetimes with the repulsive cosmological constant has been presented for properly selected values of the polytropic parameter $n$ in \cite{Stu-Hle-Nov:2016:PHYSR4:} where also the special case of uniform density spheres corresponding to the polytropes with $n = 0$ is included. It has been explicitly demonstrated that the role of the cosmological constant could be very strong for largely extended polytropic spheres.
	
It is instructive to relate the three characteristic length scales of the external vacuum spacetime to the polytrope length scale $\mathcal{L}$, the sphere radius $R = \mathcal{L}\xi_{1}$ and its central density. For spheres with very large central density, related to the central densities of very compact objects such as neutron stars or quark stars, the length scale is comparable to the black hole horizon scale (gravitational radius), with decreasing central density the polytrope length scale increases. On the other hand, for the observationally estimated cosmological constant $\Lambda \sim 1.1 \times 10^{-56}$\, cm$^{-2}$, the length scale and extension of all astrophysically relevant polytropic spheres is much lower than the length scale of the cosmological horizon.

For the polytropic spheres with extremely low central energy density that are also extremely extended for sufficiently low relativistic parameter ($\sigma \leq 10^{-4}$, i.e., in the non-relativistic regime), the crucial role plays the length scale of the static radius since it represents the uppermost limit on the extension of the general relativistic polytropic spheres with the cosmological constant \citep{Stu-Hle-Nov:2016:PHYSR4:}. Note that in the case of the relativistic polytropic spheres with large polytropic index ($n > 3.3$), the extremely extended polytropic spheres could exist also for much larger values of the central energy density, and large values of the relativistic parameter ($\sigma > 0.1$), being very close to the critical values implying the special character of the density and pressure profiles --- their extension is also restricted by the static radius \citep{Stu-Hle-Nov:2016:PHYSR4:}. Here we shall discuss in detail the possibility that such extremely extended polytropic spheres could represent the DM halos, with the extension and mass of the halos given by the standard observation estimates. The CDM halos can be represented by non-relativistic polytropes with $\sigma \ll 1$. On the other hand, the hot (warm) DM halos can be represented by relativistic polytropes with $\sigma > 0.1$, if its value is close to the critical value predicting unlimited polytropic spheres for vanishing cosmological constant \citep{Nil-Ugg:2000:ANNPH1:GRStarPoEqSt,Stu-Hle-Nov:2016:PHYSR4:}. Note that the trapping polytropes can be both compact and extremely extended, in dependence on the polytropic index $n$.

We first summarize the limits on the existence of the polytropic spheres related to the presence of the observationally restricted cosmological constant and study the circular geodesics of the internal polytrope spacetime that enable the introduction of a special class of the trapping polytropes.

%%%%%%%%%%%%%%%%%%%%%%%%%%%%%%%%%%%%%%%%%%%%%%%%%%%%%%%%%%%%%%%%%%%%%%%%%%%%%%%%%%%%%%%%%%%%%%%%%%%%%%%%%%%%%%%%%%%%%%%%%%%%%%%%%%%%%%%%%%%%

\section{Constraints on general relativistic polytropes with cosmological constant}\label{Sec:4}
We briefly discuss the construction of the models of the general relativistic polytropes with $n>0$ and demonstrate the dependence of their existence on the cosmological parameter $\lambda$; details can be found in \cite{Stu-Hle-Nov:2016:PHYSR4:}.

By integrating numerically the two differential structure equations for a fixed polytropic index $n>0$, we obtain a sequence of polytropic spheres determined by the central density $\rho_{\mathrm{c}}$, the relativistic parameter $\sigma$, and the cosmological parameter $\lambda$, if the cosmological constant is a free parameter. The observationally fixed value of vacuum energy density, $\rho_{\mathrm{vac}}$, related to the repulsive cosmological constant, $\Lambda \sim 1.1    \times 10^{-56}\,\mathrm{cm^{-2}}$, governs the value of the cosmological parameter $\lambda$ due to the given central density of the polytrope, $\rho_{\mathrm{c}}$, and then it is not a free parameter.

\begin{figure}[tb]
        \centering\includegraphics[width=\linewidth]{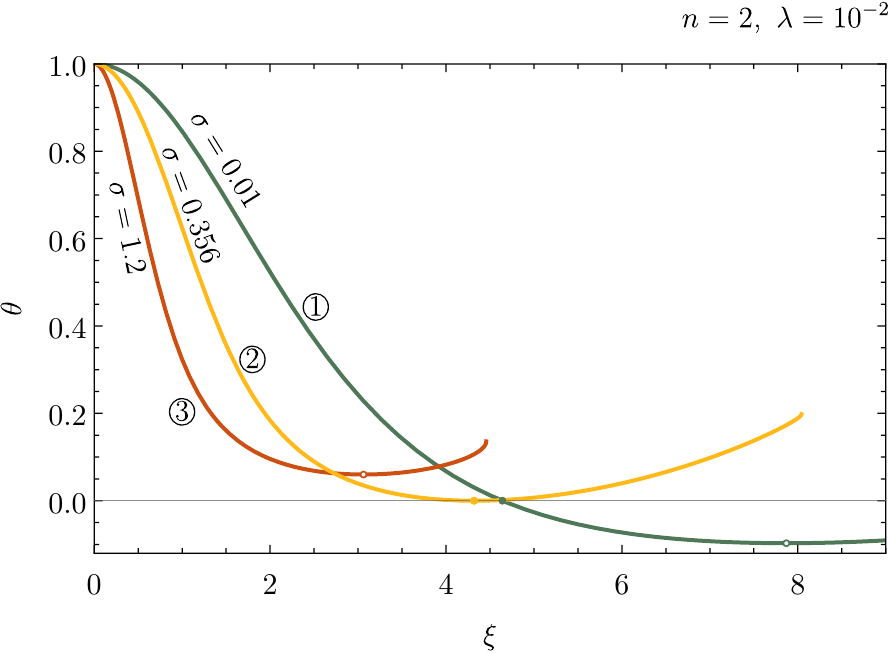}
        \caption{\label{Fig_tlak} Surface of the polytropic sphere is determined by the first zero of $\theta(\xi)$, if exists. Such points are depicted on the plot by solid dots, see curves \circledd{1} and \circledd{2}.The expected $\theta$-curve (corresponding to pressure) is a monotonically decreasing function. As a manifestation of the influence of the cosmological parameter $\lambda$, the $\theta$ function can start to increase even before the first zero is reached, see curve \circledd{3}. Change of monotonicity occurs in local minimum $\mathrm{d}\theta/\mathrm{d}\xi = 0$ and if such minimum occurs at $\xi_1$, where $\theta(\xi_1) = 0$, such point is identified as the \emph{static radius of the external spacetime} of such polytrope $\xi_\mathrm{s} = \frac{3}{2}\frac{\nu(\xi_1)}{\lambda}$, see curve \circledd{2}. If $\theta(\xi_\mathrm{min}) > 0$ at such minimum localized at $\xi_\mathrm{min}$, like on the curve \circledd{3}, the ratio $\xi_\mathrm{min}\Big/\left(\frac{3}{2}\frac{\nu(\xi_\mathrm{min})}{\lambda}\right) > 1$. At the center of each configuration, the value of $\lim_{\xi\to 0}\xi\Big/\left(\frac{3}{2}\frac{\nu(\xi)}{\lambda}\right) = \left(2\lambda\right)^{1/3}$ is smaller then 1 for $\lambda < 1/2$ (we take our attention only to positive cosmological constant). So, we have for curves like \circledd{3} at some point $0 < \xi < \xi_\mathrm{min}$, $\xi = \frac{3}{2}\frac{\nu(\xi)}{\lambda}$. We do not consider such cases in the present paper, postponing them for a future study. 
        (Notice that calculation to the negative values of $\theta$ is possible only for special values of $n$ and the corresponding part of the curve \circledd{1} is shown here for illustration only. Generally, such curves end on $\theta = 0$. Depicted ends of the curves \circledd{2} and \circledd{3} are due to the stiffness problem, which does not need to be overcome in normal situations as we stop calculation at $\xi_1$ or $\xi_\mathrm{min}$.)}
    \end{figure}

The lowest solution $\xi_{1}$ of the equation $\theta(\xi) = 0$ (if it exists, see Fig.~\ref{Fig_tlak}) determines the dimensional radius of the polytropic sphere $R = \mathcal{L} \xi_{1}$. The dimensionless mass parameter $\nu(\xi_{1})$ determines the polytrope gravitational mass $M = \mathcal{M} \nu(\xi_{1})$. The radial profiles of the mass-energy density, pressure, gravitational mass parameter, and the metric coefficients of the polytropic sphere are determined by the functions $\rho(\xi; n, \sigma, \lambda)$, $p(\xi; n, \sigma, \lambda)$, $\nu(\xi;n,\sigma,\lambda)$, $g_{tt}(\xi; n, \sigma, \lambda)$, $g_{rr}(\xi; n, \sigma, \lambda)$ and are described in detail, along with other characteristics such as gravitational and binding energy, in \cite{Stu-Hle-Nov:2016:PHYSR4:}; characteristic examples of the radial profiles of these functions can be found in Appendices~\ref{APPENDIX1} and \ref{APPENDIX2}. Here we concentrate our attention on the extension $R$ and mass $M$ of the polytropic spheres and limits on the allowed values of $R$ and $M$.

%%%%%%%%%%%%%%%%%%%%%%%%%%%%%%%%%%%%%%%%%%%%%%%%%%%%%%%%%%%%%%%%%%%%%%%%%%%%%%%%%%%%%%%%%%%%%%%%%%%%%%%%%%%%%%%%%%%%%%%%%%%%%%%%%%%%%%%%%%%%
%%%%%%%%%%%%%%%%%%%%%%%%%%%%%%%%%%%%%%%%%%%%%%%%%%%%%%%%%%%%%%%%%%%%%%%%%%%%%%%%%%%%%%%%%%%%%%%%%%%%%%%%%%%%%%%%%%%%%%%%%%%%%%%%%%%%%%%%%%%%

\subsection{Limits on the general relativistic polytropes}
\begin{figure*}[ht!]
	   \begin{center}
		  \subfloat[][]{\includegraphics[width=.5\linewidth]{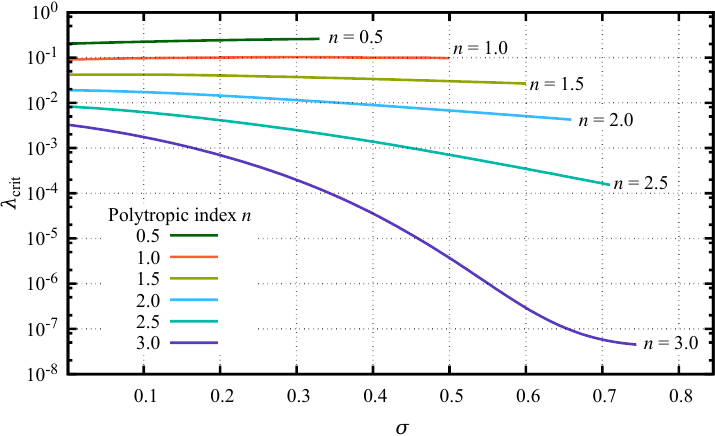}\label{SigLamCrit_1}}
		  \hfill%
		  \subfloat[][]{\includegraphics[width=.5\linewidth]{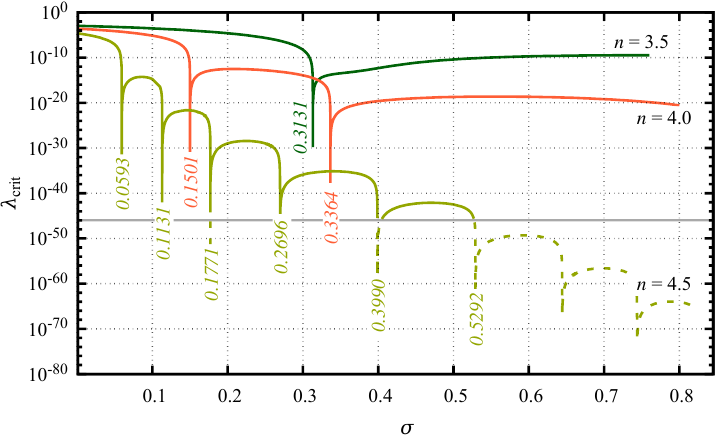}\label{SigLamCrit_2}}
		\end{center}
        \caption{\label{SigLamCrit}Dependence of critical value $\lambda_\mathrm{crit}$ on relativistic parameter $\sigma$. On the left we demonstrate the behaviour for polytropic index $n<3$ and on the right for $n>3$. The gray line on the right highlights the restriction $\lambda = 10^{-46}\,\mathrm{g\,cm^{-3}}$. The configurations for particular polytropic index $n$ can only exists below corresponding curve $\lambda_\mathrm{crit}(n;\sigma)$.}
	\end{figure*}

The role of the cosmological parameter $\lambda$ is concentrated in putting strong limits on the existence of polytropic spheres in dependence on both the polytropic index $n$ and the relativistic parameter $\sigma$. The critical values of the cosmological parameter given by the function $\lambda_{\mathrm{crit}} = \lambda_{\mathrm{crit}}(n,\sigma)$ limit the existence of polytropic spheres; they have been determined by numerical calculations and are illustrated as functions of relativistic parameter $\sigma$ for selected representative values of $n$ in Figs.\,\ref{SigLamCrit}(a) and \ref{SigLamCrit}(b). For fixed values of $n$ and $\sigma$, the critical parameter $\lambda_\mathrm{crit}(n,\sigma)$ determines the minimal value of the central energy density
    \begin{equation}\label{eqCritical}
	   \rho_\mathrm{c}(n,\sigma) = \frac{\rho_\mathrm{vac}}{\lambda_\mathrm{crit}(n,\sigma)} = \frac{1}{\lambda_\mathrm{crit}(n,\sigma)} \frac{\Lambda c^2}{8\pi G}
	\end{equation}
allowing for the existence of static polytropic spheres. For $\sigma<\sigma_\mathrm{c(min)}(n,\sigma)$ static polytropes cannot exist due to cosmic repulsion; for $\sigma = \sigma_\mathrm{c(min)}$ we have a limiting configuration with the surface located at the static radius. In the following, we assume $\Lambda \sim 10^{-56}\,\mathrm{cm}^{-2}$ given by the observational restrictions.

The polytropic spheres are allowed at regions of the parameter space located below the critical curves. Character of the critical function $\lambda_{\mathrm{crit}}(n,\sigma)$ strongly depends on the value of the polytropic index $n$. Generally, it increases with $n$ decreasing. The critical function $\lambda_{\mathrm{crit}}(\sigma;n)$ demonstrates two characteristic regimes of its behaviour, in dependence on the polytropic index $n$. Since it is known that for $n > 3.3$, the critical points of the relativistic parameter exist, denoted as $\sigma_{f(n)}$, when no limit on the extension of the polytropic sphere exists, and for $n \geq 5$ no static polytropic spheres can exist for any values of $\sigma$ \citep{Nilsson2000}, we construct examples of the critical function $\lambda_{\mathrm{crit}}(\sigma;n)$ in the first regime, for $n \leq 3$, in Fig.~\ref{SigLamCrit}(a), and examples of the $\lambda_{\mathrm{crit}}(\sigma;n)$ function in the second regime, for $3.4 < n < 5$, in Fig.~\ref{SigLamCrit}(b). In Fig.~\ref{SigLamCrit}(a) we cover the standard values of the polytropic index, for the non-relativistic fluid with $n = 3/2$, and $n = 3$ for the ultrarelativistic fluid; we add the critical functions for some values $n < 3/2$, namely $n = 0.5,\, 1$, and for $n = 2,\, 2.5$. In Fig.~\ref{SigLamCrit}(b) we give construction of the critical function $\lambda_{\mathrm{crit}}(\sigma;n)$ for the values of the polytropic index $n$ implying the existence of the critical values of the relativistic parameter, $\sigma_{f(n)}$, when the limits on the existence of polytropic spheres vanish for vanishing cosmological parameter $\lambda$; the values of $n = 3.5,\, 4,\, 4.5$ are chosen. Extension of the critical curves is in all the considered cases restricted by the value of the relativistic parameter $\sigma$ corresponding to the equality of the velocity of sound and the velocity of light (the so-called causality limit) --- for details see \cite{Stu-Hle-Nov:2016:PHYSR4:}.

In the first regime (Fig.~\ref{SigLamCrit}(a)), for $n \leq 3$, the function $\lambda_{\mathrm{crit}}(\sigma;n)$ slightly monotonically decreases with $\sigma$ increasing; it is limited by the value of $\lambda_{\mathrm{crit}} = 10^{-7}$ even in the limit of $\sigma \rightarrow 1$. In the special case of $n = 3$ it decreases from the starting point $\lambda_{\mathrm{crit}}(\sigma = 0; n = 3) = 3 \times 10^{-3}$ down to $\lambda_{\mathrm{crit}}(\sigma = 0.7; n = 3) = 10^{-7}$ and remains constant with increasing values of $\sigma$.

In the second regime (Fig.~\ref{SigLamCrit}(b)), for $n > 3.3$, the function $\lambda_{\mathrm{crit}}(n;\sigma)$ loses its monotonic character, and there are forbidden polytropes for some special values of the relativistic parameter $\sigma$ in dependence on the polytrope index, as they should have infinite extension. For $n = 3.5$, the polytropic spheres are forbidden for one specific value of $\sigma_{1\,(n = 3.5)} = 0.3131$. For $n = 4$, there are two specific forbidden values of $\sigma_{1\,(n = 4)} = 0.1501$, $\sigma_{2\,(n = 4)} = 0.3364$. A third forbidden configuration with $n = 4$ corresponds to $\sigma$ breaking the causality limit at $\sigma_{3\,(n = 4)} = 0.8341$, and is not considered here. For $n = 4.5$, there exists an ensemble of 8 critical values of the relativistic parameter, namely: $\sigma_{1\,(n = 4.5)} = 0.0593$, $\sigma_{2\,(n = 4.5)} = 0.1131$, $\sigma_{3\,(n = 4.5)} = 0.1771$, $\sigma_{4\,(n = 4.5)} = 0.2696$, $\sigma_{5\,(n = 4.5)} = 0.399$, $\sigma_{6\,(n = 4.5)} = 0.5292$, $\sigma_{7\,(n = 4.5)} = 0.6510$, $\sigma_{8\,(n = 4.5)} = 0.7440$. Note that the critical points can be considered as realistic if the critical curves located between these points give realistic values of the cosmological parameter, $\lambda > 10^{-46}$. For this reason, the last three critical points can be excluded from our consideration.

For the non-relativistic polytrope spheres with relativistic parameter $\sigma < 0.1$, under the first critical value of $\sigma$, the critical function $\lambda_{\mathrm{crit}}(\sigma;n) > 10^{-5}$ for all the polytropic indexes $n<5$. In such situations the polytropic spheres with very small central density and extension close to the static radius have their structure influenced by the repulsive cosmological constant (see \cite{Stu-Hle-Nov:2016:PHYSR4:}). For the $n = 3.5$ polytrope, at $\sigma > \sigma_{1\,(n = 3.5)}$, there is $\lambda_{\mathrm{crit}}(\sigma;n) < 10^{-9}$. For the $n = 4$ polytrope, at $\sigma_{1\,(n = 4)} < \sigma < \sigma_{2\,(n = 4)}$, the critical function $\lambda_{\mathrm{crit}}(\sigma;n) < 10^{-12}$, while at $\sigma > \sigma_{2\,(n = 4)}$, there is $\lambda_{\mathrm{crit}}(\sigma; n) < 10^{-19}$. A similar behaviour occurs for the $n = 4.5$ polytropes, when the $\sigma$-profiles of the critical function $\lambda_{\mathrm{crit}}(\sigma;n)$ demonstrate maxima in between the critical values of the relativistic parameter, having the maxima values strongly decreasing with increasing number $f(n)$ of the order of the critical value of the relativistic parameter. Because it is relevant physically to consider also the restriction on the cosmological parameter $\lambda$ related to the physically acceptable central density of the polytropic configuration that should correspond to the central density of neutron or quark stars, we have to take into account the central density $\rho_{\rm c} < \rho_\mathrm{NS} \sim 10^{17}\,$g\,cm$^{-3}$, and having $\rho_\mathrm{vac} \sim 10^{-29}\,$g\,cm$^{-3}$, we obtain the restriction $\lambda > 10^{-46}$ on the physical relevance of the critical curves.

\subsection{Extension and mass of the polytropes}
The basic global characteristics of the general relativistic polytropes are given by their extension and mass. The extension of the polytropic spheres is governed by the length scale factor $\mathcal{L}$ and the dimensionless radius $\xi_1$, while their mass is governed by the mass scale factor $\mathcal{M}$ and the dimensionless mass parameter $\nu_1 = \nu(\xi_1)$. The values of $\xi_1$ and $\nu(\xi_1)$ are solutions of the polytrope structure equations.

The dependencies of the polytrope extension parameter $\xi_{1}$ and the polytrope mass parameter $\nu_{1} = \nu(\xi_{1})$ on the parameters $n$, $\sigma$ and $\lambda$ are extensively discussed in \cite{Stu-Hle-Nov:2016:PHYSR4:}, where it has been demonstrated that the static polytropic spheres cannot have extension exceeding the static radius of the external spacetime, given by their mass and the cosmological constant through the relation
	\begin{equation}
	   r_{\mathrm{s}} = \left(\frac{3GM}{2 c^2\Lambda}\right)^{1/3}\, .
	\end{equation}
This crucial result supports the indications that the gravitationally bounded systems in the expanding universe with repulsive cosmological constant cannot exceed the static radius, obtained in the framework of the Einstein--Strauss-de Sitter vacuola model \citep{Stu:1983:BULAI:,Stu:1984:BULAI:,Stu-Hle:1999:PHYSR4:,Stu-Sch:2011:JCAP:CCMagOnCloud}.

The cosmic repulsion influences the global parameters of the polytropic spheres and the radial profiles of their energy density, pressure, and metric coefficients, for very extended polytropic spheres. Such polytropes occur in the first regime, if the cosmological parameter is high enough, namely, $\lambda > 10^{-9}$, corresponding thus to polytropes with very low central energy density that are also non-relativistic, with the relativistic parameter $\sigma < 10^{-4}$. The role of the cosmological parameter $\lambda$ increases with increasing polytropic index $n$. For the special case of the polytropic spheres under the second regime, with $n > 3.3$, the influence of the cosmic repulsion can be relevant also for polytropes with relatively large central densities, having $\lambda < 10^{-18}$, where the fluid is relativistic, with the relativistic parameter $\sigma > 0.1$ being close to the critical values implying extremely large configurations. Detailed discussion of the properties of the extremely extended and low-density polytropic spheres, along with the influence of the cosmic repulsion on their structure, can be found in \cite{Stu-Hle-Nov:2016:PHYSR4:}, where, on the other hand, also extremely compact spheres with extremely high central density are treated.

Here we focus attention on the extremely extended polytropic spheres --- we test if their extension and mass can be well fitted, for the observationally fixed value of the cosmological constant, to the extension and mass of the DM halos assumed in large galaxies or in galaxy clusters. Such extremely extended polytropes cannot cross the static radius of the external SdS spacetime, in agreement with the idea that the static radius represents the uppermost boundary of the gravitationally bounded systems in spacetimes with the repulsive cosmological constant \citep{Stu-Sla-Kov:2009:CLAQG:,Stu-Sch:2011:JCAP:CCMagOnCloud}. We expect that above the static radius cosmic repulsion prevails and any polytropic spherical configuration has to be expanding. 

We realize the fitting procedure in both regimes of the behaviour of the polytropic spheres that predict very extended polytropic configurations. However, before realizing the fitting procedures, we study two important properties of the polytropic structures. First, we discuss the geodesic structure of their internal spacetime in order to enable testing of the galaxy rotational curves by the properties of the circular geodesics of the polytrope spacetime; regions of stable circular geodesics are determined, and a detailed discussion of the null geodesics allows the determination of the trapping polytropes, containing stable null circular geodesics, across the parameter space of the polytropic spheres. Second, we discuss in detail the dynamic stability of the polytropic spheres against the radial perturbations. 

%%%%%%%%%%%%%%%%%%%%%%%%%%%%%%%%%%%%%%%%%%%%%%%%%%%%%%%%%%%%%%%%%%%%%%%%%%%%%%%%%%%%%%%%%%%%%%%%%%%%%%%%%%%%%%%%%%%%%%%%%%%%%%%%%%%%%%%%%%%%
\section{Circular geodesics of the internal polytrope spacetime}

\begin{figure*}
  \centering
  \includegraphics[width=0.33\textwidth]{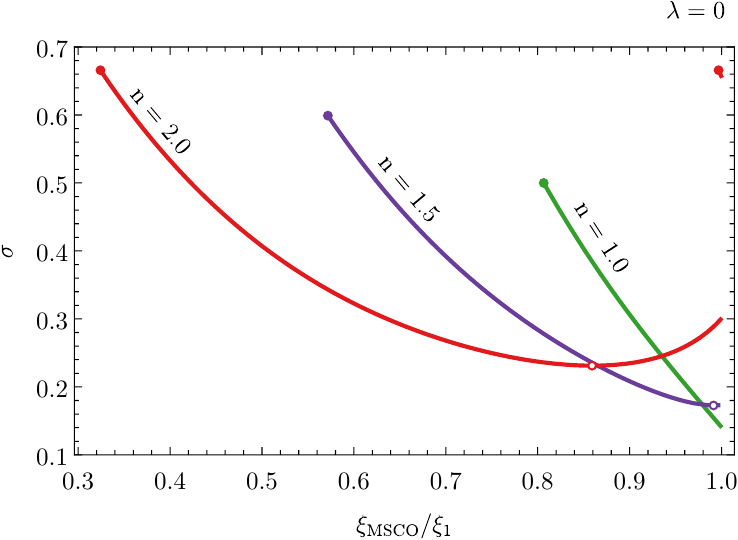}\hfill\includegraphics[width=0.33\textwidth]{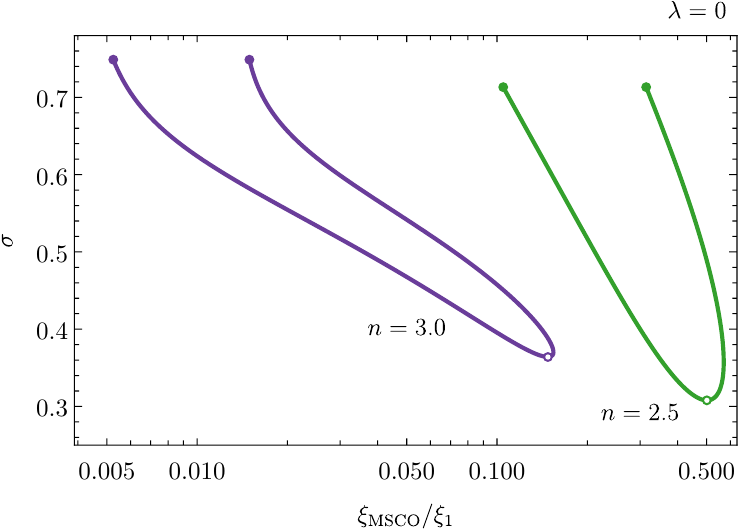}\hfill\includegraphics[width=0.33\textwidth]{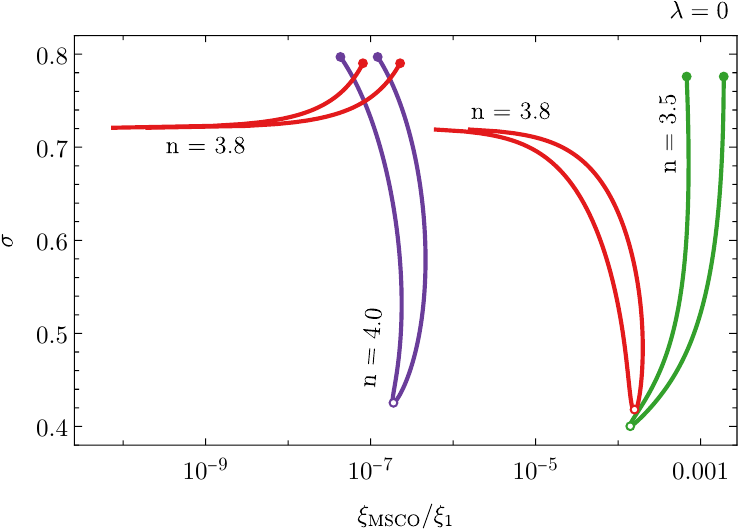}
  \caption{Relative positions of marginally stable circular orbits shown for several values of polytropic parameter $n$. For $n = 3.8$ we can see the onset of extremely large polytropes for higher values of parameter $\sigma$, since values of the absolute positions $\xi_\mathrm{MSCO}$ are only slightly affected by $\sigma$ and all remain in the interval $(0.8,\,2.4)$. Solid circles show the maximum considered value of sigma  $\sigma_\mathrm{causal}\equiv n/(n+1)$.}\label{refMSCO}
\end{figure*}

We consider the velocity curves of the stars in the galaxy plane that could be selected as the equatorial plane, $\theta = \pi/2$, of the spherically symmetric internal spacetime of the polytropic sphere. Then the circular geodesic motion of the stars can be characterized by two constants of motion --- specific energy $E$ and specific angular momentum $L$, which are defined as the ratio of the covariant energy (angular momentum) related to the rest energy of the star $m$ that is by itself a constant of the motion. The equations of the equatorial geodesic motion can then be given in general spherically symmetric static spacetime in the separated and integrated form as follows
    \begin{align}
        \frac{\dd t}{\dd w} & = \frac{E g_{\phi\phi}}{-g_{tt}g_{\phi\phi}}\, ,\\
        \frac{\dd \phi}{\dd w} & = \frac{L}{g_{\phi\phi}}\, , \\
        g_{rr}\left(\frac{\dd r}{\dd w}\right)^2 & = R(r)\, ,
    \end{align}
where $w$ is the proper time and the function governing the radial motion reads
    \begin{equation}
        R(r) = -1 + \frac{E^2 g_{\phi\phi} + L^2 g_{tt}}{-g_{tt}g_{\phi\phi}}\, .
    \end{equation}
The turning points of general radial motion can be determined by the effective potential introduced by the relation
    \begin{equation}
        E^2 = V_\mathrm{eff}(r) \equiv -g_{tt}\left(1 + \frac{L^2}{g_{\phi\phi}}\right)\, .
    \end{equation}
The circular geodesics can be determined directly from the function $R(r)$ or by the local extrema of the effective potential $V_\mathrm{eff}$.

Using the conditions of the circular motion, $R = 0$ and $\dd R/\dd r = 0$, we can express the constants of the motion $E$, $L$ and the angular velocity relative to the distant observers, $\Omega$, in terms of the metric coefficients in the form
    \begin{eqnarray}
        E & = & \frac{-g_{tt}}{\sqrt{-g_{tt}-g_{\phi\phi}\Omega^2}}\, , \\
        L & = & \frac{g_{\phi\phi}\Omega}{\sqrt{-g_{tt}-g_{\phi\phi}\Omega^2}}\, ,
    \end{eqnarray}
    \begin{equation}
        \Omega \equiv \frac{\dd \phi}{\dd t} = \frac{\sqrt{-g_{tt,r}g_{\phi\phi,r}}}{g_{\phi\phi,r}}\, ,
    \end{equation}
where $,r$ ($,rr$) denotes the first (second) derivative in the radial direction. The condition of marginal stability of the circular geodesics, $R_{,rr} = 0$, can be expressed in the form
    \begin{equation}
        E^2 g_{\phi\phi,rr} + L^2 g_{tt,rr} + (g_{tt}g_{\phi\phi})_{,rr} = 0\, .
    \end{equation}
The velocity profile of the circular geodesics can be then determined by a simple formula
    \begin{equation}
        v(r) = \frac{L}{\sqrt{g_{\phi\phi}}}\, .
    \end{equation}

In the internal polytrope spacetime, where the metric coefficient $g_{tt} = -\eto{2\Phi}$ is governed by the structure function $\theta(\xi)$ and the polytrope parameters, while $g_{\phi\phi} = \mathcal{L}^2\xi^2$, the effective potential takes the form
    \begin{multline}
        V_\mathrm{eff}(\xi) = (1 + \sigma\theta )^{-2(n + 1)} \left\{\vphantom{\frac{1}{3}} 1 - \right. \\
        \left. -2(n + 1)\sigma \vphantom{\frac{1}{2}} \left[\frac{\nu(\xi_{1})}{\xi_{1}} + \frac{1}{3}\lambda\xi_{1}^{2}\right] \right\} \left(1 + \frac{L^2}{\mathcal{L}^2\xi^2}\right) \, .
    \end{multline}

The radial profile of the specific energy of the circular geodesics takes the form
    \begin{multline}
        E^2(\xi) = (1 + \sigma\theta)^{-2(n + 1)} \left\{1 - 2(n + 1)\sigma \vphantom{\frac{1}{2}} \times\right.\\
        \times\left.\left[\frac{\nu(\xi_{1})}{\xi_{1}} + \frac{1}{3}\lambda\xi_{1}^{2} \right] \right\} \left[1 + \frac{(n + 1)\sigma\xi}{1 + \sigma\theta}\frac{\dd\theta}{\dd\xi}\right]^{-1}\, ,
    \end{multline}
the radial profile of the specific angular momentum reads
    \begin{equation}
        L^2(\xi)/\mathcal{L}^2 = -\frac{(n + 1)\sigma \xi^3\frac{\dd\theta}{\dd\xi}}{1 + \sigma\theta} \left[1 + \frac{(n + 1)\sigma \xi}{1 + \sigma\theta}\frac{\dd\theta}{\dd\xi}\right]^{-1} \, ,
    \end{equation}
and the angular velocity relative to the coordinate time that can be close to the proper time of distant static observers reads \footnote{For simplicity, the static observer is assumed close to the static radius of the polytropic spacetime, where the influence of the cosmic expansion can be considered negligible, and the proper time of the observer is close to the coordinate time. The view related to observers approaching the cosmological horizon under the cosmic repulsion can be found in \cite{Stu-Char-Sche:2018:EPJC,Stu-Char:2024:PhRvD}.}
    \begin{equation}
        \Omega^2(\xi) \mathcal{L}^2 = \frac{-\left\{1 - 2(n + 1)\sigma\left(\frac{\nu(\xi_1)}{\xi_1} + \frac{\lambda}{3}\xi_1^2\right)\right\}(n + 1)\sigma\frac{{\mathrm{d}\theta}}{{\mathrm{d}\xi}}}{\xi(1 + \sigma\theta)^{2n + 3}} \, .
    \end{equation}
Note that the angular velocity related to the proper time of the particle following a circular orbit is given by the relation 
\begin{equation}
        \omega = \frac{\dd\phi}{\dd w} = \frac{L}{r^2} = \frac{L}{\mathcal{L}^2 \xi^2}\, . 
\end{equation}

Recall that the relevant structure function satisfies the condition $\mathrm{d}\theta/\mathrm{d}\xi < 0$ that guarantees positiveness of the radial profiles of the circular geodesic functions $E^2, L^2, \Omega^2$. Characteristic examples of the specific energy and specific angular momentum radial profiles of the circular geodesics of the internal spacetime of the polytropic spheres are presented in Appendices~\ref{APPENDIX1} and \ref{APPENDIX2}. 

The condition for the marginally stable circular geodesics takes the form
    \begin{equation}\label{eqMSCO}
      \left(1 + \sigma  \theta\right) \xi \frac{\dd^2 \theta}{\dd \xi^2}  +   \left[3\left( 1+  \sigma  \theta\right) + (1 + 2n) \sigma \xi \frac{\dd\theta}{\dd \xi}\right] \frac{\dd\theta}{\dd \xi} = 0\, .
    \end{equation}
This complex condition can be treated only numerically and its solution is shown on Fig.~\ref{refMSCO} for several values of parameter $n$ where we assume for simplicity $\lambda = 0$. We can see relatively complex dependence on the polytropic parameter $n$. For sufficiently low values ($n \lesssim 1.4$) there is only one, the innermost value; for large values of $n$, two MSCO exist; the exceptional case is $n = 3.8$ where two branches of MSCO exist, each governing the inner and outer marginal orbit. The split into two branches is a consequence of the presence of $\sigma_\mathrm{crit}$ in the range of $\sigma$ for which MSCOs exist.

The radial profile of the orbital velocity of the circular geodesics that should be relevant in comparison of the polytrope model predictions with the observational data reads
    \begin{equation}
        v(\xi) = \left[\frac{-(n + 1)\sigma \xi \frac{\dd\theta}{\dd\xi}}{1 + \sigma\theta}\right]^{1/2} \left[1 + \frac{(n + 1)\sigma\xi}{1 + \sigma\theta}\frac{\dd\theta}{\dd\xi}\right]^{-1/2}\, .
    \end{equation}

Notice that all the functions characterizing the circular geodesics are governed only by the structure function $\theta(\xi)$, being independent of the ``mass'' structure function $\nu(\xi)$. They are influenced only by the parameter $\nu(\xi_1)$ characterizing the total mass of the polytropic sphere; similarly, the cosmological constant occurs explicitly only in the constant term $\lambda \xi_{1}^2$. Of course, the influence of the cosmological constant is implicitly contained in the structure function $\theta(\xi)$ that is a solution of the structure equation depending on the cosmological constant. As explicitly shown in \cite{Stu-Hle-Nov:2016:PHYSR4:}, the role of the cosmological constant is crucial for extended polytropes that are relevant in our attempt to use them in modeling dark matter halos.

\subsection{Trapping polytropes}

The analysis of the circular geodesics of the internal polytrope spacetime enables the introduction of the so-called trapping polytropic spheres, i.e., polytropes containing a central region of trapped null geodesics \citep{Stu-Hle-Nov:2016:PHYSR4:,Nov-Hla-Stu:2017:PRD:trapping}. Since the region of trapped null geodesics is unstable against gravitational perturbations, leading to gravitational collapse forming a central black hole, and is substantially smaller in comparison to the whole trapping polytrope \citep{Stu-etal:2017:JCAP:}, the trapping polytropes could potentially serve as an alternative explanation for the existence of the supermassive black holes in galaxies observed at redshifts $z > 6$.

\begin{figure}[tb]
    \centering
    \includegraphics[width=\linewidth]{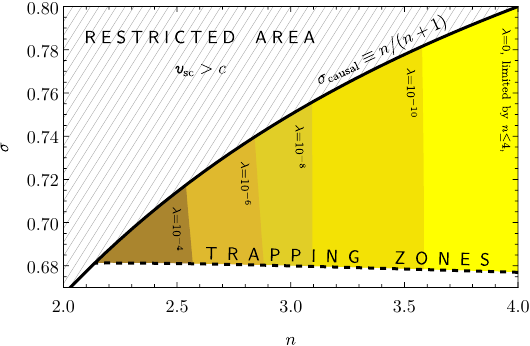}
    \caption{\label{fig:n-sigma-lambda} Trapping region in $n$--$\sigma$ parameter space. For spatially finite configurations, possible pairs $(n,\sigma)$ are additionally  bounded in dependence on the value of $\lambda$ (see Fig.~\ref{SigLamCrit}).}
\end{figure}

\begin{figure}[tb]
\centering\includegraphics[width=\linewidth]{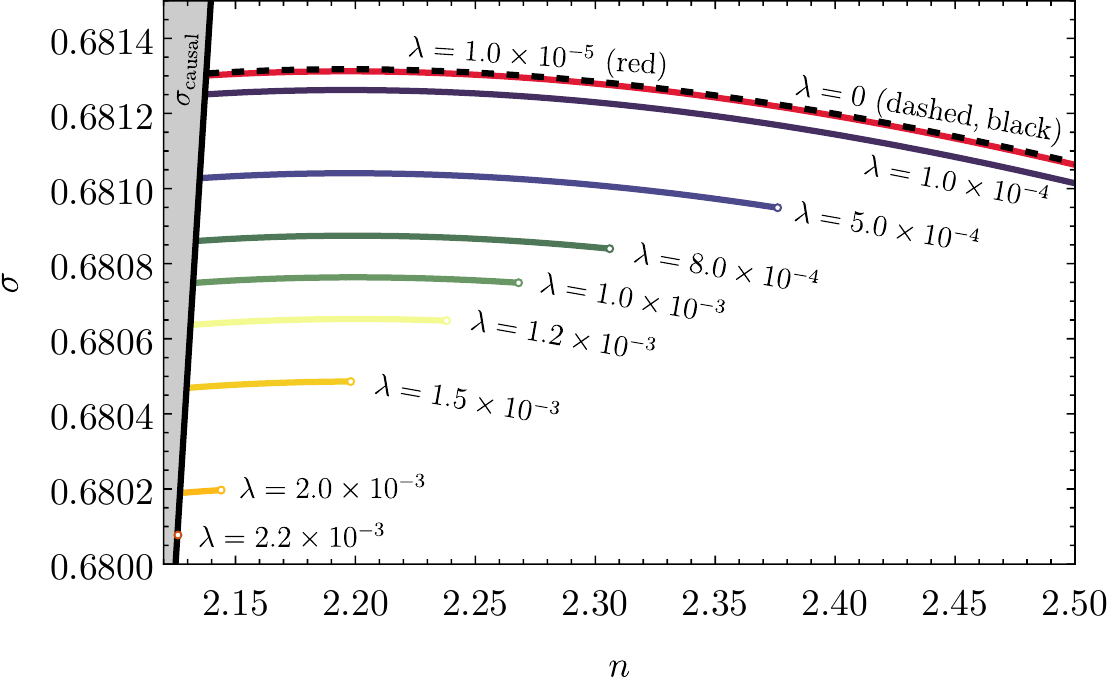}
\caption{\label{Fig_ZACHhranice} Plot represents minimal values of the parameter $\sigma$ with respect to parameters $n$ and $\lambda$ for which the trapping phenomenon occurs. Their limitations with respect to the parameter $n$ is set by considered physically plausible values of $\sigma_\mathrm{causal}$ and by the requirement of existence of polytropic configurations (rings at the end of the lines; using depicted scope visible up to $\lambda = 5\times 10^{-4}$). Borderlines for common values of $n$ for cases $\lambda < 10^{-5}$ differs only negligibly from the case $\lambda = 0$.}
\end{figure}

For null geodesics the effective potential takes the form
    \begin{multline}
        V_\mathrm{null}(\xi) = (1 + \sigma\theta)^{-2(n + 1)}\times \\
        \left\{1-2(n + 1)\sigma\left[\frac{\nu(\xi_{1})}{\xi_{1}} + \frac{1}{3}\lambda\xi_{1}^{2}\right] \right\} \frac{L^2}{\mathcal{L}^2\xi^2}\, .
    \end{multline}
The local extrema of this effective potential are given by the condition
    \begin{equation}
        1 + \frac{(n + 1)\sigma\xi}{1 + \sigma\theta}\frac{\dd\theta}{\dd\xi} = 0
    \end{equation}
that, as expected, corresponds to the loci of simultaneous divergence of the specific energy and specific angular momentum of circular geodesics. These extrema govern the existence of circular null geodesics that in the case of the internal polytrope spacetimes come generally as twins --- for properly chosen values of the parameters $n$, $\sigma$, $\lambda$ a stable inner (unstable outer) null circular geodesics exist in the trapping polytropes. The null circular geodesics first occur when $\dd^2V_\mathrm{null}/\dd\xi^2 = 0$, implying the condition
    \begin{multline}
         3 + \sigma\left\{3\sigma\theta^2 + (n + 1)\xi\left[\frac{\dd\theta}{\dd\xi}\left(4 + (2n + 3)\sigma\xi\frac{\dd\theta}{\dd\xi}\right) - \xi\frac{\dd^2\theta}{\dd\xi^2}\right]\right. + \\
        \left. + \theta\left[6 + (n + 1)\sigma\xi\left(4\frac{\dd\theta}{\dd\xi} - \xi\frac{\dd^2\theta}{\dd\xi^2}\right)\right] \right\}= 0\, .
    \end{multline}
We determine by using a computational code the region in the parameter space where the trapping polytropes exist. Numerical calculations reflecting the influence of the cosmological parameter on the trapping polytropes are represented in Fig.~\ref{fig:n-sigma-lambda}  and in Fig.~\ref{Fig_ZACHhranice}. As expected, with increasing parameter $\lambda$ the restrictions are corresponding to decreasing polytropic parameter $n$ and decreasing relativistic parameter $\sigma$. However, strong differences in restrictions on the trapping polytropes in the $\sigma$--$n$ plane arise for the values of $\lambda > 10^{-5}$; above this line the results are indistinguishable from the limit $\lambda = 0$ case --- see Fig.~\ref{Fig_ZACHhranice}. The detailed study of the trapping polytropes in dependence on the cosmological parameter is planned for the future paper. Here we make comments on the application of the results obtained in the limiting case $\lambda = 0$ in \cite{Stu-etal:2017:JCAP:} on the possibility of gravitational collapse of the trapping region and its consequences. First, we summarize the previous results. 

In the case of $\lambda = 0$, the regions of the parameter space $n$--$\sigma$ corresponding to the trapping polytropes were obtained in \cite{Nov-Hla-Stu:2017:PRD:trapping} --- the regions start at $n = 2.12$ and $\sigma =0.680$ and are considered up to $n = 4$ where the range of the relativistic parameter $0.667 < \sigma < 0.8$. In \cite{Stu-etal:2017:JCAP:} it was demonstrated that in all of these trapping polytropes the trapping regions are unstable relative to gravitational perturbations and gravitational collapse leading to the creation of a black hole. The fate of the system of the created black hole and the remaining polytrope depends on the extension of the trapping region in relation to the complete polytrope sphere. If the black hole represents only a minor part of the polytrope, we can expect possible stabilization and modification of the polytrope remnants by rotational effects or other phenomena. On the other hand, if the trapping region represents a relatively large part of the polytrope sphere, we can expect fast collapse of the polytrope remnant onto the created black hole. Detailed analysis in \cite{Stu-etal:2017:JCAP:} shows that for polytropes with polytropic index $n < 3.5$ the ratio of the trapping region extension to the whole polytrope extension is decreasing from $0.1$ for $n \sim 2.2$ to $0.001$ for $n \sim 3.5$, indicating possible fast conversion of the whole polytrope to a black hole, while for $n \geq 3.8$ the ratio is falling under the value of $10^{-7}$ suggesting the possibility of the stabilization of the remaining part of the polytrope. (In the intermediate region of polytropes with $3.5 < n < 3.8$ the ratio is falling from $10^{-3}$ to $10^{-7}$ --- for details see \cite{Stu-etal:2017:JCAP:}.) The ratio of the mass inside the trapping region to the total mass of the polytrope demonstrates two sections \cite{Stu-etal:2017:JCAP:}: in the first section, the ratio evolves from $0.55$ for $n\sim2.2$ to $0.03$ for $n \sim 3.75$ suggesting the conversion of the whole polytrope to a black hole, while in the second section the ratio is $\sim 10^{-3}$ for $3.8 < n < 4$. The second section is thus interesting for the explanation of supermassive black holes in the active nuclei in the very early galaxy structures. 

On the base of the summary of results obtained for the case $\lambda=0$, we are able to give estimates of the influence of the parameter $\lambda$ presented in our results summarized in Fig.~\ref{fig:n-sigma-lambda} and Fig.~\ref{Fig_ZACHhranice}. As the equations governing local extrema of the effective potential of null geodesics, and the extension of the trapping region, do not explicitly contain the cosmological constant, being dependent on $\lambda$ only implicitly due to the function $\theta(\xi)$, we can assume, because of the results obtained above, that for sufficiently low values of the parameter $\lambda$ the extension of the trapping region and the ratio of the mass contained in the trapping region and the total polytrope mass are given with high precision by the analysis done for the $\lambda = 0$ case. We can thus conclude that in the physically interesting cases connected with the trapping polytropes having polytrope index $3.8<n<4$, the relation of the mass $M_\mathrm{tr}$ contained in the trapping region and the mass of the whole polytrope mass $M$, the DM halo mass $M_\mathrm{Halo}$ is estimated as $M_\mathrm{tr}/M\sim{} M_\mathrm{tr}/M_\mathrm{Halo}\sim{}10^{-3}$ as demonstrated in detail in \cite{Stu-etal:2017:JCAP:}). Numerical calculations for some cases with inclusion of the sufficiently low cosmological parameter $\lambda$ confirm this expectation of validity of creation of supermassive black holes in extremely extended dark matter halos and their clusters. Therefore, we can expect supermassive black holes of mass going from $M_\mathrm{BH-galaxy}\sim 10^{9}M_{\odot}$ in large galaxy halos of mass $M_\mathrm{galaxy}\sim 10^{12}M_{\odot}$ to the black hole mass $M_\mathrm{BH-cluster}\sim 10^{12}M_{\odot}$ in halos of large galaxy clusters with mass $M_\mathrm{cluster}\sim 10^{15}M_{\odot}$. Of course, a more detailed study of the role of parameter $\lambda$ across the whole region of trapping polytropes is planned for future work.

\subsection{Geodesic structure of the polytropes in dependence on their parameters}
Now we have to find families of the general relativistic polytropes allowing for the existence of stable circular geodesics and the possibility of related unstable circular geodesics, governing the circular motion of test particles. These can correspond to the circular motion of baryonic matter in the spacetime background given by the polytropes.

We give an analysis of the timelike circular geodesics in dependence on the three free parameters of the polytropic models: polytropic index $n$, relativistic parameter $\sigma$, and central density $\rho_\mathrm{c}$ or equivalently the parameter $\lambda$ related to the observationally constrained cosmological constant (vacuum energy) considered in the present paper. Note that the circular geodesics give an illustrative insight into the character of the gravitational field of the internal polytrope spacetime, especially the specific energy of the circular geodesics representing the magnitude of the gravitational binding at a given radius of the polytropic configurations.

\begin{figure}
    \centering
    \includegraphics[width=\linewidth]{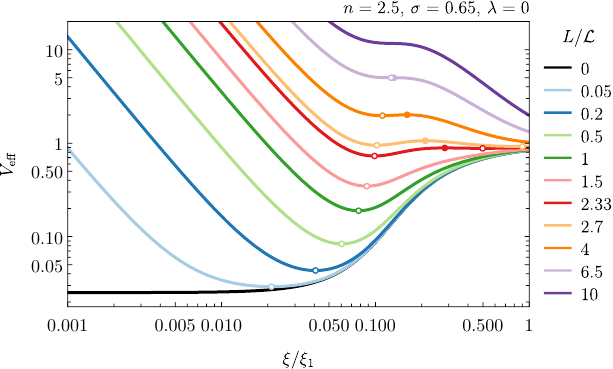}
    \caption{\label{fig:effective-potentila-L} Radial profile of the function $V_\mathrm{eff}$. Notice the change of its behaviour relative to the value of parameter $L/\mathcal{L}$.}
\end{figure}

\begin{figure}
    \centering
    \includegraphics[width=\linewidth]{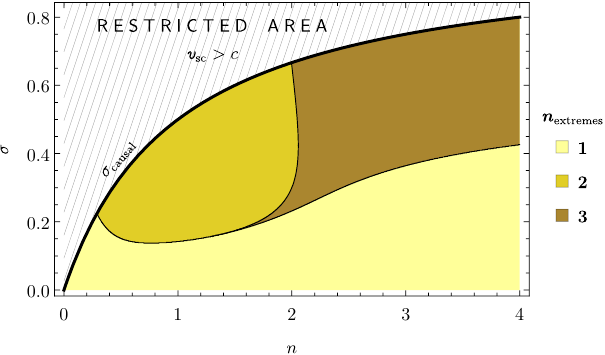}
    \caption{\label{fig:nsigma-types} Figure depicting the maximum number of extremes of the function $V_\mathrm{eff}$ with respect to the value of parameter pair $(n,\sigma)$. Parameter $L$ (on which $V_\mathrm{eff}$ is dependent also) has been taken in such way to maximize this number.}
\end{figure}

We first give the possible types of the behaviour of the effective potential that imply classification of the polytrope spacetime according to the character of distribution of the circular orbits across the spacetime. The representative sequences of the effective potential are presented in Fig.~\ref{fig:effective-potentila-L}. The effective potentials are constructed for a polytrope selected to contain all the possible cases of the behaviour of the effective potential. Therefore, there are potential curves having one (stable) equilibrium point, two equilibrium points (the inner stable and outer unstable), three equilibrium points (inner and outer stable, mediate unstable), or no equilibrium point. The parameter space of the polytropes can be separated into three regions: the first corresponds to polytropes allowing only for one stable point, having one region of stable circular orbits; the second allows for the existence of two equilibrium points where the region of stable circular orbits is restricted by the limiting outer marginally stable orbit, and the third allows even for three equilibrium points, where two separated regions of stable circular orbits can exist, being separated by a region of unstable orbits. The results of computational code for separating the different types of polytropes from the viewpoint of the properties of circular geodesics are presented in Fig.~\ref{fig:nsigma-types}; the separation procedure requests very precise calculations because of very small differences in the magnitude of the effective potential local extrema. Therefore, we give the separation in the parameter space for a fixed value of $\lambda = 0$.  

The representative radial profiles of $E$ and $L$ are given later in the discussion of the polytrope properties (in the Appendix --- Figs.~\ref{fig:sp_case1}, \ref{fig:sp_case2}, \ref{fig:sp_case1_reg2}, and \ref{fig:sp_case2_reg2}).

Of crucial interest is the mapping of the boundary in the parameter space, separating the polytropes where only the stable circular geodesics exist from those where also unstable circular geodesics are allowed. Of course, from the astrophysics point of view, the regions of stable circular geodesics are only relevant for the motion of matter in dark matter halos. The region of parameters allowing for the existence of marginally stable (and consequently also unstable circular geodesics) is generated by a numerical code, and the results are reflected in Fig.~\ref{fig:nsigma-types}.

As expected intuitively, the non-relativistic polytropes allow only stable circular geodesics across their whole internal spacetime; for the existence of regions of unstable circular geodesics, the relativistic parameter $\sigma$ has to be sufficiently high.

It should be stressed that the radial profiles of the velocity of circular geodesics can be directly related to the velocity profiles observed at galaxies (or galaxy clusters) only at their outer regions. The inner regions could give relevant information too, but we have to keep in mind that in these regions they are masked by the significant contribution of visible, or more generally, standard baryonic matter to the gravitational potential of the galaxy (cluster). In the case of the trapping polytropes, and their relation to the high-redshift galaxies, an even more complex situation occurs because of both the expected formation of the central black hole and the lack of observational data that could be improved by the planned space mission ATHENA.

%%%%%%%%%%%%%%%%%%%%%%%%%%%%%%%%%%%%%%%%%%%%%%%%%%%%%%%%%%%%%%%%%%%%%%%%%%%%%%%%%%%%%%%%%%%%%%%%%%%%%%%%%%%%%%%%%%%%%%%%%%%%%%%%%%%%%%%%%%%%
\section{Polytrope stability against radial pulsations}\label{sec_2}
In the context of the gravitational instability of the central regions of trapping polytropes against collapse \citep{Stu-etal:2017:JCAP:}, it is crucial to test the stability of the polytropes with the vacuum energy described by the cosmological constant in relation to the radial pulsations that can be taken as another relevant test of the polytrope instability. We study the dynamical stability of the polytropes against radial pulsations applying the standard method of \cite{Cha:1964:ASTRJ2:}, modified in \cite{Mis-Tho-Whe:1973:Gra:} and extended for inclusion of the cosmological constant \citep{Stu-Hle:2005:RAGtime:,Pos-Hla-Stu:2020:PHYS4:DynStability:}. We extend the study of polytrope stability against radial pulsations presented in \cite{Pos-Hla-Stu:2020:PHYS4:DynStability:} to those having the polytropic index $n > 3$, covering thus the whole region of the trapping polytropes with $n \leq 4$.

\subsection{Einstein equations governing radial pulsations}
In the standard Schwarzschild coordinates ($t$, $r$, $\theta$, $\varphi$), the spacetime element Eq.~\eqref{tpa007e1} of the radially pulsating, spherically symmetric polytropic configuration has to be governed by the metric coefficients taken in the general form including the time dependence  
    \begin{equation}
        \Psi = \Psi(r,t)\, ,\qquad\Phi = \Phi(r,t)\, .
    \end{equation}
The matter inside the pulsating configuration is assumed to be a perfect fluid with $\rho(r, t)$ being the energy density and $p(r,t)$ being the isotropic pressure.

The equilibrium (unperturbed) state of the poly\-trope, about which the configuration pulsations are realized, is the static solution given in the previous section that is characterized by the functions $\Phi(r)$, $\Psi(r)$, $\rho(r)$, $p(r)$, $M(r)$ determined by Eqs.~\eqref{grp49}--\eqref{grp43a}.

The pulsating polytropic configuration (with perturbed quantities depending on time) is determined by the Einstein gravitational equations taking the form
    \begin{align}
        \frac{1}{r^2}\frac{\partial}{\partial r}\left(r\epow{-2\Psi}\right)  - \frac{1}{r^2} + \Lambda &= \frac{8\pi G}{c^4}\,T_{(t)}^{(t)}\, ,\\
        \frac{\epow{-2\Psi}- 1}{r^2} + \frac{2\Phi'}{r}\epow{-2\Psi} + \Lambda &= \frac{8\pi G}{c^4}\,T_{(r)}^{(r)}\, ,  \label{eq7}\\
        -\epow{-2\Phi}\left(\ddot{\Psi} + \dot{\Psi}^2 - \dot{\Psi}\dot{\Phi} \right)- \epow{-2\Psi}\left(\Phi'' + \Phi'^2  + \right. \nonumber \\
         - \left.\Phi'\Psi' + \frac{\Phi'-\Psi'}{r}\right) + \Lambda &= \frac{8\pi G}{c^4}\,T_{(\theta)}^{(\theta)}\, ,\\
        \frac{2\dot{\Psi}}{r}\epow{-2\Psi} &= \frac{8\pi G}{c^4}\,T_{(t)}^{(r)}\, .\label{rce_9}
    \end{align}
Here, the prime denotes the partial derivative with respect to the radial coordinate, and the dot denotes the partial derivative with respect to the time coordinate.

For pulsations of a small amplitude, the metric coefficients $\Psi(r,t)$ and $\Phi(r,t)$, and the thermodynamic variables $\rho(r,t)$, $p(r,t)$ and $n(r,t)$ ($n$ being the number density of baryons in the fluid) measured in the fluid rest frame, can be described by their small Euler variations
    \begin{equation}
        q(r,t) = q_0(r) + \delta q(r,t)\, ,
    \end{equation}
where the general variable is related to quantities $\delta q \equiv(\delta\Phi$, $\delta\Psi$, $\delta\rho$, $\delta p$, $\delta n)$. The subscript $0$ is related to these variables in the equilibrium state.

The pulsation is governed by the radial displacement $\xi$ of the fluid from the equilibrium position
    \begin{equation}
        \xi = \xi(r,t)\, .
    \end{equation}
Then the Euler perturbations $\delta q$ are connected to the Lagrangian perturbations $\Delta q$ measured by an observer co-moving with the pulsating fluid by the relation
    \begin{equation}
        \Delta q (r,t) = q(r + \xi(r, t),t) - q_0 (r)\approx\delta q + q_0'\xi\, .
\end{equation}

The pulsation dynamics, i.e., the time evolution of the perturbation function, is governed by the Einstein gravitational equations which have to be combined with the energy-momentum conservation, baryon conservation, and the thermodynamic laws. All the relevant equations have to be \emph{linearized} relative to the displacement from the static equilibrium configuration. We have to obtain the dynamic equation for the evolution of the fluid displacement $\xi(r,t)$ and the set of \emph{initial-value equations} expressing the perturbation functions $\delta\Phi$, $\delta\Psi$, $\delta\rho$, $\delta p$, $\delta n$ in terms of the displacement function $\xi(r,t)$. The cosmological constant (connected to the vacuum state of the spacetime) is not perturbed, as we do not assume any relation of the cosmological constant (vacuum energy) to matter. We perform the dynamical stability analysis following the method introduced in \cite{Cha:1964:ASTRJ2:} and developed in \cite{Mis-Tho-Whe:1973:Gra:}, see also \citep{Boh:2005:PRD:,Stu-Hle:2005:RAGtime:}. Detailed derivation of the equations governing the 1-st order perturbations of polytropes reflecting the influence of the cosmological constant can be found in \cite{Pos-Hla-Stu:2020:PHYS4:DynStability:}; here we give a summary of the perturbation equations and extend significantly calculations and discussion.

\subsection{Pulsation dynamics and Sturm--Liouville equation}\label{pudyneq}
Introducing a renormalized displacement function \citep{Mis-Tho-Whe:1973:Gra:}
    \begin{equation}
        \zeta\equiv r^2 \epow{-\Phi_0}\xi\, ,
    \end{equation}
the dynamic equation governing the polytrope pulsations takes the form
    \begin{equation}
        W\ddot{\zeta} = \left(P\zeta'\right)' + Q\zeta .
    \end{equation}
Here $.$ ($,$) means derivation related to time (radial) coordinate. The functions $W(r)$, $P(r)$, $Q(r)$ are determined for the equilibrium polytropic configuration by the relations including the influence of the cosmological constant 
    \begin{gather}
        W \equiv (\rho_0 + p_0)\frac{1}{r^2}\,\epow{3\Psi_0 + \Phi_0}\, ,\\
        P \equiv \gamma p_0 \frac{1}{r^2}\,\epow{\Psi_0 + 3\Phi_0}\, ,\\
        Q \equiv \epow{\Psi_0 + 3\Phi_0}\left[\frac{(p_0')^2}{\rho_0 + p_0}\frac{1}{r^2} - \frac{4p_0'}{r^3} - (\rho_0 + p_0)\left(\frac{8\pi G}{c^4}p_0-\Lambda\right)\frac{\epow{2\Psi_0}}{r^2}\right]\, , 
    \end{gather}
where the adiabatic index $\gamma$ is determined by the relation (for details see \cite{Pos-Hla-Stu:2020:PHYS4:DynStability:})
    \begin{displaymath}
        \gamma = \left(p\,\frac{\partial n}{\partial p}\right)^{-1} \left[n-(\rho + p)\frac{\partial n}{\partial\rho}\right]\, .
    \end{displaymath}
This $\gamma$ is not necessarily identical with the adiabatic index related to the equation of state, as discussed in \citep{Hla-Pos-Stu:2020:Modern}.

The boundary conditions guarantee that the displacement function cannot imply a divergent energy density and pressure perturbations at the centre of the sphere --- therefore, $\xi/r$ is finite for $r\rightarrow 0$. The Lagrange variations of the pressure must keep the condition $p = 0$ at the surface of the polytropic configuration at the radius $R$. Therefore,
    \begin{equation}
        \Delta p = -\gamma p_0\frac{\epow{\Phi_0}}{r^2}\left(r^2\epow{-\Phi_0}\xi\right)'\rightarrow 0\qquad \text{as}\qquad r\rightarrow R\, .
    \end{equation}

The linear dynamical stability analysis is realized by the standard assumption of the displacement decomposition
    \begin{equation}
    	\zeta(r,t) = \zeta(r)\epow{-\mathrm{i}\omega t}\, 
    \end{equation}
implying reduction of the dynamic equations to the Sturm--Liouville equation
    \begin{equation}
        \left(P\zeta'\right)' + (Q + \omega^2W)\zeta = 0\, , \label{eSL}
    \end{equation}
with boundary conditions 
    \begin{alignat}{2}
        &\frac{\zeta}{r^3}&\qquad\text{is finite as}&\qquad \label{e34} r \rightarrow 0\, ,\\
        &\gamma p_0 \frac{\epow{\Phi_0}}{r^2}\zeta' \rightarrow 0 &\text{as}&\qquad r\rightarrow R\, .
    \end{alignat}

The Sturm--Liouville equation (\ref{eSL}) and the boundary conditions determine eigenfrequencies $\omega_j$ and corresponding eigenmodes $\zeta_i(r)$, where $i = 1,2,\ldots ,n$. The stable polytropic configurations have a discrete spectrum of the normal radial modes. The $i$-th mode has $i$ nodes between the centre and the surface of the polytrope. The eigenvalue Sturm--Liouville (SL) problem can be represented in the variational form as the extremal values of
    \begin{equation}
        \omega^2 = \frac{\int_0^R\left(P\zeta'^2 - Q\zeta^2\right)\,\dd r}{\int_0^R W\zeta^2\,\dd r} \label{e36}
    \end{equation}
determine the eigenfrequencies $\omega_i$ and the corresponding functions $\zeta_i (r)$ are the eigenfunctions that have to satisfy the orthogonality relation \cite{Mis-Tho-Whe:1973:Gra:}
    \begin{equation}
        \int_0^R \epow{3\Psi_0-\Phi_0}(p_0 + \rho_0)r^2\xi_{i}\xi_{j}\,\dd r = 0\, ; \quad i \neq j\, .
    \end{equation}

The absolute minimum value of Eq.\,(\ref{e36}) represents the squared frequency of the fundamental mode of the radial pulsations. The negative values correspond to the unstable configurations, as $\epow{\mathrm{i}\omega t}$ grows exponentially with time; the positive values correspond to the configurations stable against adiabatic radial perturbations. Therefore, a sufficient condition for the dynamical instability is the vanishing of the right-hand side of Eq.\,(\ref{e36}) for a trial function satisfying the boundary conditions \citep{Mis-Tho-Whe:1973:Gra:,Cha:1964:ASTRJ2:}.

The condition $\omega^2 = 0$ for the marginally stable configurations enables us to deduce from Eq.\,(\ref{e36}) a formula giving the critical value of the adiabatic index $\gamma_{\mathrm{c}}$, assuming it is constant through the configuration in accord with \cite{Cha:1964:ASTRJ2:}; for a detailed discussion of the issue of the adiabatic index, see \cite{Hla-Pos-Stu:2020:Modern}. The critical value of the adiabatic index is thus given by the formula 
    \begin{multline}
        \gamma_{\mathrm{c}} = \left(\int_0^R \frac{\epow{\Psi_0 + 3\Phi_0}}{r^2} \left[\frac{(p_0')^2}{\rho_0 + p_0} - \frac{4p_0'}{r} - (\rho_0 + p_0)\left(\frac{8\pi G}{c^4}p_0 - \right.\right.\right.\\
        \left.\left. -\Lambda\Big)\epow{2\Psi_0}\right] {\zeta}^2\,\dd r\right) \Bigg/ \left({\int_0^R \frac{p_0}{r^2}\epow{\Psi_0 + 3\Phi_0}\zeta^{'2} \,\dd r} \right)\, .
    \end{multline}
Nevertheless, the Chandrasekhar approach can be generalized, as will be demonstrated in the following subsection. 

\subsection{Dynamical stability of polytropes}
The Sturm-Liouville equation can be used to determine the dynamical instability of spherical configurations of perfect fluid with any equation of state. Here, we first present results of the stability study in the special case of the uniform density configurations (see also \citet{Boh:2005:PRD:,Stu-Hle:2005:RAGtime:}) and then concentrate on the general polytropic spheres.

In the special case of the polytropes with $n = 0$, corresponding to spheres with uniformly distributed energy density \citep{Stu:2000:ACTPS2:,Stu-Hle-Nov:2016:PHYSR4:}, the metric coefficients can be given by elementary functions and the stability problem can be solved relatively easily. It was demonstrated \citep{Boh:2005:PRD:,Stu-Hle:2005:RAGtime:} that the instability of the uniform spheres occurs for
    \begin{equation}
        \gamma<\gamma_{\mathrm{c}} \equiv\frac{2(2-\lambda)}{3(1-2\lambda)} + \frac{19-4\lambda(13-7\lambda)}{42(1-2\lambda)(1 + \lambda)}\frac{r_{\mathrm{g}}}{R}\, .  \label{rce_c}
    \end{equation}

For vanishing cosmological constant ($\lambda = 0$) the condition reduces to the result presented in \citep{Cha:1964:ASTRJ2:}
    \begin{equation}
        \gamma < \gamma_{\mathrm{c}} \equiv \frac{4}{3} + \frac{19}{42}\frac{r_{\mathrm{g}}}{R}\, .
    \end{equation}

In the case of the relativistic polytropes, we can directly apply the formulae for the radial profiles of the metric coefficients, the energy density, and pressure as introduced in \citep{Stu-Hle-Nov:2016:PHYSR4:}. Then the variational Sturm-Liouville equation for dynamical stability of the polytropic spheres with respect to radial pulsations takes, under the assumption of the adiabatic index $\gamma$ being constant across the sphere, the form
    \begin{align}\label{eqSturm}
        \omega^2L^2\rho_{\mathrm{c}}c^2 \int_0^{x_1}\epow{3\Psi + \Phi}\theta^n(1 + \sigma\theta) \bar\zeta^2\,\frac{\dd x}{x^2} = \nonumber\\
        \gamma\sigma\rho_{\mathrm{c}}c^2 \int_0^{x_1} \epow{\Psi + 3\Phi}\theta^{n + 1} \left(\oder{\bar\zeta}{x}\right)^2\frac{\dd x}{x^2} -\sigma(n + 1)\rho_{\mathrm{c}}c^2 \times\nonumber\\
        \times\int_0^{x_1}\epow{\Psi + 3\Phi}\Bigg\{\theta^n\oder{\theta}{x}\frac{4}{x}\left[\frac{\sigma(n + 1)x}{4(1 + \sigma\theta)}\oder{\theta}{x} - 1\right] - \nonumber\\
        -2(1 + \sigma\theta)\theta^n \left(\sigma\theta^{n + 1}-\lambda \right)\epow{2\Psi}\vphantom{\left(\oder{\bar\zeta}{x}\right)^2}\Bigg\}\bar\zeta^2\,\frac{\dd x}{x^2}\, .
    \end{align}
We have cancelled now the subscript 0 as it is not necessary. For the polytropes under consideration, the adiabatic index can be given in the form
    \begin{equation}
        \gamma = \left(1 + \frac{1}{n}\right)\left(1 + \sigma\theta\right)\, ,
    \end{equation}
that is generally a function of radius.

%%%%%%%%%%%%%%%%%%%%%%%%%%%%%%%%%%%%%%%%%%%%%%%%%%%%%%%%%%%%%%%%%%%%%%%%%%%%%%%%%%%%%%%%%%%%%%%%%%%%%%%%%%%%%%%%%%%%%%%%%%%%%%%%%%%%%%%%%%%%
    \begin{figure*}[htb]
        \centering\includegraphics[width=.7\linewidth]{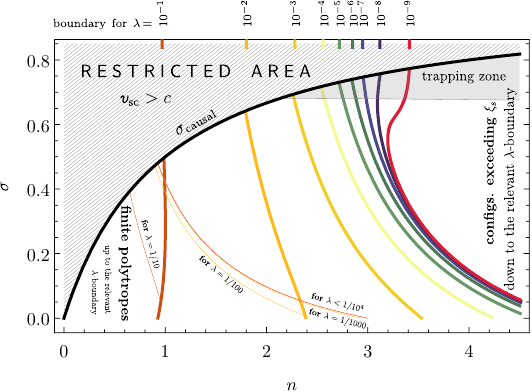}
        \caption{\label{Fig_2} Restrictions on the polytrope parameters. Unified ($n$--$\sigma$) plot for possible finite polytropic configurations distinguished by the cosmological parameter $\lambda$. Considered polytropes have the parameter pairs $n$--$\sigma$ limited by the $\sigma_\mathrm{causal}$ boundary (black line) and the boundaries determined by the static radius given by a chosen $\lambda$ (coloured thick lines); monotonically decreasing pressure is expected in the whole interior of the configuration. Thin coloured borderlines depict transition between polytropes stable and unstable against the radial oscillations. For given $n$ the parameter $\sigma$ needs to be lower than that depicted by this border. For $\lambda < 1/10^4$ these borderlines are all hardly distinguishable from the case $\lambda = 0$. They have been determined using both trial functions and the plotted ones are the more restrictive for the chosen $\lambda$. Generally, no stable configurations having $\sigma < \sigma_\mathrm{casual}$ and $n > 3$ exist. Detailed view of the boundaries for lower $\lambda$ are on the next Figure.}
    \end{figure*}
%%%%%%%%%%%%%%%%%%%%%%%%%%%%%%%%%%%%%%%%%%%%%%%%%%%%%%%%%%%%%%%%%%%%%%%%%%%%%%%%%%%%%%%%%%%%%%%%%%%%%%%%%%%%%%%%%%%%%%%%%%%%%%%%%%%%%%%%%%%%

In the following, we consider the specifics related to the adiabatic index in treating the dynamical stability of polytropes. We apply the approach of our previous paper \citep{Pos-Hla-Stu:2020:PHYS4:DynStability:}, extending its results to values of the polytropic index $n > 3$ relevant for extremely extended polytropes. We discuss in detail the dependence of the solutions on the parameter $\lambda$ governing the role of the cosmological constant in relation to the central density of the configuration. 

We use a more general approach in comparison with Chandrasekhar's method \citep{Cha:1964:ASTRJ2:} -- the adiabatic index in \eqref{eqSturm} is considered as an effective one, following the method introduced in \cite{Merafina:1989} 
    \begin{equation}
        \langle \gamma \rangle = \frac{\displaystyle \int\limits_0^{x_1}\,\frac{\gamma\,\theta^{n + 1}}{x^2}\left(\oder{\zeta}{x}\right)^2 \epow{\Psi + 3\Phi}\,\dd x}{\displaystyle \int\limits_0^{x_1}\frac{\theta^{n + 1}}{x^2}\left(\oder{\zeta}{x}\right)^2 \epow{\Psi + 3\Phi}\,\dd x}\, .
    \end{equation}
The stability condition then takes the form
    \begin{equation}
        \left<\gamma\right> > \gamma_\mathrm{cr}\, .
    \end{equation}

The relation of the radial derivatives of $p$ and $\Phi$ is transferred into the form
    \begin{equation}
        \oder{\Phi}{x} = -\frac{(n + 1)\sigma}{1 + \sigma\theta} \oder{\theta}{x}\, ,
    \end{equation}
and the Sturm--Liouville equation \eqref{eSL} can be expressed in the form
    \begin{equation}
        \frac{\dd}{\dd x}\left(P(x)\frac{\dd\zeta}{\dd x}\right) + \mathcal{L}^2\left(Q(x) + \omega^2 W(x)\right)\zeta = 0\, ,
    \end{equation}
where  we define
    \begin{align}
        P(x) \equiv &  \frac{\left<\gamma\right> \rho_{\mathrm c}\sigma \theta^{n + 1}}{\mathcal{L}^2 x^2}\,\epow{\Psi_0 + 3\Phi_0}\, ,\\
        Q(x) \equiv &  \frac{\rho_{\mathrm c}\sigma (n+1)\theta^n \epow{\Psi_0 + 3\Phi_0}}{\mathcal{L}^4 x^2}\left[\frac{\sigma (n+1)}{1+\sigma\theta}\left(\frac{\dd \theta}{\dd x}\right)^2 + \right. \nonumber \\  & - \left.\frac{4}{x}\left(\frac{\dd \theta}{\dd x}\right) - 2(1 + \sigma\theta)(\sigma \theta^{n+1}-\lambda)\epow{2\Psi_0} \right]\, ,\\
        W(x) \equiv &  \frac{\rho_{\mathrm c} \theta^n (1 + \sigma\theta)}{\mathcal{L}^2 x^2}\,\epow{3\Psi_0 + \Phi_0}\, .
    \end{align}

Short discussion of the dynamical stability of the polytropes with non-zero cosmological constant has been presented in \citep{Pos-Hla-Stu:2020:PHYS4:DynStability:}, however, with restriction to polytropic indexes $n\leq 3$. Therefore, the case of very extended trapping polytropes with $n > 3.5$ was omitted. For this reason we are presenting in this paper a detailed testing with inclusion of the polytropes with $n > 3$, considering the polytropic index up to $n = 4$. We focus on details of the existence of such polytropes, especially on boundaries of so-called forbidden regions concentrated around some special values of the relativistic parameter $\sigma$, and on the behaviour of the very extended trapping polytropes demonstrating instability against gravitational perturbations leading to gravitational collapse of their central region which were studied in \cite{Stu-etal:2017:JCAP:}. The discussion is separated for selected values of the cosmological parameter $\lambda = \frac{\rho_{vac}}{\rho_{\mathrm c}}$ relating the vacuum energy density to the central density of the polytrope. We demonstrate that such the gravitationally unstable polytropic structures are also dynamically unstable against radial pulsations. 

For calculations, we use the standard trial functions
    \begin{equation}
        \xi_1 = x \epow{\Phi/2}\, ,\qquad \xi_2 = x\, , \label{Chandratrial}
    \end{equation}
yielding
    \begin{equation}
        \bar{\zeta}_1 = x^3\epow{-\Phi/2}\, ,\qquad \bar{\zeta}_2 = x^3 \epow{-\Phi}\, .
    \end{equation}

In the most interesting case of very extended trapping polytropes, the cosmological parameter $\lambda = \rho_\mathrm{vac}/\rho_\mathrm{c}$ should be very low, as we expect a large central density of matter; for illustrative reasons we make calculations for appropriately selected values of the cosmological parameter $\lambda$.
 
The critical value of the adiabatic index can be determined by numerical integration only. We consider values of the repulsive cosmological constant that could clearly demonstrate its role in the stability of polytropic spheres. On the other hand, we study polytropic spheres with a magnitude of the cosmological parameter large enough to enable the existence of stable circular geodesics in the exterior region of the polytropic sphere. The results are given in Fig.\,\ref{Fig_2}.

\begin{figure*}[htb]
        \begin{center}
            \includegraphics[width=0.48\linewidth]{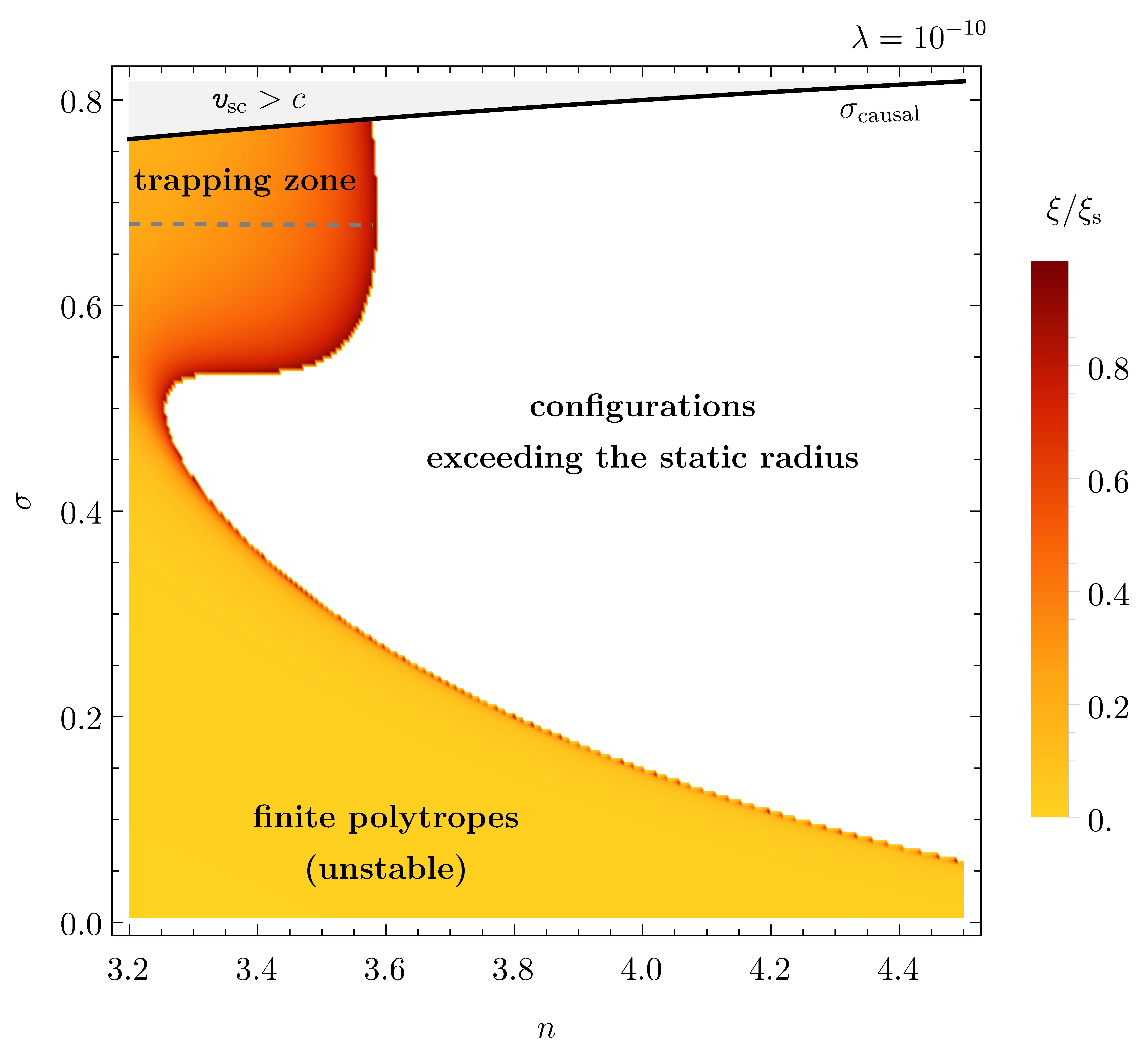}\qquad\includegraphics[width=0.48\linewidth]{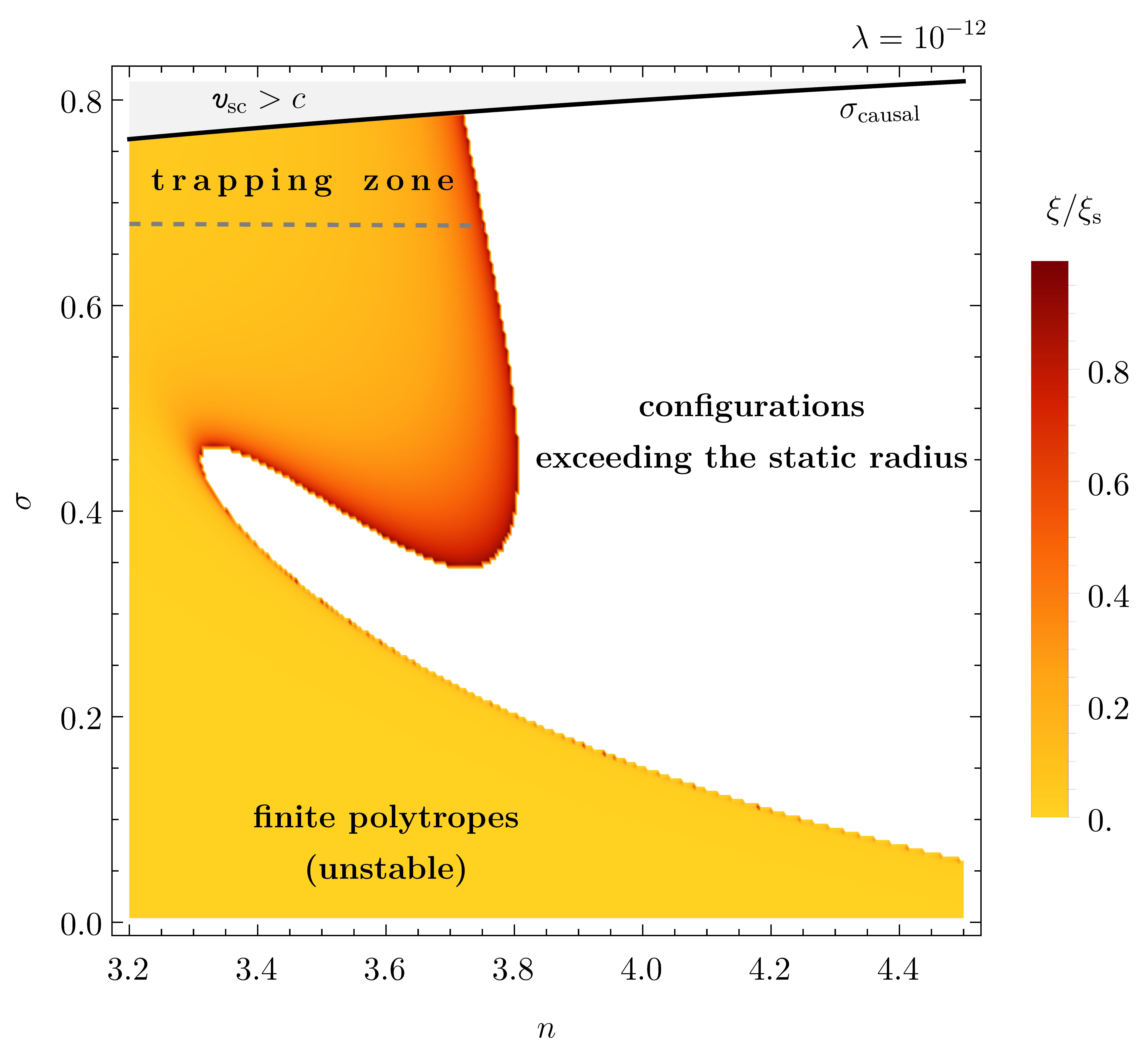}
            \includegraphics[width=0.48\linewidth]{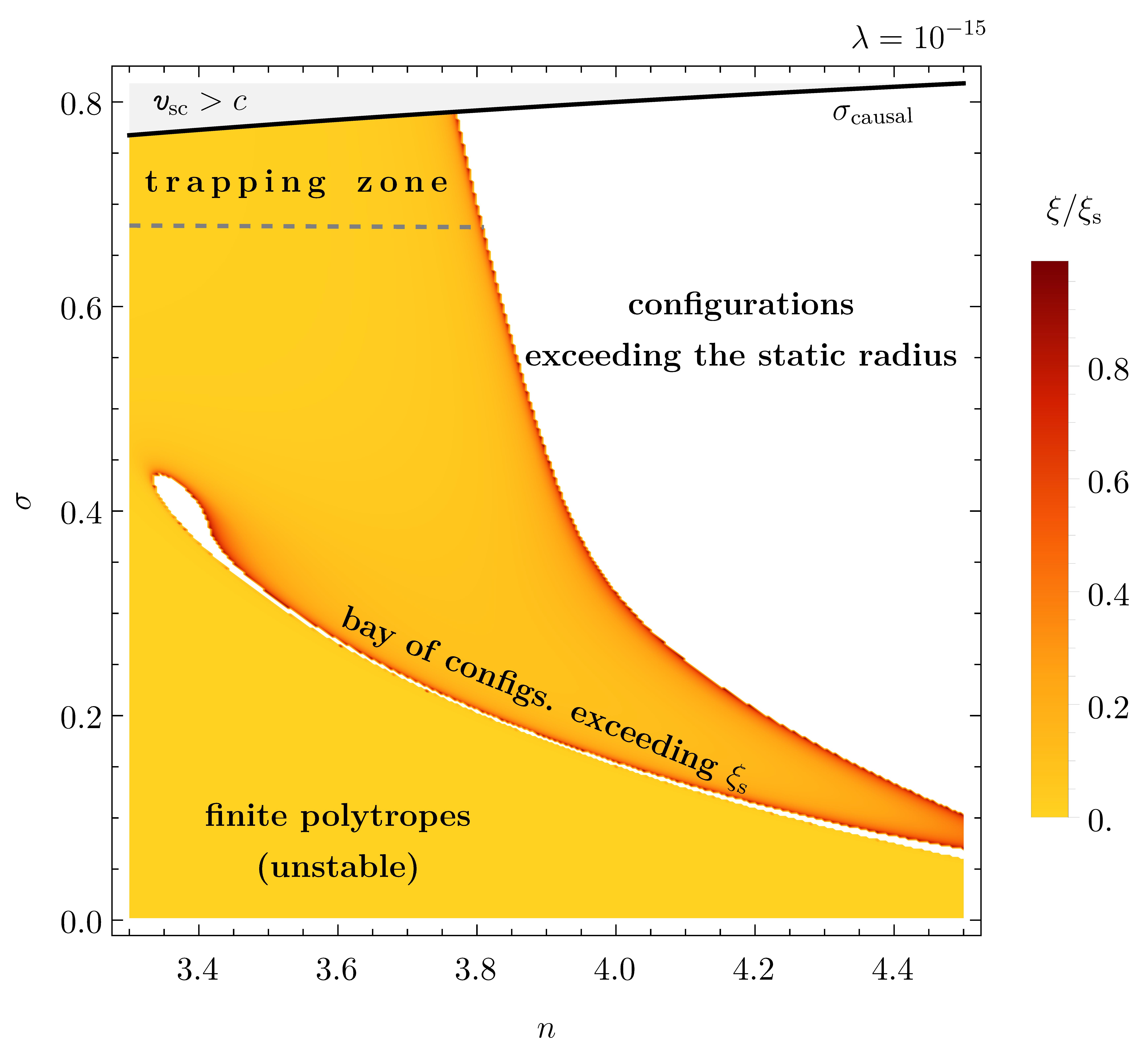}\qquad\includegraphics[width=0.48\linewidth]{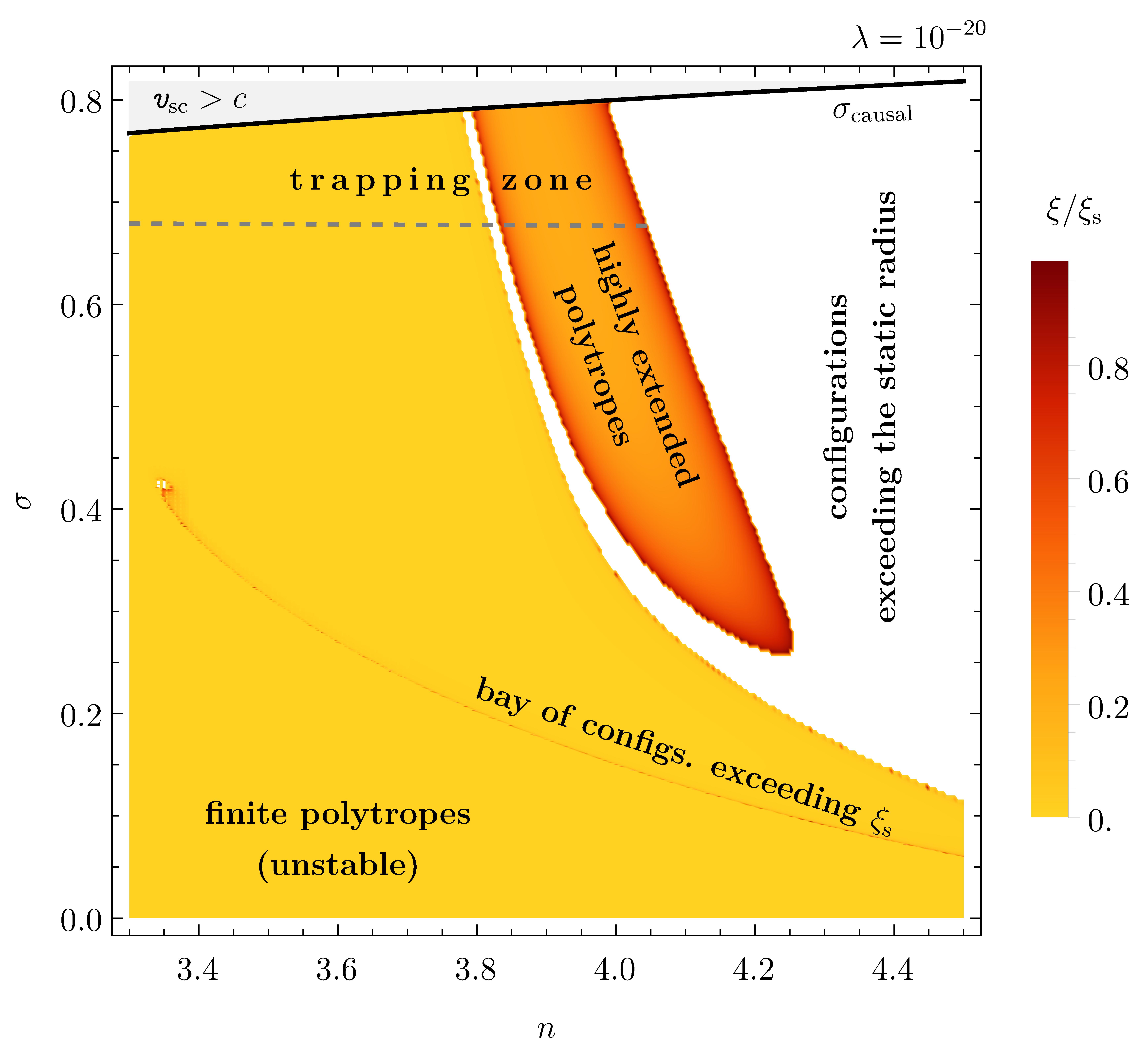}
        \end{center}
        \caption{\label{Fig_RSS} Details of the ($n$--$\sigma$) plain for higher values of $n$ for several values of cosmological parameter $\lambda$. Depicted is the ratio of the radius $\xi_1$ and the static radius $\xi_\mathrm{s}\equiv \left(3\nu(\xi_1)/2\lambda\right)^{1/3}$ for the finite configurations. In the plotted region, all considered configurations are unstable to radial perturbations.}
    \end{figure*}

We can see that the trapping region transforms into the forbidden region where no acceptable polytropic spheres can exist due to the effect of the cosmic repulsion. The extension of the polytropes cannot exceed the static radius. For a clear illustration of this effect, we present the extension of the polytropes as related to the static radius in Fig.~\ref{Fig_RSS}, where regions of the most extended polytropes are depicted. For large values of the polytropic index ($n > 3$), the polytropes are unstable against radial perturbations even for large values of $\lambda$, including the region of the trapping polytropes where the instability against gravitational perturbations is also present \cite{Stu-etal:2017:JCAP:}.

%%%%%%%%%%%%%%%%%%%%%%%%%%%%%%%%%%%%%%%%%%%%%%%%%%%%%%%%%%%%%%%%%%%%%%%%%%%%%%%%%%%%%%%%%%%%%%%%%%%%%%%%%%%%%%%%%%%%%%%%%%%%%%%%%%%%%%%%%%%%
%%%%%%%%%%%%%%%%%%%%%%%%%%%%%%%%%%%%%%%%%%%%%%%%%%%%%%%%%%%%%%%%%%%%%%%%%%%%%%%%%%%%%%%%%%%%%%%%%%%%%%%%%%%%%%%%%%%%%%%%%%%%%%%%%%%%%%%%%%%%

%%%%%%%%%%%%%%%%%%%%%%%%%%%%%%%%%%%%%%%%%%%%%%%%%%%%%%%%%%%%%%%%%%%%%%%%%%%%%%%%%%%%%%%%%%%%%%%%%%%%%%%%%%%%%%%%%%%%%%%%%%%%%%%%%%%%%%%%%%%%
\section{Models of DM halos}\label{mogahalo}

In order to test the possibility to model galactic dark matter halos by the general relativistic polytropic spheres in spacetimes with the repulsive cosmological constant, it is useful to express the length and mass scales of the relativistic polytropes in the form adjusted to the astrophysically relevant, galactic conditions. We consider DM halos related to typical galaxies, comparable to the Milky Way, or to extremely large galaxies, or to the clusters of galaxies. The length scale of galactic halos related to typical galaxies, similar to the Milky Way galaxy, is estimated to be 100--200\,kpc, while the estimated mass of the halo is considered to be about 1--5$\times{}10^{12} M_{\odot}$. Of course, in the case of extremely large and massive galaxies and galaxy clusters, the extension of the halo can increase up to 1~Mpc or slightly more, and the halo mass could be as large as $10^{15} M_{\odot}$ \citep{Zio:2005:NUOC2:GalCollObj}.

Therefore, we give the length scale as
    \begin{equation}
        \mathcal{L} = 1.061 [\sigma (n + 1)]^{1/2}\frac{(10^{-20}\,\mathrm{g\, cm^{-3}})^{1/2}}{\rho_{\mathrm{c}}^{1/2}}\,100\,\mathrm{kpc}\, ,
    \end{equation}
and the mass scale as
    \begin{equation}
        \mathcal{M} = 2.217 [\sigma (n + 1)]^{3/2}\frac{(10^{-20}\, \mathrm{g\, cm^{-3}})^{1/2}}{\rho_{\mathrm{c}}^{1/2}}\,10^{18}\, \mathrm{M_{\odot}}\, .
    \end{equation} 

The polytropic spheres with given mass and length scales are precisely determined by the solution of the structure equations given by the radial coordinate $\xi_{1}(n, \sigma, \lambda)$ and the related mass parameter $\nu(\xi_{1})(n,\sigma, \lambda)$. Generally, the exact solutions can strongly modify the length and mass scales; however, for non-relativistic dark matter with $\sigma \ll 1$, the length and mass scales are decisive. Then we can obtain the polytrope sphere with extension and mass in agreement with the galactic halo estimates for $[\sigma (n + 1)] < 10^{-4}$. However, the central density of such polytropes has to be very small.

\subsection{Rough mapping based on scaling estimates}

We first give the estimates related to the extension and mass scales for the limiting DM halos under consideration. At the lower end we consider their extension $R \sim 100$\,kpc and mass $M \sim 10^{12}\,M_{\odot}$, to the upper end where $R \sim 1$\,Mpc and $M \sim 10^{15}M_{\odot}$.

Rough estimates on the extension scale give the relation
    \begin{equation}
        \frac{\mathcal{L}}{100\,\mathrm{kpc}} = \left(\frac{\sigma}{\tilde{\rho}_\mathrm{c}}\right)^{1/2}\, ,
    \end{equation}
where
    \begin{equation}
        \tilde{\rho}_\mathrm{c} = \frac{\rho_\mathrm{c}}{10^{-20}\mathrm{g\, cm^{-3}}}\, ,
    \end{equation}
while the rough mass estimate relation takes the form
    \begin{equation}
        \frac{\mathcal{M}}{10^{18}M_{\odot}} = \sigma \left(\frac{\sigma}{\tilde{\rho}_\mathrm{c}}\right)^{1/2} = \sigma\frac{\mathcal{L}}{100\,\mathrm{kpc}}\, .
    \end{equation}
For the lower end DM halos ($M \sim 10^{12} M_{\odot}$) we then find the estimates for the relativistic parameter and central density
    \begin{equation}
       \sigma \sim 10^{-6}\, ,~\rho_\mathrm{c} \sim 10^{-23} \mathrm{g\, cm}^{-3}\, .
    \end{equation}
For the upper end CDM halos ($M \sim 10^{15} M_{\odot}$) we find the estimates
    \begin{equation}
       \sigma \sim 10^{-4}\, ,~\rho_\mathrm{c} \sim 10^{-22} \mathrm{g\, cm}^{-3}\, .
    \end{equation}
Clearly, the rough estimates based on the length and mass scaling predict only strongly non-relativistic, $\sigma < 10^{-4}$, and highly diluted fluid with central density $\rho_\mathrm{c} \sim 10^{-22} \mathrm{g\, cm}^{-3}$. Of course, our estimates are relevant for the fluids with non-relativistic (first) regime of the behaviour of the dimensionless polytrope parameters $\xi_1$ and $\nu(\xi_{1})$.

\subsection{Exact mappings based on exact models of the polytropic spheres}
Detailed analysis of the correspondence of the extension and mass of the general relativistic polytropic spheres and the observationally restricted DM halos is realized by numerical methods using the exact solutions of the polytrope structure equations. In the matching procedure for the exact models of polytropes, we use the extension $R = \mathcal{L}\xi_{1}$ and mass $M = \mathcal{M} \nu(\xi_{1})$. We separate the procedure of the exact mapping of the polytropic spheres into two cases, corresponding to the two regimes giving extremely extended and massive polytropic spheres. The first one is related to the non-relativistic polytropic fluid with the relativistic parameter $\sigma < 10^{-3}$ when we expect the extended configurations with a very low central density, predicted by the rough scale estimates. The second case corresponds to the very extended polytropic configurations with relativistic parameter $\sigma$ very close to the critical values, when the dimensionless radius $\xi_1$ can be extremely large.

\subsubsection{Mapping in the non-relativistic regime}
\begin{figure}[ht]
	   \begin{center}
	       \includegraphics[width=\linewidth]{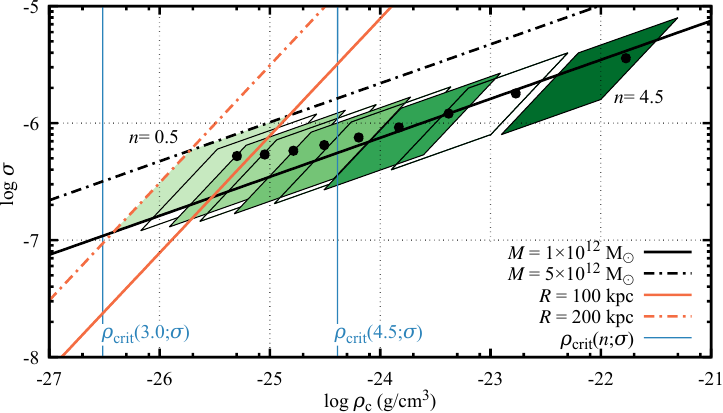}\\[3mm]
		   \includegraphics[width=\linewidth]{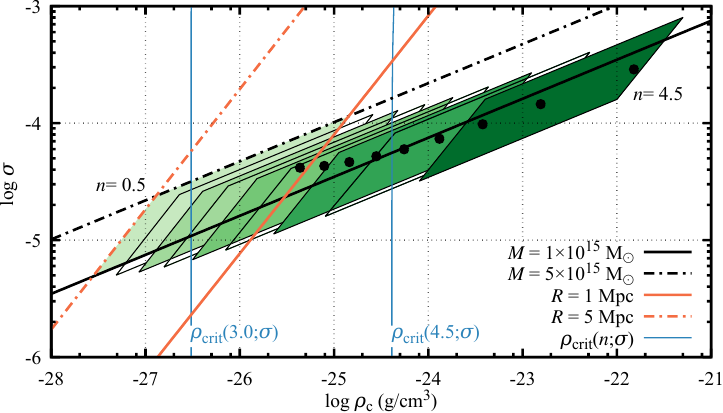}
            \caption{\label{regime_1}Allowed regions of galaxy-like and cluster-like polytropes in the non-relativistic regime. The Figure shows the polytrope extension $R$ and mass $M$ at the edges of the selected intervals $R = 100$--$200$\,kpc \& $M = 1$--$5\times10^{12}\mathrm{M}_\odot$ (top figure) and $R = 1$--$5$\,Mpc \& $M = 1$--$5\times10^{15}\mathrm{M}_\odot$ (bottom figure). We are drawing only the extensions $R$ and $M$ for the polytropic index $n = 0.5$ (4 lines). However the ``polygon shape'' intersections are demonstrated for $n = 0.5$ up to $n = 4.5$ with step size $0.5$. The intersection for $n = 0.5$ is the most left one, for $n = 1.0$ the intersection region is located to the right from the previous and so on. Also the non-integer $n$ are filled with single hue colour (increasing the hue for higher polytropic index) for better overview. The two vertical lines corresponds to the critical value $\rho_\mathrm{crit}(n;\sigma)$, i.e., the $\lambda_\mathrm{crit}(n;\sigma)$ (see Section~\ref{Sec:4}) for two polytropic index ($4.5$,$3.0$). The configurations for the particular $n$ can only exists on the right side of the curve. The dots represents the selected points for the polytropic spheres whose profiles are given in the next section.}
	   \end{center}
    \end{figure}

Recall that in the case of non-relativistic polytropes we model the CMD halos. In Figure~\ref{regime_1} the constant values of the polytrope extension $R$ and mass $M$ are given as the functions of the parameters $\sigma$ and $\rho_\mathrm{c}$ that uniquely determine also the cosmological parameter $\lambda$. Here we follow the first regime, the case of non-relativistic polytropes, $\sigma < 10^{-3}$, with the whole range of considered polytropic indexes for which the critical function $\lambda_\mathrm{crit}$ was constructed. In the parameter space $\sigma$--$\rho_{\rm c}$ we give the regions allowing for polytropic spheres with extension and mass comparable to those of the observed large galaxies and the galaxy clusters for the polytropic indexes $n$ = 0.5, 1, 1.5, 2, 2.5, 3, 3.5, 4, 4.5. In all nine cases, the allowed regions are numerically computed and constructed for the galaxy-like regime with 100\,kpc $\leq R \leq$ 200\,kpc, $10^{12}M_{\odot} \leq M \leq 5\times{}10^{12}M_{\odot}$, and the cluster-like regime with 1\,Mpc $\leq R \leq$ 5\,Mpc, $10^{15}M_{\odot} \leq M \leq 5\times{}10^{15}M_{\odot}$.

The results reflected in Fig.~\ref{regime_1} demonstrate for both the galaxy-like and cluster-like allowed regions surprisingly large scatter of the regions in both the parameters $\sigma$ and $\rho_\mathrm{c}$ from their values predicted by the rough estimates based on the extension and mass scales. The scatter is strong especially in the case of the central density. Such a scatter is caused by the significant dependence of the dimensionless solutions $\xi_1$ and $\nu(\xi_1)$ of the polytrope structure equations on the polytropic index $n$, shown explicitly in \cite{Stu-Hle-Nov:2016:PHYSR4:}. Generally, in both the galaxy-like and cluster-like regimes the allowed regions are shifted to larger values of both $\sigma$ and $\rho_\mathrm{c}$ with increasing polytropic index $n$.

The allowed regions for the galaxy-like polytropes have the relativistic parameter in the interval of $10^{-7} < \sigma < 10^{-5}$, symmetric around rough estimate $\sigma \sim 10^{-6}$, but the central density varies in the interval $10^{-26}\,\mathrm{g\, cm}^{-3} < \rho_\mathrm{c} < 10^{-21}\,\mathrm{g\, cm}^{-3}$ that is shifted to the lower densities from the rough estimate $\rho_\mathrm{c} \sim 10^{-23}\,\mathrm{g\, cm}^{-3}$. Notice that the allowed regions of polytropes with $n \leq 2.5$ must have $\rho_\mathrm{c} < 10^{-24}\,\mathrm{g\, cm}^{-3}$ and $\sigma < 0.6$. On the other hand, the $n = 4.5$ polytropes must have $\rho_\mathrm{c} > 10^{-23}\,\mathrm{g\, cm}^{-3}$ and $\sigma > 10^{-6}$. We shall see in the next section that the density and mass radial profiles of the polytropes belonging to the allowed regions differ significantly for different values of $n$, indicating large variability for matching the observed velocity curves.

The allowed regions for the cluster-like polytropes have the relativistic parameter in the interval of $10^{-5} < \sigma < 10^{-3}$, that is again symmetric around rough estimate $\sigma \sim 10^{-4}$. However, the central density varies in the interval $10^{-27}\,\mathrm{g\, cm}^{-3}  < \rho_\mathrm{c} < 10^{-21}\,\mathrm{g\, cm}^{-3}$ that is shifted to the lower densities from the rough estimate $\rho_\mathrm{c} \sim 10^{-22}\,\mathrm{g\, cm}^{-3}$ even more significantly than in the case of galaxy-like polytropes. In fact, the upper limit on the central density in the case of the $n = 4.5$ cluster-like polytropes is at the same value as for the galaxy-like polytropes, but the relativistic parameter $\sigma$ differs by two orders. Notice that in the case of the $n = 0.5$ cluster-like polytropes, the central density of such extremely extended structures has to be as small as $\rho_\mathrm{c} \sim 5\times 10^{-28}\,\mathrm{g\, cm}^{-3}$, while $\sigma < 10^{-5}$.

%%%%%%%%%%%%%%%%%%%%%%%%%%%%%%%%%%%%%%%%%%%%%%%%%%%%%%%%%%%%%%%%%%%%%%%%%%%%%%%%%%%%%%%%%%%%%%%%%%%%%%%%%%%%%%%%%%%%%%%%%%%%%%%%%%%%%%%%%%%%
%%%%%%%%%%%%%%%%%%%%%%%%%%%%%%%%%%%%%%%%%%%%%%%%%%%%%%%%%%%%%%%%%%%%%%%%%%%%%%%%%%%%%%%%%%%%%%%%%%%%%%%%%%%%%%%%%%%%%%%%%%%%%%%%%%%%%%%%%%%%

\subsubsection{Mapping in the relativistic regime}
\begin{figure*}[ht]
	   \begin{center}
            \includegraphics[width=.495\linewidth]{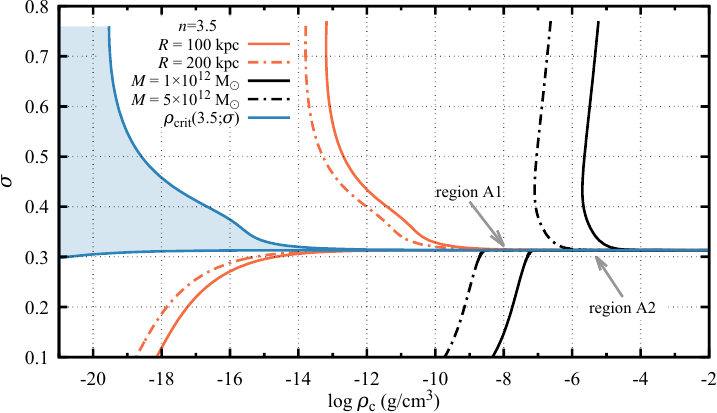}\hfill\includegraphics[width=.495\linewidth]{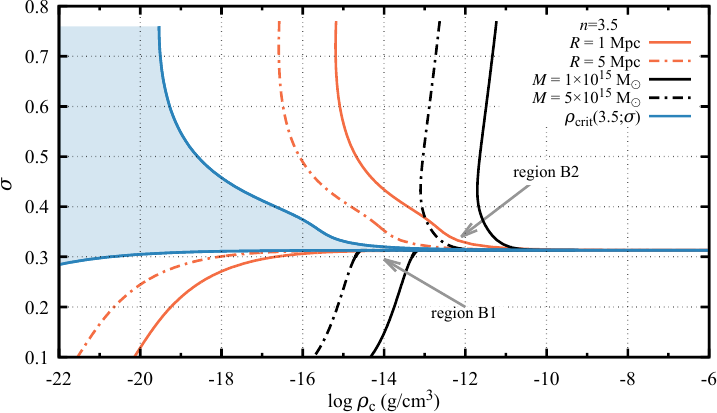}\\
            \includegraphics[width=.495\linewidth]{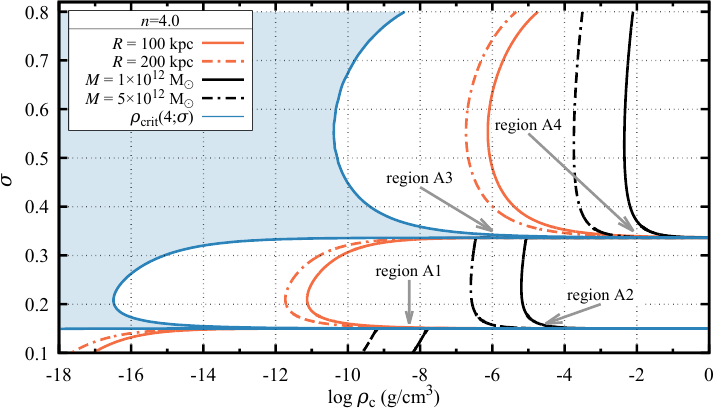}\hfill\includegraphics[width=.495\linewidth]{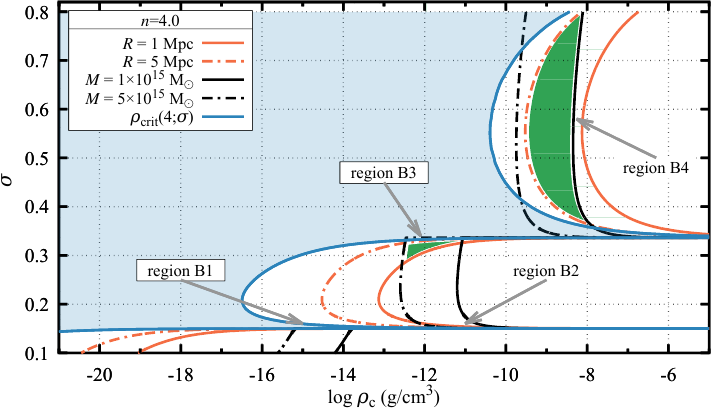}\\
			\includegraphics[width=.495\linewidth]{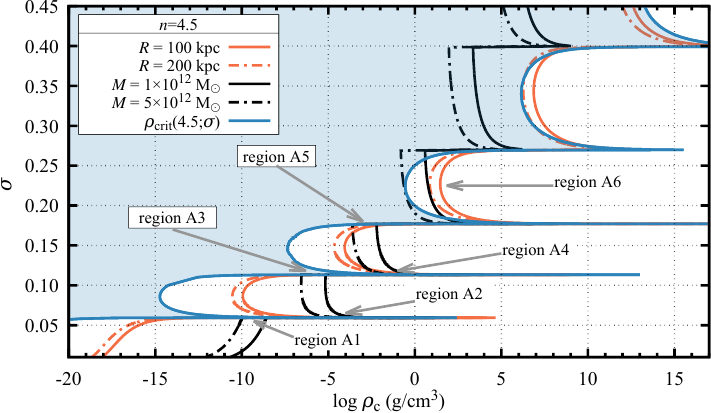}\hfill\includegraphics[width=.495\linewidth]{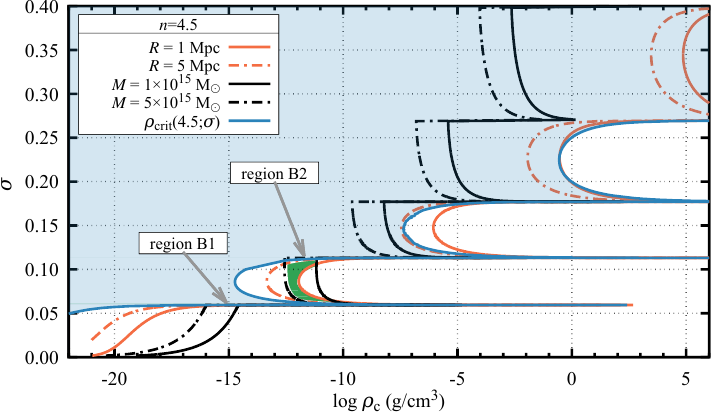}
	   \end{center}
        \caption{\label{regime_2}Allowed regions of galaxy-like (left column) and cluster-like (right column) polytropes in the relativistic regime. Dependence of the polytrope extension $R$ and mass $M$ on the parameters $\sigma $ and $\rho_{\mathrm{c}}$. For the polytropic indexes $n = 3.5,\, 4.0,\, 4.5$ (from top to bottom) we highlight the regions of later interest with letter and number (letter A is used for the galaxy-like polytropes, while letter B is used for the cluster-like polytropes). Furthermore the function $\rho_{\mathrm{crit}}$ is shown, where the shaded is the area of not existing configuration for particular $n$.}
    \end{figure*}

Recall that in the relativistic regime we model non-cold DM halos, e.g., the WDM halos. In Figure~\ref{regime_2} the constant values of the polytrope extension $R$ and mass $M$ are given as the functions of the parameters $\sigma$ and $\rho_\mathrm{c}$ in the second regime, for the vicinity of the critical values of the relativistic parameter $\sigma$, in the case of the polytropic spheres with the polytropic indexes $n = 3.5,\, 4,\, 4.5$. As in the previous case of the non-relativistic regime, the allowed regions are numerically constructed for the galaxy-like regime with 100\,kpc $\leq R \leq$ 200\,kpc, $10^{12}M_{\odot} \leq M \leq 5\times10^{12}M_{\odot}$, and the cluster-like regime with 1\,Mpc $\leq R \leq$ 5\,Mpc, $10^{15}M_{\odot} \leq M \leq 5\times10^{15}M_{\odot}$. The numerical procedure requests very precise methods as we are generally searching in extremely narrow regions of the parameter space $\sigma$--$\rho_\mathrm{c}$, and in the regions where the dimensionless solutions for the polytropes behave in a non-standard way in the vicinity of the critical values of parameter $\sigma$.

Figure~\ref{regime_2} represents a general overview of the searching for the allowed regions. Because of the general overview, details cannot be presented here, with the exception of some special cases when the allowed regions are sufficiently extended. The detailed shape of the allowed regions will be in these cases presented in the next section along with the representative radial profiles of the density, mass, and metric coefficients of the internal polytrope spacetime.

%%%%%%%%%%%%%%%%%%%%%%%%%%%%%%%%%%%%%%%%%%%%%%%%%%%%%%%%%%%%%%%%%%%%%%%%%%%%%%%%%%%%%%%%%%%%%%%%%%%%%%%%%%%%%%%%%%%%%%%%%%%%%%%%%%%%%%%%%%%%
%%%%%%%%%%%%%%%%%%%%%%%%%%%%%%%%%%%%%%%%%%%%%%%%%%%%%%%%%%%%%%%%%%%%%%%%%%%%%%%%%%%%%%%%%%%%%%%%%%%%%%%%%%%%%%%%%%%%%%%%%%%%%%%%%%%%%%%%%%%%

For the $n = 3.5$ polytropes, there exist two narrow allowed regions of the galaxy-like type (denoted as A1, A2) and two narrow allowed regions of the cluster-like type (denoted as B1, B2), all of them in the vicinity of the critical $\sigma_{1\,(n = 3.5)}$.

For the $n = 4$ polytropes, two galaxy-like narrow allowed regions A1, A2 exist around the first critical point $\sigma_{1\,(n = 4)}$, and two narrow allowed regions A3, A4 exist around the second critical point $\sigma_{2\,(n = 4)}$. Further, two cluster-like narrow allowed regions B1, B2 exist around the first critical point $\sigma_{1\,(n = 4)}$, while one extended allowed region B3 exists under the second critical point $\sigma_{2\,(n = 4)}$ and the other extended allowed region B4 exists above this critical point.

For the $n = 4.5$ polytropes, two galaxy-like narrow allowed regions A1, A2 exist around the first critical point $\sigma_{1\,(n = 4.5)}$, and two narrow allowed regions A3, A4 exist around the second critical point  $\sigma_{2\,(n = 4.5)}$. One narrow allowed region A5 exists near the third critical point $\sigma_{3\,(n = 4.5)}$, and one allowed region A6 exists between the critical points  $\sigma_{3\,(n = 4.5)}$ and  $\sigma_{4\,(n = 4.5)}$. One narrow cluster-like allowed region B1 is located near the critical point $\sigma_{1\,(n = 4.5)}$, and an extended allowed region B2 is located between the critical points $\sigma_{1\,(n = 4.5)}$ and $\sigma_{2\,(n = 4.5)}$. No other allowed regions were found for the $n = 4.5$ polytropes near the higher-order critical points of the relativistic parameter existing for this kind of polytropes.

%%%%%%%%%%%%%%%%%%%%%%%%%%%%%%%%%%%%%%%%%%%%%%%%%%%%%%%%%%%%%%%%%%%%%%%%%%%%%%%%%%%%%%%%%%%%%%%%%%%%%%%%%%%%%%%%%%%%%%%%%%%%%%%%%%%%%%%%%%%%
\section{Radial profiles of density, pressure and metric coefficients in selected polytropic spheres}
We give the radial profiles of the energy density, pressure, mass, and metric tensor coefficients for typical polytropic spheres with polytrope parameters taken from the central part of the galaxy-like and cluster-like allowed regions selected by the previous procedure for all the considered values of the polytropic index $n$ in both the non-relativistic and relativistic regimes of the fitting procedure. We thus obtain the characteristic radial profiles of density, pressure, and metric coefficients related to all the polytropic spheres that could potentially represent the DM halos due to their extension and mass.

The polytropic spheres are governed by the structure functions $\theta(\xi)$ and $\nu(\xi)$ of the dimensionless coordinate $\xi$, by the length scale $\mathcal{L}$ and the mass scale $\mathcal{M}$. Their extension reads $R = \mathcal{L}\xi_{1}$ and gravitational mass $M = \mathcal{M} \nu(\xi_1) = {c^{2}}\mathcal{L}\sigma(n + 1) \nu(\xi_{1})/{G}$. We obtain the polytrope radial profiles using Eqs. \eqref{grp49}--\eqref{grp43}.

We again separate our discussion according to the fitting regime. While considering the galaxy-like and cluster-like polytropic spheres in the non-relativistic regime, we keep the selection of the parameters $\sigma$ and $\rho_\mathrm{c}$ in the allowed regions to obtain in both cases polytropes with similar dimensionless characteristics $\xi_1$ and $\nu_1 = \nu(\xi_1)$ for a fixed polytropic index $n$.

\subsection{Profiles of polytropes in the non-relativistic regime}
We first construct the radial profiles of the non-relativistic polytropic spheres when all the polytropic indexes $0 < n \leq 4.5$ can be relevant. For each considered $n$, we have selected typical values of the other parameters characterizing the polytropes, $\sigma$ and $\rho_\mathrm{c}$, corresponding to the center of the allowed regions obtained by the previous numerical procedures fixing the radius and mass of the polytropic spheres. The results are presented for the selections obtained for the non-relativistic polytropes with $\sigma < 10^{-3}$. We give the results separately for the polytropic spheres related to typical galaxies, keeping them nearly equal for all values of $n$ to the extension $R \sim 115$\,kpc and mass $M \sim 2.6\times 10^{12} M_{\odot}$, and for the polytropic spheres corresponding to typical clusters of galaxies, keeping for all values of $n$ the extension $R \sim$ 1100\,kpc and mass $M \sim 2\times10^{15} M_{\odot}$. Notice that for the non-relativistic polytropic spheres we have to introduce a special scale to represent the metric coefficients, as they are very close to unity in the non-relativistic configurations. The selected parameters are summarized for both galaxy-like and cluster-like polytropes in Table~\ref{tab:centre_small}.

Notice that in the non-relativistic regime the allowed regions depend rather smoothly on the polytropic index, and the required extension and mass parameters. Extension to modified values of $R$ and $M$ is straightforward.

%%%%%%%%%%%%%%%%%%%%%%%%%%%%%%%%%%%%%%%%%%%%%%%%%%%%%%%%%%%%%%%%%%%%%%%%%%%%%%%%%%%%%%%%%%%%%%%%%%%%%%%%%%%%%%%%%%%%%%%%%%%%%%%%%%%%%%%%%%%%
    \begin{table}[ht]%
        \centering%
        \caption{\label{tab:centre_small}Selected points of the parameters $\sigma$ and $\rho_\mathrm{c}$ from the galaxy-like and cluster-like polytrope allowed regions that govern the polytropic configurations given in  Figs.~\ref{fig:profiles_case1} \& \ref{fig:profiles_case2} for all the considered polytropic indexes.}%
        \resizebox{\hsize}{!}{%
            \begin{tabular}{lclllll}\hline\hline%
                Scheme & Index $n$ & $\sigma$ & $\rho_\mathrm{c}$ ($\mathrm{g\, cm}^{-3}$) & $\lambda$  & $R$ (kpc) & $M$ ($M_{\odot}$) \\ \hline
                \multirow{9}{*}{Galaxy-like}	& 0.5 & 5.23E-7 & 5.03E-26 & 1.99E-4 & 115.4 & 2.60E12\\
	 			                       			& 1.0 & 5.39E-7 & 8.99E-26 & 1.11E-4 & 115.4 & 2.60E12\\
	 						                    & 1.5 & 5.80E-7 & 1.64E-25 & 6.11E-5 & 115.3 & 2.59E12\\
	 						                    & 2.0 & 6.48E-7 & 3.11E-25 & 3.21E-5 & 115.5 & 2.60E12\\
	 						                    & 2.5 & 7.53E-7 & 6.38E-25 & 1.57E-5 & 115.5 & 2.60E12\\
	 						                    & 3.0 & 9.20E-7 & 1.48E-24 & 6.77E-6 & 115.4 & 2.60E12\\
	 						                    & 3.5 & 1.21E-6 & 4.17E-24 & 2.40E-6 & 115.6 & 2.61E12\\
	 						                    & 4.0 & 1.79E-6 & 1.70E-23 & 5.89E-7 & 115.3 & 2.59E12\\
	 						                    & 4.5 & 3.59E-6 & 1.69E-22 & 5.92E-8 & 115.5 & 2.60E12\\ \hline
                \multirow{9}{*}{Cluster-like}   & 0.5 & 4.16E-5 & 4.35E-26 & 2.30E-4 & 1106 & 1.98E15\\
	 						  	                & 1.0 & 4.30E-5 & 7.87E-26 & 1.27E-4 & 1102 & 1.98E15\\
							  	                & 1.5 & 4.65E-5 & 1.45E-25 & 6.88E-5 & 1098 & 1.98E15\\
	 						  	                & 2.0 & 5.21E-5 & 2.81E-25 & 3.56E-5 & 1089 & 1.97E15\\
	 						  	                & 2.5 & 5.98E-5 & 5.54E-25 & 1.81E-5 & 1105 & 1.97E15\\
	 						  	                & 3.0 & 7.35E-5 & 1.31E-24 & 7.62E-6 & 1096 & 1.97E15\\
	 						  	                & 3.5 & 9.77E-5 & 3.77E-24 & 2.65E-6 & 1093 & 1.99E15\\
	 						  	                & 4.0 & 1.45E-4 & 1.56E-23 & 6.43E-7 & 1084 & 1.97E15\\
	 						  	                & 4.5 & 2.88E-4 & 1.52E-22 & 6.58E-8 & 1095 & 1.97E15\\ \hline
            \end{tabular}%
        }
    \end{table}
%%%%%%%%%%%%%%%%%%%%%%%%%%%%%%%%%%%%%%%%%%%%%%%%%%%%%%%%%%%%%%%%%%%%%%%%%%%%%%%%%%%%%%%%%%%%%%%%%%%%%%%%%%%%%%%%%%%%%%%%%%%%%%%%%%%%%%%%%%%%

\subsubsection{Galaxy-like profiles}
For the scales related to the polytropic spheres that could represent galaxies, the characteristic radial profiles are presented in Fig.~\ref{fig:profiles_case1} for each of the considered polytropic indexes that were used in the fitting procedure for the non-relativistic polytropes. We can immediately see the crucial influence of the polytropic index $n$ in the radial profiles of all the characteristic quantities. For low values of the index, $n\leq 1$, density and pressure (mass) slowly decrease (increase) up to the polytrope edge, and the metric coefficient $g_{tt}$ also slowly decreases up to the edge. Starting from the $n = 1.5$ polytropic spheres, the decrease of density and pressure (increase of mass) is getting sharper with increasing $n$, and a local minimum of the $g_{tt}$ radial profile occurs, being deeper and more shifted to the polytrope centre with increasing $n$. For the polytropes with $n>3$, the mass of the polytrope is concentrated in the central regions, and the density (pressure) rapidly drops to very low values.

\subsubsection{Cluster-like profiles}
For the scales related to the polytropic spheres that could represent clusters of galaxies, the characteristic radial profiles are presented in Fig.~\ref{fig:profiles_case2} for each of the considered polytropic indexes, for which the fitting procedure has been realized. The character of the radial profiles in dependence on the polytropic index $n$ is the same as in the case of the galaxy-like typical polytropes, as can be expected due to the same dimensionless characteristics of the polytropes. Again, the uniformity of the energy density and mass distribution across the polytropic sphere strongly decreases with increasing $n$. Also, the gravitational potential well given by the metric coefficient $g_{tt}$ has a similar character as for galaxy-like configurations; however, it is deeper for the cluster-like configurations. Thus, the relativistic effects can be stronger in this case.

\subsubsection{Variability across the allowed region}
\begin{figure*}[ht]
       \begin{center}
            \includegraphics[width=.32\linewidth]{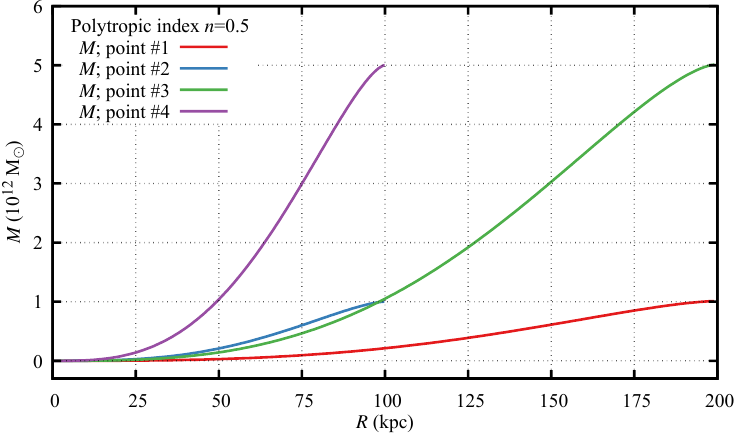}\hfill\includegraphics[width=.32\linewidth]{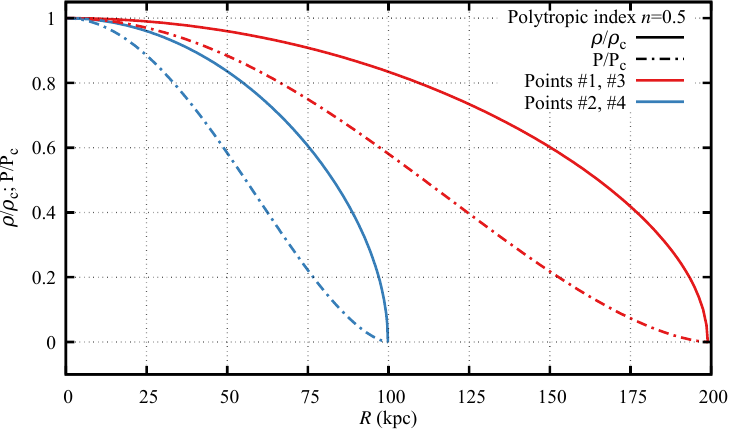}\hfill\includegraphics[width=.33\linewidth]{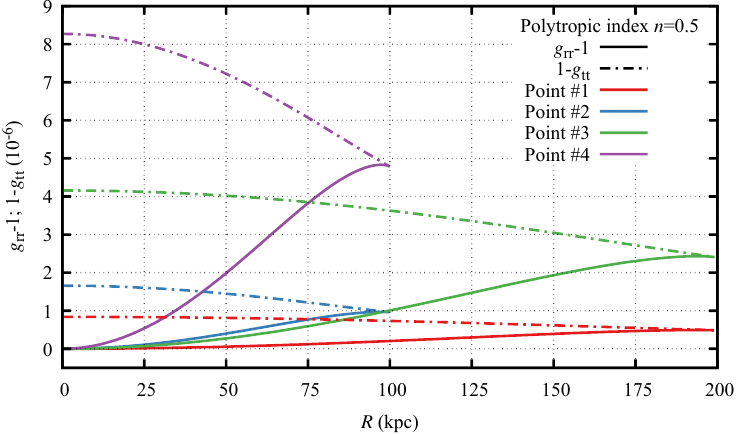}\\[3mm]
			\includegraphics[width=.32\linewidth]{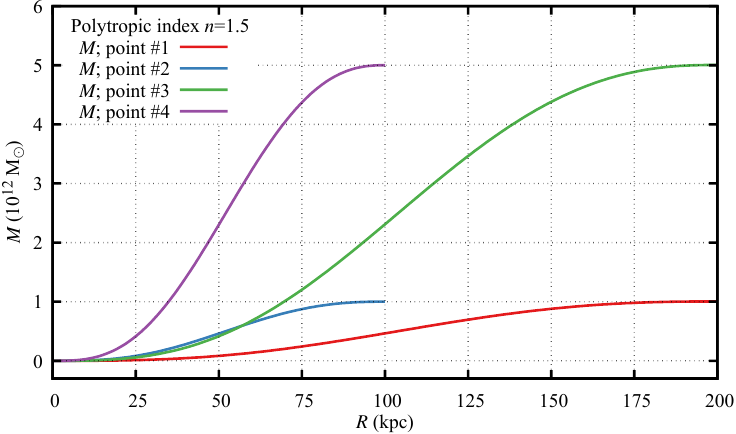}\hfill\includegraphics[width=.32\linewidth]{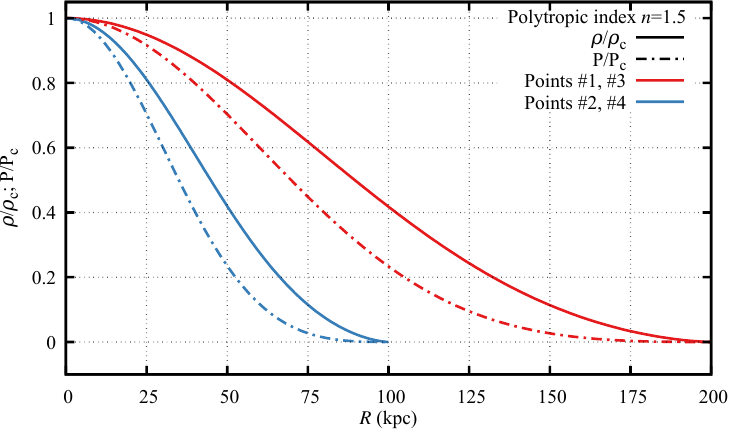}\hfill\includegraphics[width=.33\linewidth]{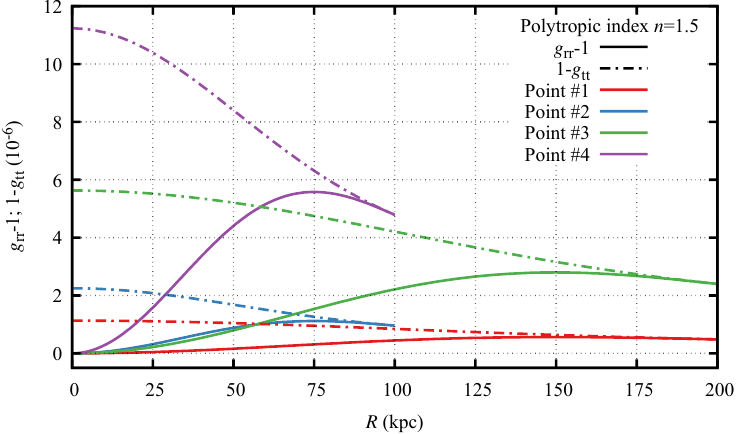}
            \caption{\label{fig:profile_edges_0} Radial profiles of the mass $\nu$ (left column), energy density $\rho$ (middle column), pressure $P$ (middle column) and metric coefficients $g_{tt}$, $g_{rr}$ (right column) constructed for the parameters $\sigma$ and $\rho_\mathrm{c}$ on the edges of the allowed regions. The first row of the figure corresponds to the polytropic sphere with $n = 0.5$, second row corresponds to $n = 1.5$ polytropes. For better visualization of the metric coefficients we introduced the special scale convenient for the non-relativistic configurations.}
	   \end{center}
    \end{figure*}

In order to demonstrate the role of the variations of the parameters $\sigma$ and $\rho_\mathrm{c}$ across the allowed regions, we construct the polytrope profiles for the special selection of polytropic indexes $n = 0.5,\, 1.5$, considering for comparison the radial profiles of the polytropic characteristic functions for the parameters $\sigma$ and $\rho_\mathrm{c}$ with values reflecting the extension of the allowed region for galaxy-like polytropes. Therefore, the representing parameters $\sigma$ and $\rho_\mathrm{c}$ (black dots in Figures) are computed from the values of the four allowed region corners. The resulting radial profiles are presented in Fig.~\ref{fig:profile_edges_0}. (For the cluster-like polytropes, the results are of the same character.)

%%%%%%%%%%%%%%%%%%%%%%%%%%%%%%%%%%%%%%%%%%%%%%%%%%%%%%%%%%%%%%%%%%%%%%%%%%%%%%%%%%%%%%%%%%%%%%%%%%%%%%%%%%%%%%%%%%%%%%%%%%%%%%%%%%%%%%%%%%%%
%%%%%%%%%%%%%%%%%%%%%%%%%%%%%%%%%%%%%%%%%%%%%%%%%%%%%%%%%%%%%%%%%%%%%%%%%%%%%%%%%%%%%%%%%%%%%%%%%%%%%%%%%%%%%%%%%%%%%%%%%%%%%%%%%%%%%%%%%%%%

We can see that the qualitative character of all the profiles remains the same for all four selected limiting points of the allowed regions. Only the mass scale and the polytrope extension are modified accordingly. Of course, the height of the gravitational potential reflected by the metric coefficient $g_{tt}$ is different in the selected polytropes.

\subsection{Profiles of polytropes in the relativistic regime}
In the case of the relativistic polytropic spheres that could represent large galaxies or clusters of galaxies, only the polytropic indexes $n>3.3$ are allowed. For the selections obtained in the second regime of relativistic polytropes with near-critical values of the relativistic parameter $\sigma \sim \sigma_\mathrm{crit}$, we separate the discussion according to the considered values of the polytropic index: $n$ = 3.5, 4, 4.5. Because of the rather complex form of the allowed regions of the parameter space $\sigma$--$\rho_\mathrm{c}$, selected by the fitting procedure on the extension and mass of the polytropes, we restrict again our attention to just one choice of the parameters $\sigma$ and $\rho_\mathrm{c}$ in each selected allowed region.

For polytropes of the relativistic regime we can see that the matching of the parameters $\sigma$ and $\rho_\mathrm{c}$ for various values of $R$ and $M$ is non-trivial, and strongly dependent on polytropic index $n>3.3$. Moreover, we observe also an extremely large spread of the central density between various allowed regions. In some allowed regions, very small central density is implied by the matching procedure, say $\rho_\mathrm{c} \sim 10^{-16}\,\mathrm{g\, cm}^{-3}$, while in a neighbouring one, very large central density is required, say $\rho_\mathrm{c} \sim 10^{-2}\,\mathrm{g\, cm}^{-3}$.

For each of the selected values of the polytropic index $n = 3.5,\, 4,\, 4.5$, and each of the corresponding A, B regions of the $\sigma$ and $\rho_\mathrm{c}$ parameters, we choose typical values of these parameters that we use in the construction of velocity profiles of the circular geodesic motion in the related polytropic spheres. Usually, the regions A, B are concentrated in the vicinity of the critical points $\sigma_\mathrm{crit}(n)$, but there exists also an important exception, namely B4 occurring for $n = 4$ case, where the extension of $\sigma$ parameter is very large, allowing even for the existence of the so-called trapping polytropes whose astrophysical relevance is exposed in \cite{Stu-etal:2017:JCAP:}. In all the considered cases we also depict (in \emph{blue}) region forbidden by the critical value of $\lambda$, $\lambda_\mathrm{crit}$, by the curve $\rho_\mathrm{crit}(n, \sigma)$ defined by Eq.~(\ref{eqCritical}).

\subsubsection{Profiles of polytropes with $n = 3.5$}
Now all the selected allowed regions are related to the single critical point of the relativistic parameter, and we choose $\sigma$ always in its close vicinity. The selected values of the parameters $\sigma$ and $\rho_\mathrm{c}$ are given in Table~\ref{tab:centre_35} for both the galaxy-like and cluster-like polytropes. The resulting polytrope radial profiles are demonstrated in the left column of Fig.~\ref{fig:profiles_regime2_case2_n35} for the galaxy-like polytropes, and in the right column of Fig.~\ref{fig:profiles_regime2_case2_n35} for the cluster-like polytropes.
In the left columns of these figures, we present details of the corresponding polytrope allowed regions. In the $n = 3.5$ polytrope case, all the allowed regions are closely related to the critical relativistic parameter.

%%%%%%%%%%%%%%%%%%%%%%%%%%%%%%%%%%%%%%%%%%%%%%%%%%%%%%%%%%%%%%%%%%%%%%%%%%%%%%%%%%%%%%%%%%%%%%%%%%%%%%%%%%%%%%%%%%%%%%%%%%%%%%%%%%%%%%%%%%%%
    \begin{table}[ht]
        \caption{\label{tab:centre_35}Selected points for the polytropic sphere $n = 3.5$ used in the Figs.~\ref{fig:profiles_regime2_case1} \& \ref{fig:profiles_regime2_case2_n35}.}
        \resizebox{\columnwidth}{!}{%
        \begin{tabular}{cclllll}\hline\hline
            \multicolumn{1}{l}{Scheme} & Region & $\sigma$ & $\rho_\mathrm{c}$ ($\mathrm{g\, cm}^{-3}$) & $\lambda$  & $R$ (kpc) & $M$ ($M_{\odot}$)\\ \hline
            \multirow{2}{*}{Galaxy-like}   & A1 & 0.313091 & 1.4E-8  & 7.14E-22 & 145.4 & 2.22E12\\
                                           & A2 & 0.313133 & 4.0E-5  & 2.50E-25 & 144.2 & 2.39E12\\ \hline
            \multirow{2}{*}{Cluster-like}  & B1 & 0.312770 & 1.3E-14 & 7.69E-16 & 2558  & 2.25E15\\
                                           & B2 & 0.316300 & 3.2E-12 & 3.11E-18 & 2523  & 2.47E15\\ \hline
        \end{tabular}
        }
    \end{table}
%%%%%%%%%%%%%%%%%%%%%%%%%%%%%%%%%%%%%%%%%%%%%%%%%%%%%%%%%%%%%%%%%%%%%%%%%%%%%%%%%%%%%%%%%%%%%%%%%%%%%%%%%%%%%%%%%%%%%%%%%%%%%%%%%%%%%%%%%%%%

%%%%%%%%%%%%%%%%%%%%%%%%%%%%%%%%%%%%%%%%%%%%%%%%%%%%%%%%%%%%%%%%%%%%%%%%%%%%%%%%%%%%%%%%%%%%%%%%%%%%%%%%%%%%%%%%%%%%%%%%%%%%%%%%%%%%%%%%%%%%
    \begin{figure*}[ht]
        \begin{center}
            \includegraphics[width=.49\linewidth]{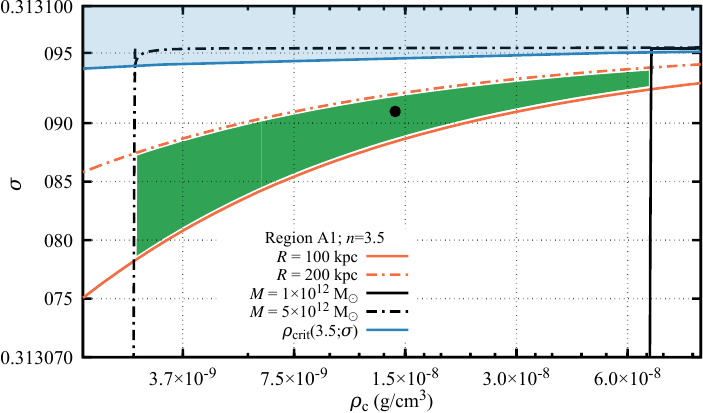}\hfill\includegraphics[width=.49\linewidth]{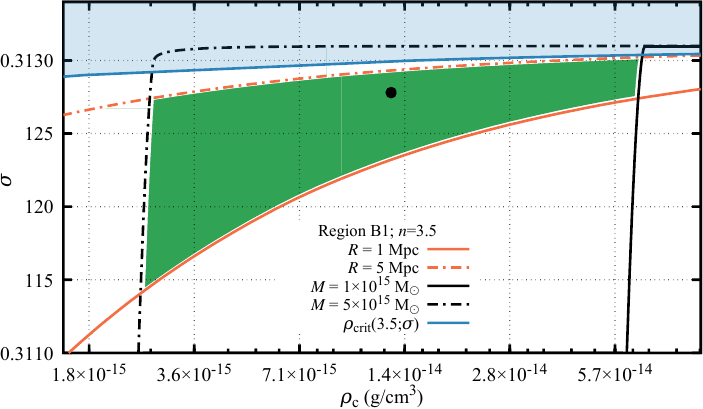}\\[3mm]
			\includegraphics[width=.49\linewidth]{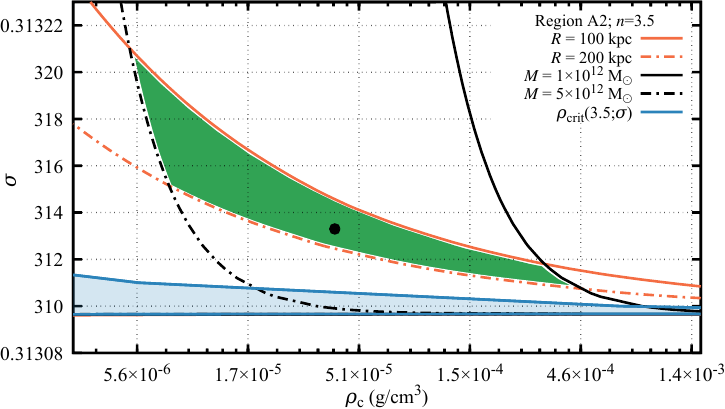}\hfill\includegraphics[width=.49\linewidth]{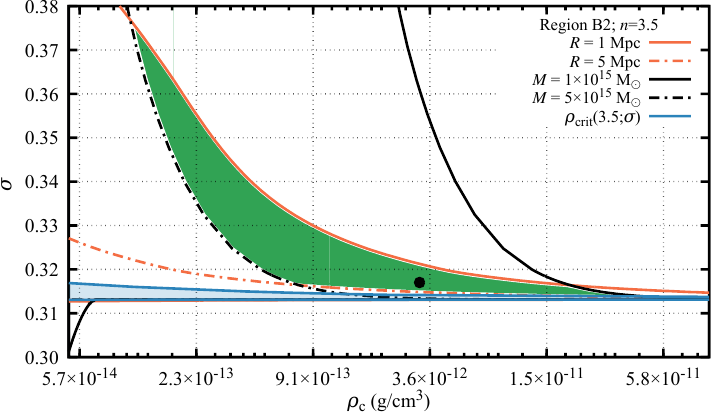}
            \caption{\label{fig:profiles_regime2_case1} Allowed regions for the $n = 3.5$ galaxy-like (A1 and A2 regions; left column) and cluster-like  (B1 and  B2 regions; right column) polytropic spheres. The corresponding radial profiles for the selected values (see Table~\ref{tab:centre_35}), depicted here by the black dots, are given in Fig.~\ref{fig:profiles_regime2_case2_n35}.}
        \end{center}
    \end{figure*}
%%%%%%%%%%%%%%%%%%%%%%%%%%%%%%%%%%%%%%%%%%%%%%%%%%%%%%%%%%%%%%%%%%%%%%%%%%%%%%%%%%%%%%%%%%%%%%%%%%%%%%%%%%%%%%%%%%%%%%%%%%%%%%%%%%%%%%%%%%%%

As intuitively expected, all the extended relativistic polytropes have extremely large dimensionless radius $\xi_1$, but the dimensionless mass parameter $\nu_1$ can be both large and small.

We can immediately observe a significant difference in the behaviour of the radial profiles in the cases A1 and B1 in comparison to the cases A2 and B2. The A1 (B1) polytropes can be characterized as extremely concentrated ones. All their mass is essentially concentrated in the central region, being essentially constant across the whole polytropic configuration. This is reflected also by the behaviour of the metric coefficient $1 - g_{tt}$ that steeply increases from the centre. For the A2 (B2) polytropes, we observe qualitatively different behaviour in comparison to the extremely concentrated polytropes, where the energy density is not decreasing so steeply, and mass increases along the whole configuration. The gravitational potential well, represented by the metric coefficient $1 - g_{tt}$ demonstrated a rather complex behaviour with one local maximum followed by a local minimum. In such extended relativistic polytropes, their mass is distributed along the whole configuration. Notice that the extremely concentrated polytropes have their mass parameter $\nu_1 < 1$, while the spread polytropes have a relatively large mass parameter $\nu_1 > 10$.

\subsubsection{Profiles of polytropes with $n = 4$}
\begin{figure*}[p]
        \begin{center}
            \includegraphics[width=.49\linewidth]{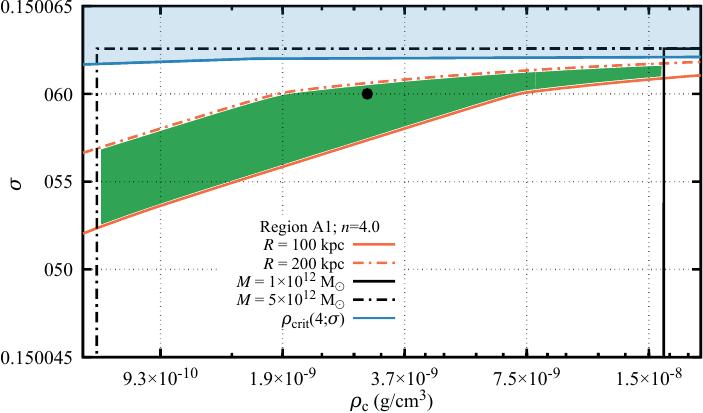}\hfill\includegraphics[width=.49\linewidth]{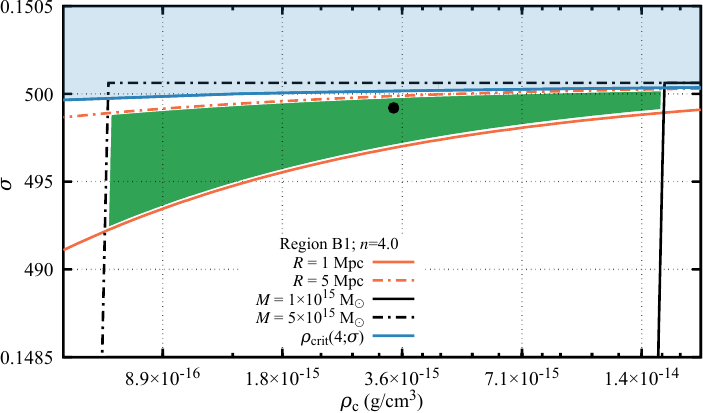}\\[3mm]
			\includegraphics[width=.49\linewidth]{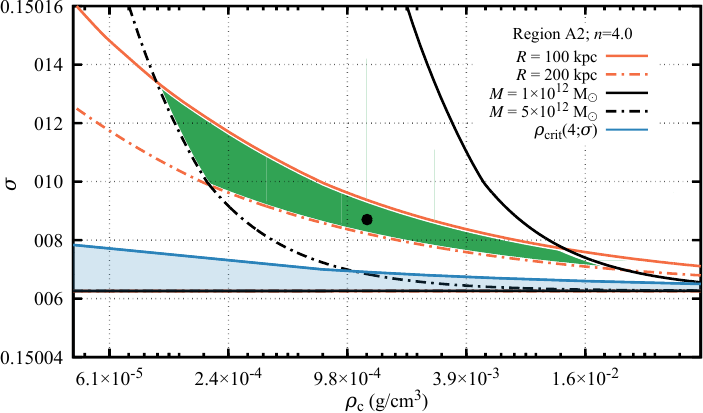}\hfill\includegraphics[width=.49\linewidth]{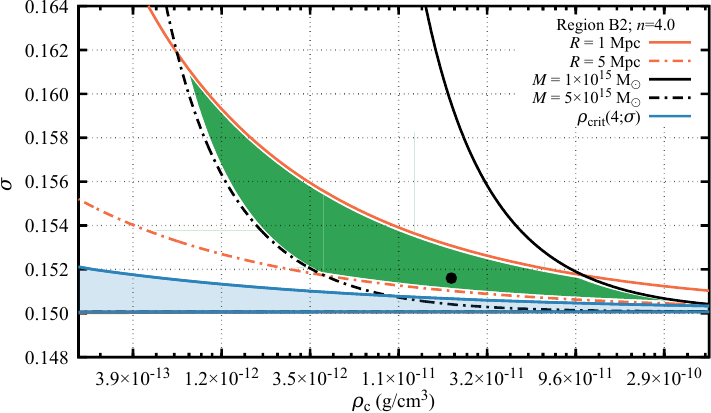}\\[3mm]
			\includegraphics[width=.49\linewidth]{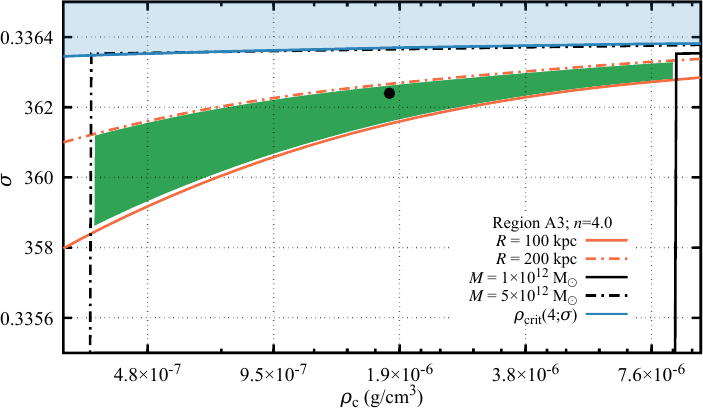}\hfill\includegraphics[width=.49\linewidth]{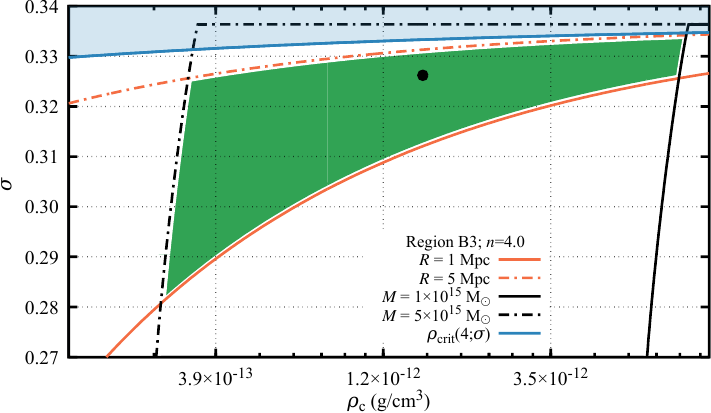}\\[3mm]
			\includegraphics[width=.49\linewidth]{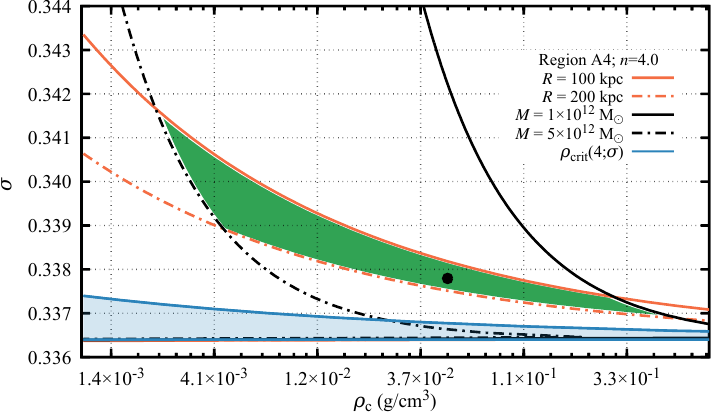}\hfill\includegraphics[width=.49\linewidth]{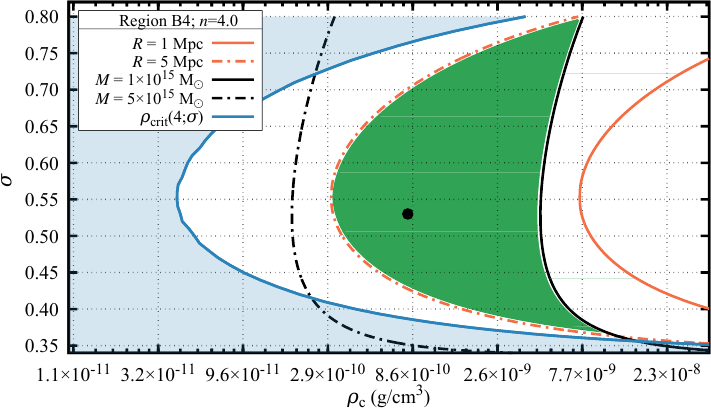}
            \caption{\label{fig:profiles_regime2_case1_n40}  The allowed regions of galaxy-type A (A1--A4, right column) and cluster-like B (B1--B4, left column) for polytropic spheres modeled using $n = 4.0$. The corresponding radial profiles for the selected values (see Table~\ref{tab:centre_40}), depicted here by the black dots, are given in Fig.~\ref{fig:profiles_regime2_case2_n40}.}
	   \end{center}
    \end{figure*}

The selected allowed regions are related to the two critical points of the relativistic parameter. The selected values of the parameters $\sigma$ and $\rho_\mathrm{c}$ are given in Table~\ref{tab:centre_40} for both the galaxy-like and cluster-like polytropes. The resulting polytrope radial profiles are demonstrated in the right column of Fig.~\ref{fig:profiles_regime2_case1_n40} for the galaxy-like polytropes (four regions A1--A4), and in Fig.~\ref{fig:profiles_regime2_case2_n40} for the cluster-like polytropes (four regions B1--B4).
In the left columns of these figures, we present details of the corresponding polytrope allowed regions. Notice that the B4 region is distributed between the second critical point of $\sigma$ parameter and its causal limit $\sigma = 0.8$.

All the extended relativistic polytropes have again extremely large dimensionless radius $\xi_1$, while the dimensionless mass parameter $\nu_1$ can be both large and small.

There is the standard distribution of the galaxy-like polytropes, as the A1 and A3 polytropes are the typical extremely concentrated polytropes with low parameter $\nu_1$, while the A2 and A4 polytropes are the typical spread polytropes with large parameter $\nu_1$. However, the cluster-like polytropes demonstrate a non-standard behaviour. The B1 polytrope is a typical extremely concentrated one, with very small $\nu_1 \lesssim 1$, and the B2 polytrope is a typical spread polytrope with large $\nu_1 \sim 100$. On the other hand, the polytropes B3 and B4 demonstrated an irregular behaviour as they are strongly concentrated, but not extremely. They have rather large dimensionless mass parameter $\nu_1 \sim 6$ for B3 and $\nu_1 \sim 70$ for B4, but the radial profile of the metric coefficient $1-g_{tt}$ demonstrates no local extreme.

Notice that both B3 and especially B4 are largely extended in the relativistic parameter $\sigma$, and in the case of B4, $\sigma$ reaches the regions where the polytropic spheres contain zones of trapped null geodesics where fast gravitational instabilities could be relevant, as shown in \citep{Stu-etal:2017:JCAP:}.

%%%%%%%%%%%%%%%%%%%%%%%%%%%%%%%%%%%%%%%%%%%%%%%%%%%%%%%%%%%%%%%%%%%%%%%%%%%%%%%%%%%%%%%%%%%%%%%%%%%%%%%%%%%%%%%%%%%%%%%%%%%%%%%%%%%%%%%%%%%%
    \begin{table}[ht]
        \centering
        \caption{\label{tab:centre_40}Selected points for the polytropic sphere $n = 4.0$ used in the Figs.~\ref{fig:profiles_regime2_case1_n40} \& \ref{fig:profiles_regime2_case2_n40}.}
        \resizebox{\columnwidth}{!}{%
		  \begin{tabular}{cclllll}\hline\hline%
                \multicolumn{1}{l}{Scheme} & Region & $\sigma$ & $\rho_\mathrm{c}$ ($\mathrm{g\, cm}^{-3}$) & $\lambda$  & $R$ (kpc) & $M$ ($M_{\odot}$) \\ \hline
		        \multirow{4}{*}{Galaxy-like}   & A1 & 0.15006  & 3.0E-9  & 3.32E-21 & 152.3 & 2.32E12 \\
                		                       & A2 & 0.150087 & 1.2E-3  & 8.14E-27 & 143.8 & 2.43E12 \\
		                                       & A3 & 0.33624  & 1.8E-6  & 5.54E-24 & 149.9 & 2.20E12 \\
		                                       & A4 & 0.337792 & 4.9E-2  & 2.04E-28 & 142   & 2.03E12 \\ \hline
		         \multirow{4}{*}{Cluster-like} & B1 & 0.14992  & 3.4E-15 & 2.96E-15 & 2620  & 2.19E15 \\
		                                       & B2 & 0.1516   & 2.0E-11 & 4.90E-19 & 2548  & 2.36E15 \\
		                                       & B3 & 0.3262   & 1.5E-12 & 6.54E-18 & 2462  & 2.32E15 \\
		                                       & B4 & 0.53     & 8.1E-10 & 1.24E-20 & 3082  & 2.36E15 \\ \hline
		\end{tabular}
	}
\end{table}
%%%%%%%%%%%%%%%%%%%%%%%%%%%%%%%%%%%%%%%%%%%%%%%%%%%%%%%%%%%%%%%%%%%%%%%%%%%%%%%%%%%%%%%%%%%%%%%%%%%%%%%%%%%%%%%%%%%%%%%%%%%%%%%%%%%%%%%%%%%%

%%%%%%%%%%%%%%%%%%%%%%%%%%%%%%%%%%%%%%%%%%%%%%%%%%%%%%%%%%%%%%%%%%%%%%%%%%%%%%%%%%%%%%%%%%%%%%%%%%%%%%%%%%%%%%%%%%%%%%%%%%%%%%%%%%%%%%%%%%%%
%%%%%%%%%%%%%%%%%%%%%%%%%%%%%%%%%%%%%%%%%%%%%%%%%%%%%%%%%%%%%%%%%%%%%%%%%%%%%%%%%%%%%%%%%%%%%%%%%%%%%%%%%%%%%%%%%%%%%%%%%%%%%%%%%%%%%%%%%%%%

\subsubsection{Profiles of polytropes with $n = 4.5$}
\begin{figure}[h]
	   \begin{center}
            \includegraphics[width=\linewidth]{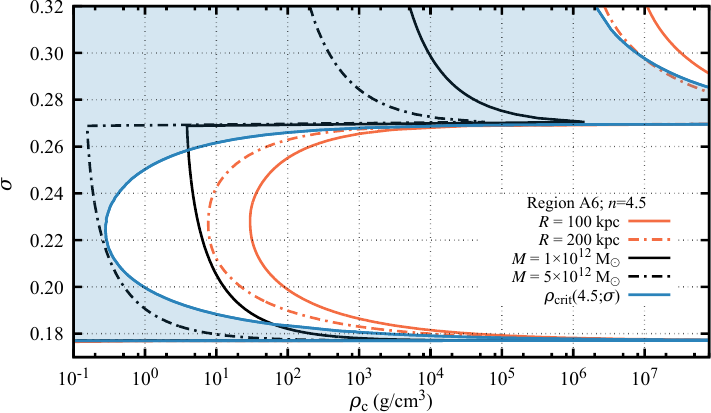}
			\caption{\label{fig:profiles_regime2_case1_n45_a6}Close-up of the region A6 for the polytropic sphere $n = 4.5$ representing galaxy-like scheme. Here the set conditions for $R$ and $M$ have no common region.}
	   \end{center}
    \end{figure}

The allowed regions are related to three critical points of the relativistic parameter. The selected values of the parameters $\sigma$ and $\rho_\mathrm{c}$ are given in Table~\ref{tab:centre_45} for both the galaxy-like and cluster-like polytropes. There are five allowed regions for the galaxy-like polytropes, but only two of them for the cluster-like polytropes. The A1, A2 regions are related to the first critical point of $\sigma$, the regions A3, A4 are related to the second critical point, and the A5 region relates to the third critical point. The A6 region is of special character as it is not representing an allowed region but serves as an example of the situation where the conditions for the extension and mass of the polytrope have no intersection, giving an allowed region --- see Fig.~\ref{fig:profiles_regime2_case1_n45_a6}.

\begin{figure*}[ht]
	   \begin{center}
            \includegraphics[width=.49\linewidth]{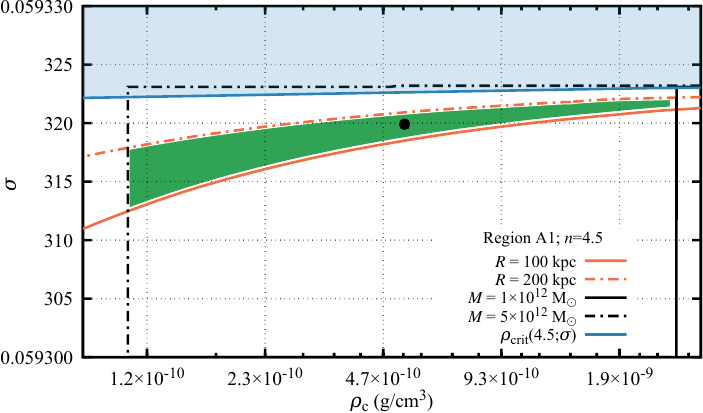}\hfill\includegraphics[width=.49\linewidth]{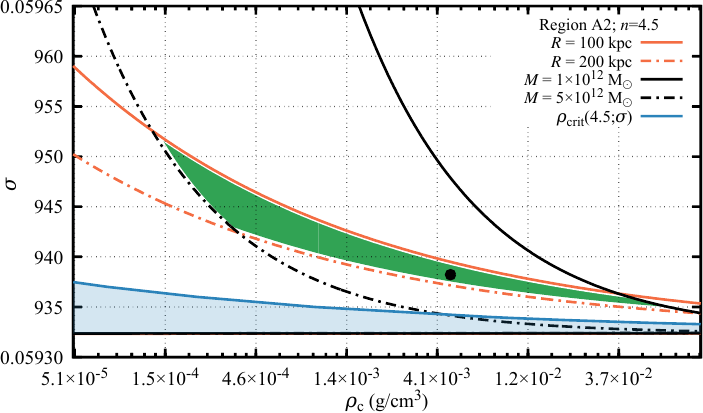}\\[3mm]
			\includegraphics[width=.49\linewidth]{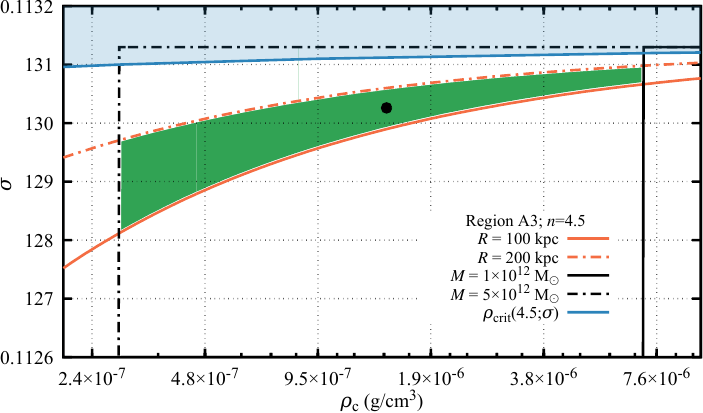}\hfill\includegraphics[width=.49\linewidth]{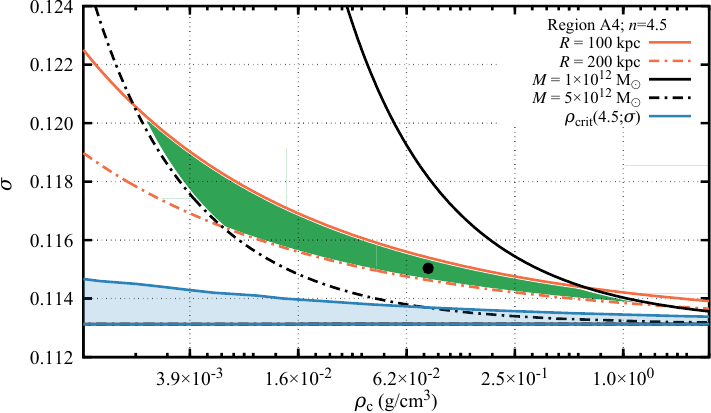}\\[3mm]
			\includegraphics[width=.49\linewidth]{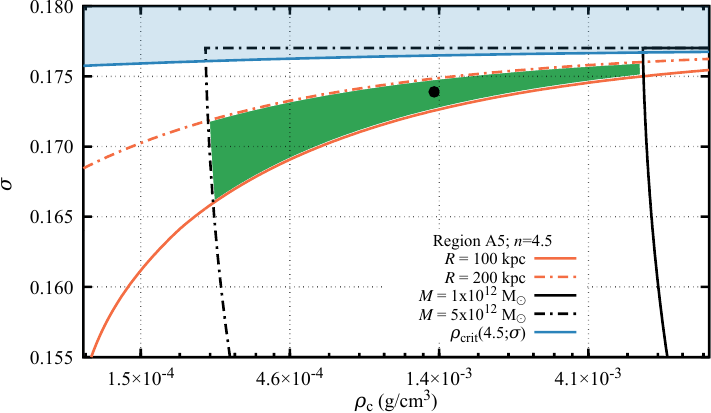}\hfill
			\caption{\label{fig:profiles_regime2_case1_n45} Details of the allowed regions A1--A5 for the polytropes with index $n = 4.5$. The corresponding radial profiles for the selected values (see Table~\ref{tab:centre_45}), depicted here by the black dots, are given in Fig.~\ref{fig:profiles_regime2_case2_n45}}
	   \end{center}
    \end{figure*}

\begin{figure*}[ht]
       \begin{center}
           \includegraphics[width=.49\linewidth]{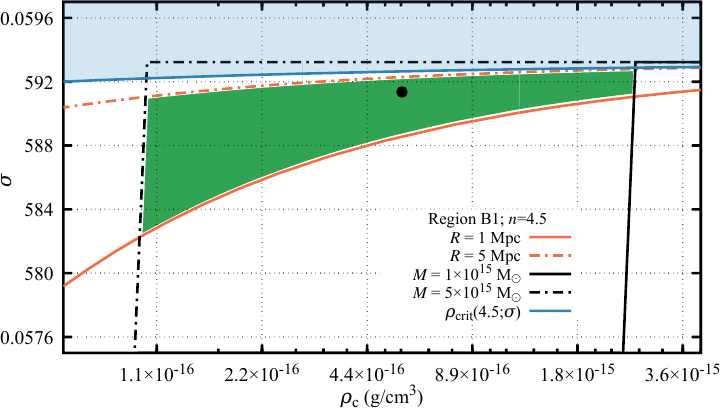}\hfill%
		   \includegraphics[width=.49\linewidth]{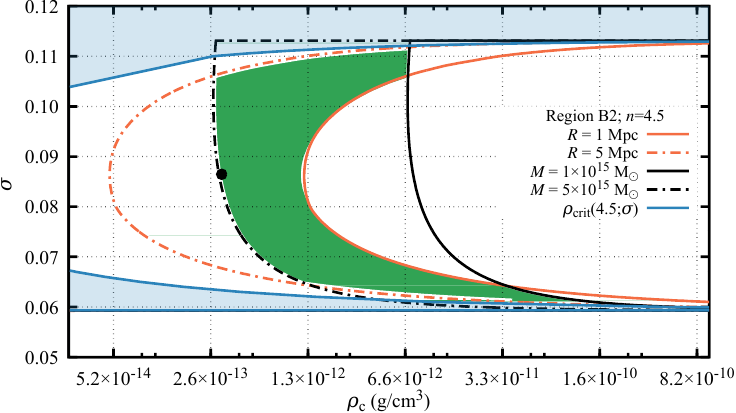}
            \caption{\label{fig:profiles_regime2_case2_n45_a} Details of the allowed regions B1--B2 for the polytropes with index $n = 4.5$. The corresponding radial profiles for the selected values (see Table~\ref{tab:centre_45}), depicted here by the black dots, are given in Fig.~\ref{fig:profiles_regime2_case2_n45_b}}
	   \end{center}
\end{figure*}

The Figures \ref{fig:profiles_regime2_case1_n45}, \ref{fig:profiles_regime2_case2_n45_a} introduce the detailed shape of the corresponding polytrope allowed regions. The related radial profiles are shown in Fig.~\ref{fig:profiles_regime2_case2_n45} for the galaxy-like polytropes (five regions A1--A5), and Fig.~\ref{fig:profiles_regime2_case2_n45_b} for the cluster-like polytropes (two regions B1--B2). 

All the extended relativistic polytropes have again extremely large dimensionless radius $\xi_1$, while the dimensionless mass parameter $\nu_1$ can be both large and small.

In the group of the galaxy-like polytropes, the A1 and A3 polytropes are the typical extremely concentrated polytropes, but for the A3 case the mass parameter is rather high, $\nu_1 \sim 21$. On the other hand, the polytropes A2, A4 and A5 are of the irregular type, having large values of $\nu_1 \sim 10^3$. The B1 polytrope is typical extremely concentrated one, with very small $\nu_1 < 1$, and the B2 polytrope is an irregular one with large $\nu_1 \sim 37$.

%%%%%%%%%%%%%%%%%%%%%%%%%%%%%%%%%%%%%%%%%%%%%%%%%%%%%%%%%%%%%%%%%%%%%%%%%%%%%%%%%%%%%%%%%%%%%%%%%%%%%%%%%%%%%%%%%%%%%%%%%%%%%%%%%%%%%%%%%%%%
    \begin{table}[ht]
        \centering
        \caption{\label{tab:centre_45}Selected points for the polytropic sphere $n = 4.5$ used in the Figs.~\ref{fig:profiles_regime2_case1_n45} \& \ref{fig:profiles_regime2_case2_n45_a} \& \ref{fig:profiles_regime2_case2_n45} \& \ref{fig:profiles_regime2_case2_n45_b} (right column).}

    	\resizebox{\columnwidth}{!}{%
	       \begin{tabular}{cclllll}\hline\hline
		      \multicolumn{1}{l}{Scheme} & Region & $\sigma$ & $\rho_\mathrm{c}$ ($\mathrm{g\, cm}^{-3}$) & $\lambda$  & $R$ (kpc) & $M$ ($M_{\odot}$)\\ \hline
		      \multirow{5}{*}{Galaxy-like}	& A1 & 0.0593199 & 5.3E-10 & 1.89E-20 & 139.4 & 2.22E12\\
			                     			& A2 & 0.0593822 & 4.8E-3  & 2.07E-27 & 140.9 & 2.07E12\\
						                    & A3 & 0.113026  & 1.5E-6  & 6.88E-24 & 134.7 & 2.20E12\\
						                    & A4 & 0.115036  & 8.2E-2  & 1.21E-28 & 133.7 & 2.00E12\\
						                    & A5 & 0.17389   & 1.3E-3  & 7.56E-27 & 140.4 & 2.16E12\\ \hline
		      \multirow{2}{*}{Cluster-like}	& B1 & 0.0591359 & 5.6E-16 & 1.79E-14 & 2507  & 2.15E15\\
			                     			& B2 & 0.086458  & 3.1E-13 & 3.22E-17 & 1918  & 4.88E15\\ \hline
		\end{tabular}
	}
\end{table}
%%%%%%%%%%%%%%%%%%%%%%%%%%%%%%%%%%%%%%%%%%%%%%%%%%%%%%%%%%%%%%%%%%%%%%%%%%%%%%%%%%%%%%%%%%%%%%%%%%%%%%%%%%%%%%%%%%%%%%%%%%%%%%%%%%%%%%%%%%%%

%%%%%%%%%%%%%%%%%%%%%%%%%%%%%%%%%%%%%%%%%%%%%%%%%%%%%%%%%%%%%%%%%%%%%%%%%%%%%%%%%%%%%%%%%%%%%%%%%%%%%%%%%%%%%%%%%%%%%%%%%%%%%%%%%%%%%%%%%%%%
%%%%%%%%%%%%%%%%%%%%%%%%%%%%%%%%%%%%%%%%%%%%%%%%%%%%%%%%%%%%%%%%%%%%%%%%%%%%%%%%%%%%%%%%%%%%%%%%%%%%%%%%%%%%%%%%%%%%%%%%%%%%%%%%%%%%%%%%%%%%

%%%%%%%%%%%%%%%%%%%%%%%%%%%%%%%%%%%%%%%%%%%%%%%%%%%%%%%%%%%%%%%%%%%%%%%%%%%%%%%%%%%%%%%%%%%%%%%%%%%%%%%%%%%%%%%%%%%%%%%%%%%%%%%%%%%%%%%%%%%%
%%%%%%%%%%%%%%%%%%%%%%%%%%%%%%%%%%%%%%%%%%%%%%%%%%%%%%%%%%%%%%%%%%%%%%%%%%%%%%%%%%%%%%%%%%%%%%%%%%%%%%%%%%%%%%%%%%%%%%%%%%%%%%%%%%%%%%%%%%%%

%%%%%%%%%%%%%%%%%%%%%%%%%%%%%%%%%%%%%%%%%%%%%%%%%%%%%%%%%%%%%%%%%%%%%%%%%%%%%%%%%%%%%%%%%%%%%%%%%%%%%%%%%%%%%%%%%%%%%%%%%%%%%%%%%%%%%%%%%%%%
%%%%%%%%%%%%%%%%%%%%%%%%%%%%%%%%%%%%%%%%%%%%%%%%%%%%%%%%%%%%%%%%%%%%%%%%%%%%%%%%%%%%%%%%%%%%%%%%%%%%%%%%%%%%%%%%%%%%%%%%%%%%%%%%%%%%%%%%%%%%

%%%%%%%%%%%%%%%%%%%%%%%%%%%%%%%%%%%%%%%%%%%%%%%%%%%%%%%%%%%%%%%%%%%%%%%%%%%%%%%%%%%%%%%%%%%%%%%%%%%%%%%%%%%%%%%%%%%%%%%%%%%%%%%%%%%%%%%%%%%%

\section{Velocity curves in the selected polytropic spheres}
For the selected polytropic spheres that could be considered as potential DM halo models, we also construct the model of the velocity curves of stars orbiting in the galactic plane. For this purpose, we give the velocity radial profiles of the circular geodesics in the equatorial plane of the polytrope internal spacetime. In the case of the polytropic spheres, the metric coefficients are constructed numerically; therefore, the velocity radial profiles of circular geodesics will also be determined numerically and roughly compared to the observed velocity profiles.

\subsection{Velocity radial profiles}
For the selected polytropic spheres that could serve as the DM halo models, we give the specific energy and specific axial angular momentum radial profiles in Appendix~\ref{APPENDIX1} for the case of non-relativistic polytropes (first regime of fitting procedure), covering thus both the galaxy-like polytropes and the cluster-like polytropes, and in Appendix~\ref{APPENDIX2} for the relativistic polytropes (second regime).

We can see that in the non-relativistic polytropes the flatness of the specific energy radial profile increases with decreasing polytropic index $n$. In all the cases the lowest energy is at the centre, giving the binding energy of the circular geodesics. The binding energy increases with increasing $n$, and it is substantially higher for the cluster-like polytropes in comparison with the galaxy-like ones if $n$ is fixed. In the relativistic extended polytropes the specific energy radial profile decreases extremely fast near the centre. The binding energy increases with increasing $n$ and it is larger for the cluster-like polytropes in comparison to the galaxy-like polytropes.

We construct the representative radial profiles of the velocity measured by distant observers and related to the circular geodesics inside the selected polytropic spheres that could represent the CDM halos in large galaxies or galaxy clusters. As in the previous sections, we separate the construction of the velocity radial profiles for the non-relativistic and relativistic polytropic spheres.

\subsubsection{Non-relativistic polytropes}
\begin{figure*}[ht]
	   \begin{center}
			\includegraphics[width=.49\linewidth]{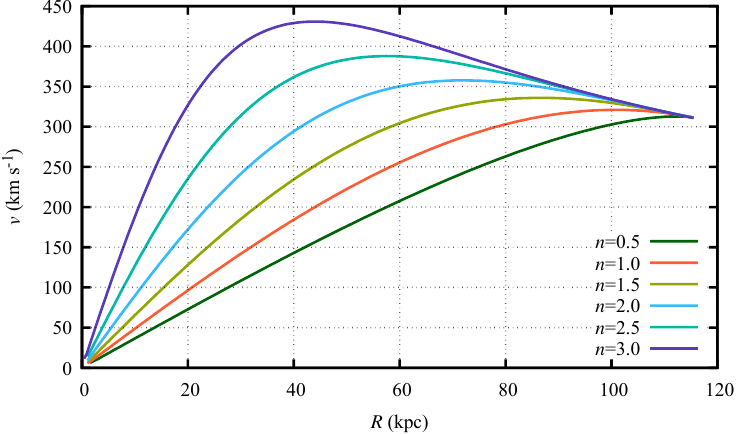}\hfill\includegraphics[width=.49\linewidth]{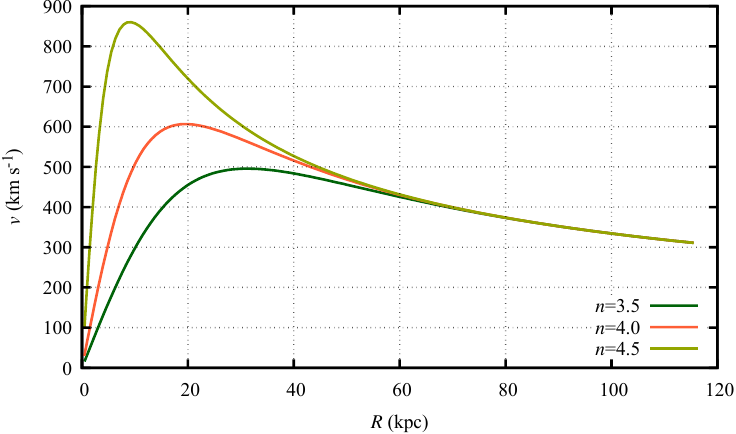}\\
			\includegraphics[width=.49\linewidth]{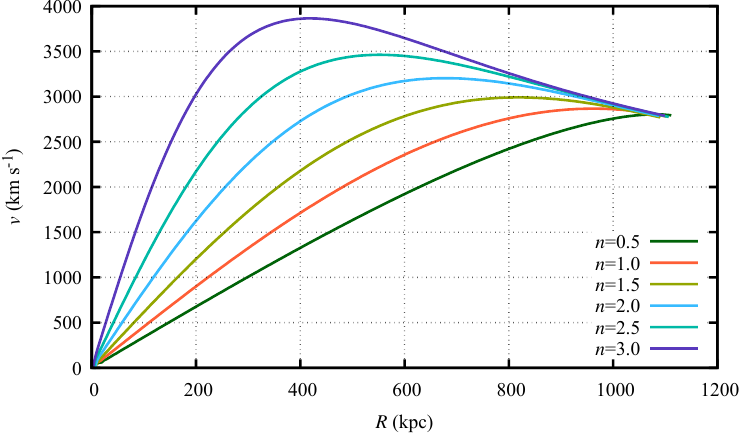}\hfill\includegraphics[width=.49\linewidth]{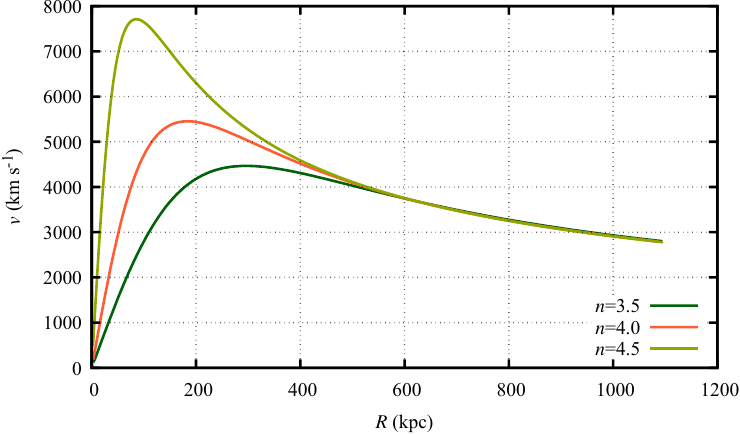}			
			\caption{\label{fig:velocity_case1_regime1}Velocity profiles of the galaxy-like (first row) and the cluster-like (second row) cases for the non-relativistic polytropes.}
	   \end{center}
    \end{figure*}

The velocity radial profiles of the circular geodesics are represented in Figs.~\ref{fig:velocity_case1_regime1} for all the considered values of the polytropic index, namely $n = 0.5,\, 1,\, 1.5,\, 2,\, 2.5,\, 3.0,\, 3.5,\, 4,\, 4.5$. We give them for both the galaxy-like polytropes and the cluster-like polytropes.

We can see that the non-relativistic polytropes with $n > 1.5$ give velocity profiles with a local maximum. For the polytropes with $n > 3$ the profiles behind the maximum have essentially Keplerian character, excluding them qualitatively as candidates for the galactic halos. However, for the polytropes with $1.5 \leq n \leq 3$, the profiles have quasi-Keplerian character in the outer regions behind the maximum, being sufficiently flat to enable their consideration as potential candidates for the galactic halos. Moreover, for the polytropes with $n < 1.5$, the rotational velocity radial profiles are purely increasing with radius and can be considered as candidates for halos of dwarf galaxies.

%%%%%%%%%%%%%%%%%%%%%%%%%%%%%%%%%%%%%%%%%%%%%%%%%%%%%%%%%%%%%%%%%%%%%%%%%%%%%%%%%%%%%%%%%%%%%%%%%%%%%%%%%%%%%%%%%%%%%%%%%%%%%%%%%%%%%%%%%%%%
%%%%%%%%%%%%%%%%%%%%%%%%%%%%%%%%%%%%%%%%%%%%%%%%%%%%%%%%%%%%%%%%%%%%%%%%%%%%%%%%%%%%%%%%%%%%%%%%%%%%%%%%%%%%%%%%%%%%%%%%%%%%%%%%%%%%%%%%%%%%

\subsubsection{Relativistic polytropes}
In this case, we give the velocity radial profiles separately for the three considered cases of polytropic index $n = 3.5,\, 4,\, 4.5$.

\begin{figure*}[ht]
	   \begin{center}
            \includegraphics[width=.49\linewidth]{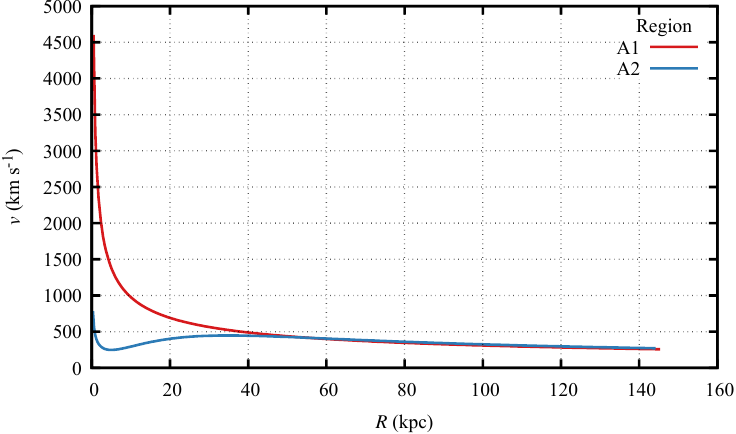}\hfill\includegraphics[width=.49\linewidth]{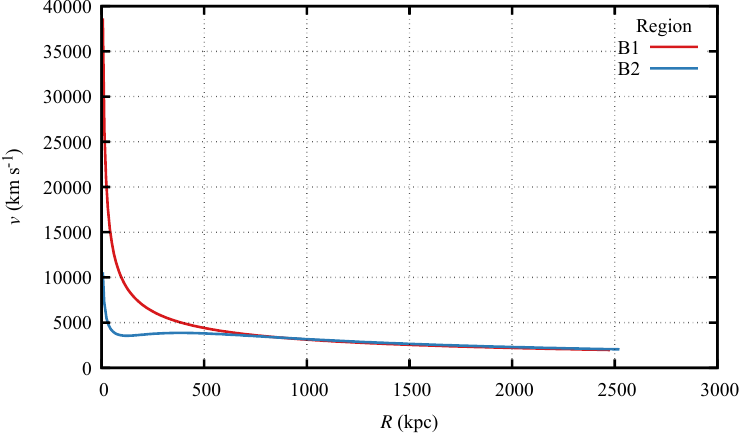}
			\caption{\label{fig:velocity_regime2_n35} Velocity profiles for the polytropic spheres with the index $n = 3.5$, related to the large galaxies (left column) and clusters of galaxies (right column).}
	   \end{center}
    \end{figure*}

In the case of the polytropes with $n = 3.5$, the velocity profiles are illustrated in Fig.~\ref{fig:velocity_regime2_n35} (left column) where we consider the galaxy-like polytropes, and in Fig.~\ref{fig:velocity_regime2_n35} (right column) for the cluster-like polytropes. In both cases, there are two regions of the relativistic parameter $\sigma$ that should be considered. The selected parameters $\sigma, \rho_\mathrm{c}$ are listed in Table~\ref{tab:centre_35}.

%%%%%%%%%%%%%%%%%%%%%%%%%%%%%%%%%%%%%%%%%%%%%%%%%%%%%%%%%%%%%%%%%%%%%%%%%%%%%%%%%%%%%%%%%%%%%%%%%%%%%%%%%%%%%%%%%%%%%%%%%%%%%%%%%%%%%%%%%%%%
%%%%%%%%%%%%%%%%%%%%%%%%%%%%%%%%%%%%%%%%%%%%%%%%%%%%%%%%%%%%%%%%%%%%%%%%%%%%%%%%%%%%%%%%%%%%%%%%%%%%%%%%%%%%%%%%%%%%%%%%%%%%%%%%%%%%%%%%%%%%

\begin{figure*}[ht]
	   \begin{center}
			\includegraphics[width=.49\linewidth]{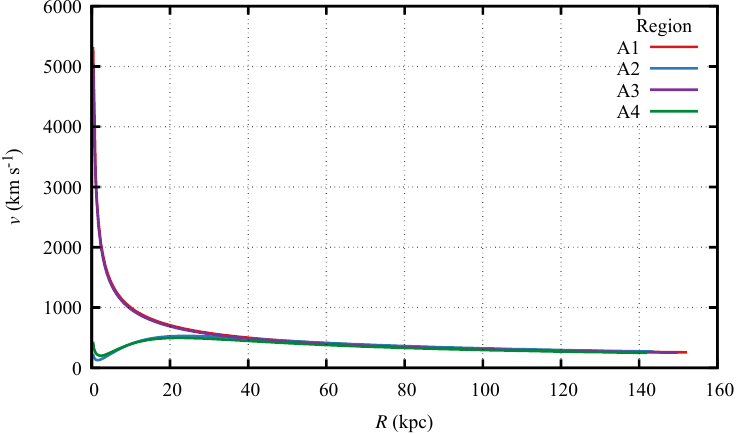}\hfill\includegraphics[width=.49\linewidth]{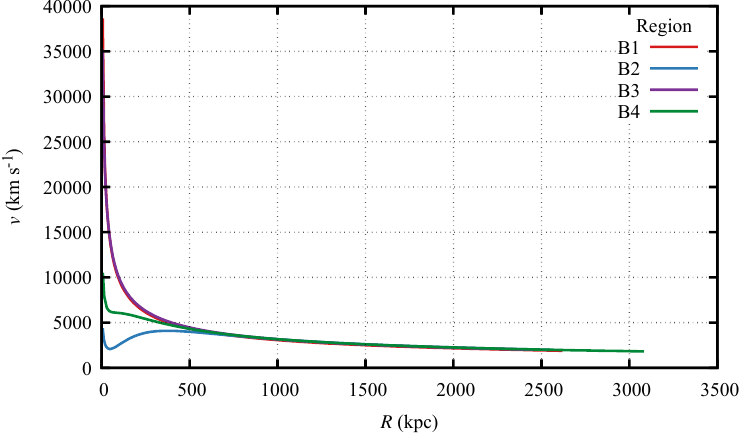}			
			\caption{\label{fig:velocity_regime2_n40} Velocity profiles for the polytropic spheres with the index $n = 4.0$, related to the large galaxies (left column) and clusters of galaxies (right column).}
	   \end{center}
    \end{figure*}

For polytropic spheres with $n = 4.0$ we follow the previous case of $n = 3.5$, but in this case, there are four regions of the relativistic parameter that have to be considered. The selected parameters $\sigma$, $\rho_\mathrm{c}$ are listed in Table~\ref{tab:centre_40}. The results are given in Fig.~\ref{fig:velocity_regime2_n40}.

%%%%%%%%%%%%%%%%%%%%%%%%%%%%%%%%%%%%%%%%%%%%%%%%%%%%%%%%%%%%%%%%%%%%%%%%%%%%%%%%%%%%%%%%%%%%%%%%%%%%%%%%%%%%%%%%%%%%%%%%%%%%%%%%%%%%%%%%%%%%
%%%%%%%%%%%%%%%%%%%%%%%%%%%%%%%%%%%%%%%%%%%%%%%%%%%%%%%%%%%%%%%%%%%%%%%%%%%%%%%%%%%%%%%%%%%%%%%%%%%%%%%%%%%%%%%%%%%%%%%%%%%%%%%%%%%%%%%%%%%%

\begin{figure*}[htb]
	   \begin{center}
            \includegraphics[width=.49\linewidth]{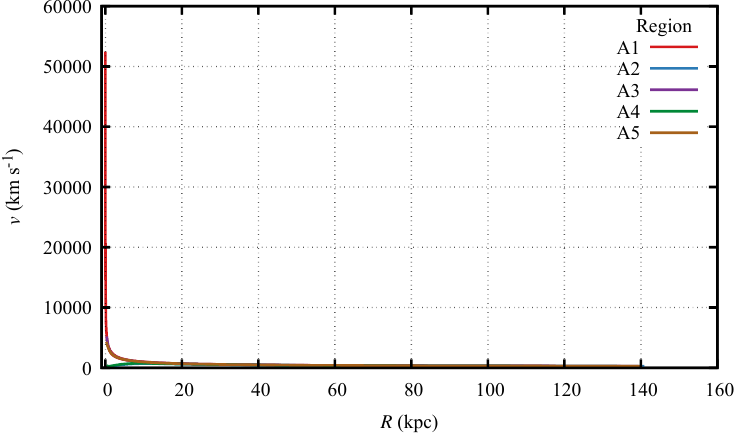}\hfill\includegraphics[width=.49\linewidth]{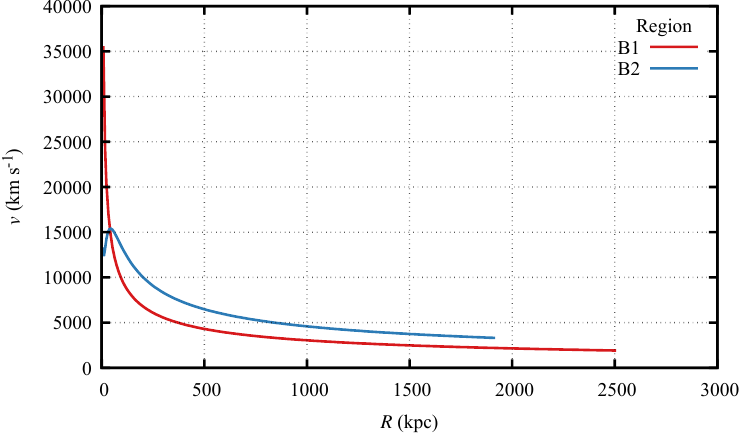}			
			\caption{\label{fig:velocity_regime2_n45} Velocity profiles for the polytropic spheres with the index $n = 4.5$, related to the large galaxies (left column) and clusters of galaxies (right column).}
	\end{center}
\end{figure*}

For polytropic spheres with $n = 4.5$ the situation is more complex; there are six regions of interest. The selected parameters $\sigma$, $\rho_\mathrm{c}$ are listed in Table~\ref{tab:centre_45}.
The resulting velocity profiles are illustrated in Fig.~\ref{fig:velocity_regime2_n45}.

%%%%%%%%%%%%%%%%%%%%%%%%%%%%%%%%%%%%%%%%%%%%%%%%%%%%%%%%%%%%%%%%%%%%%%%%%%%%%%%%%%%%%%%%%%%%%%%%%%%%%%%%%%%%%%%%%%%%%%%%%%%%%%%%%%%%%%%%%%%%

The extremely concentrated relativistic polytropes have strongly decreasing velocity profiles and can be excluded as candidates for the galactic halos. However, the spread relativistic polytropes demonstrate humpy velocity profiles that are very flat at the outer regions and could thus be considered as possible candidates for the halos.

%%%%%%%%%%%%%%%%%%%%%%%%%%%%%%%%%%%%%%%%%%%%%%%%%%%%%%%%%%%%%%%%%%%%%%%%%%%%%%%%%%%%%%%%%%%%%%%%%%%%%%%%%%%%%%%%%%%%%%%%%%%%%%%%%%%%%%%%%%%%
\section{Simple tests of applicability of the polytrope rotational velocity profiles}
In order to test the polytropic spheres as possible models of galactic or galaxy cluster halos, we have to use the observational data of velocity profiles obtained for individual galaxies (clusters) and match the data using the velocity profiles related to the polytropic spheres in combination with the velocity profiles related to the galaxy disc (bulge) or gas connected to them. Such detailed matching procedures are, of course, rather complex and time-consuming, and we plan them for a future paper. Here we realize as a simple preliminary test of our models a comparison of the velocity profiles predicted by the polytropic models with the velocity profiles determined by the standard approximate models of the DM halos applied in some of the previous studies of matching the observational data related to large galaxies and galaxy clusters \footnote{Note that the case of the dwarf galaxies will be treated separately, as the velocity profiles related to the polytropic spheres can be used for matching to the observational data solely.}.

In order to test our polytropic spheres, we apply the obtained geodesic rotational velocity radial profiles of the polytropes from the regions of allowed, restricted from above by the influence of the cosmological constant on the polytrope extension and mass, and from below by the extension of the strong influence of the galaxy disc. Note that the polytrope sphere in regions corresponding to the dominance of the galaxy disc should be applied in the direct matching to the observational data.

\subsection{Galaxy case}
In order to simplify the testing of our results, we make a comparison to the results obtained by the standard approximate models of the CDM galaxy halos. We take the two frequently used models, namely the Navarro--Frenk--White (NFW) density profile \cite{Nav-Fre-Whi:1997:ASTRJ2:UniDeProHiCl} and a core model with the density profile proposed by \cite{Bur:1995:AAL:}, using the velocity profiles corresponding to their contribution in matching the observational data. A similar approach is applied in \cite{Cab-etal:2004:GRG:NFWpoly}.

The NFW density profile is given by the relation \cite{Nav-Fre-Whi:1997:ASTRJ2:UniDeProHiCl}
    \begin{equation}
	   \rho_{\rm NFW}(R) = \frac{\rho_0}{X(1 + X)^2}\, ,
    \end{equation}
where $X = R/h$, and $\rho_0$ is the representative (scale) density and $h$ is the scale radius of the CDM halo. The mass within radius $R$ and the circular rotation velocity profile $v_{\rm NFW}$ are given by the relations
    \begin{align}
        M_{\rm NFW}(R) & = 4\pi \rho_0 h^3  \left[\ln \left(1 + X\right) - \frac{X}{1 + X}\right]\, ,\\
        v_{\rm NFW}(R) & = \sqrt{G M_{\rm NFW}(R)/R}\, .
    \end{align}
For the matching procedure, we have to choose the scale factors $\rho_0$ and $h$. For our testing of the polytrope spheres, we used the results obtained for the galaxy M33 in the paper of \cite{Cor-etal:2014:AA:Dyn:}. Note that in the mentioned paper, the authors obtained the characteristic value of the so-called concentration parameter $c = (9.5\pm{}1.5)$ and the value of the virial mass $M_{\rm vir} = (4.3\pm{}1.0)\,\Msun$. We have to relate the mass of our polytrope spheres to this virial mass.

The Burkert density profile is given by
    \begin{equation}
	   \rho_{\rm B}(R) = \frac{\rho_{\rm B;0}}{(1 + Y)(1 + Y^2)}
    \end{equation}
where the parameter $Y = R/R_{\rm B}$ and $\rho_{\rm B;0}$ is the dark matter density of the core; the characteristic radius $R_{\rm B}$ means the core radius. The rotational velocity radial profile related to the Burkert density profile then reads
    \begin{multline}
        v_{\rm B}^2(R) = \frac{2\pi G \rho_{\rm B;0} R^3_{\rm B}}{R}\left(\vphantom{Y^2}\ln(1 + Y) + \right.\\
        \left. 0.5\ln(1 + Y^2)-\tan^{-1}(Y)\right).
    \end{multline}
For the matching procedure we again use the M33 galaxy study presented in \cite{Cor-etal:2014:AA:Dyn:}.

In \cite{Cor-etal:2014:AA:Dyn:}, the NFW or Burkert halo velocity profiles were combined in the standard Newtonian framework with the velocity profiles related to those corresponding to the galaxy disc (bulge) to match the observational data. We match here the polytrope velocity radial profiles only to the velocity profiles related to the NFW and Burkert density profile obtained for the halo of the galaxy M33. In such a way, we overcome the difficulty of mixing the influence of our polytrope models that are fully general relativistic with the influence of the visible parts of the galaxy due to a simple comparison of the velocity curves of our polytrope model with the velocity curve given by the NFW (Burkert) model of the density profile. Both the velocity profiles are related to the distant (Newtonian) observers. In the matching procedure, we use as free parameters the polytrope index $n$ and the relativistic parameter $\sigma$, realizing the matching for the case of the M33 galaxy in the interval of radii 40--200\,kpc.

\begin{figure}[htb]
	   \begin{center}
			\includegraphics[width=\linewidth]{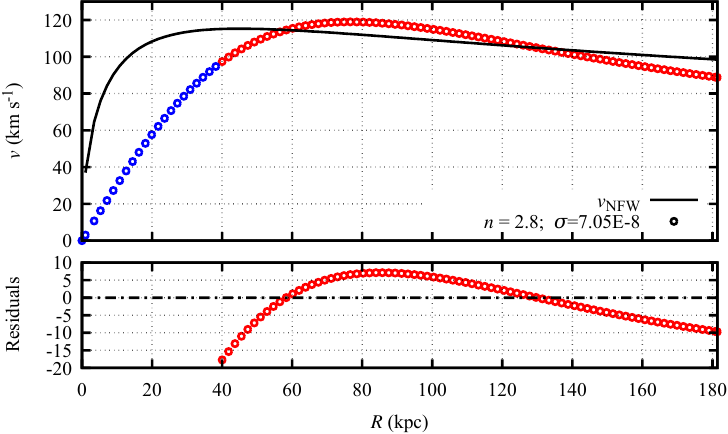}\\[3mm]
			\includegraphics[width=\linewidth]{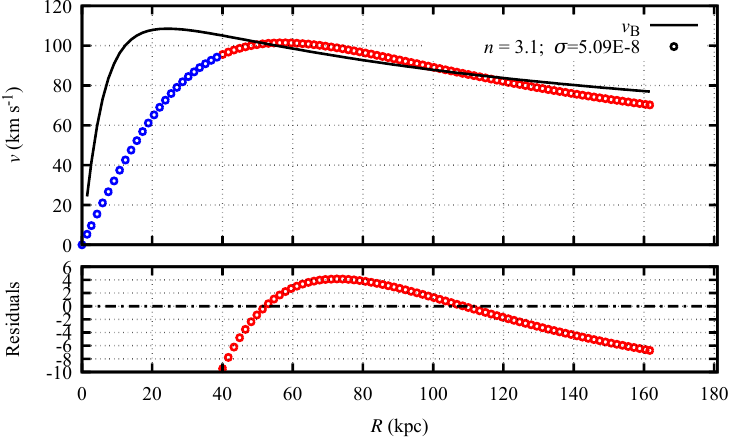}			
            \caption{\label{fig:vel_fit_galaxies}Best match of the polytrope velocity profiles (circles) to the velocity profiles related to the NFW model (top; solid line) and the Burkert model (bottom; solid line). The blue circles in both cases represent the points not included into the matching procedure, while the red ones are included. For completeness, we include also the residuals in both cases in dependence the radius.}
	   \end{center}
    \end{figure}

The results of the matching procedure of the velocity curves of our polytrope model with both velocity curves given by the NFW and Burkert density models are represented in Fig.~\ref{fig:vel_fit_galaxies}. We have to stress that in order to get significant matches, we have to remove from the matching procedure the region of 0--40\,kpc where the galaxy disc influence gives a dominant contribution. We thus compare the regions where the halo velocity contribution dominates.

The best match to the NFW model is obtained for the polytropic sphere with the polytropic index $n = 2.8$ and the relativistic parameter $\sigma = 7.05 \times 10^{-8}$. In this case, we obtain the total halo mass $M = 3.4\,\times 10^{11}\,\Msun$, which is in good correspondence with the mass obtained by \cite{Cor-etal:2014:AA:Dyn:}, i.e., $M_{\rm h} = (4.3 \pm 1.0) \times 10^{11}\Msun$. In the case of the Burkert profile, we found the best match for the polytrope having $n = 3.1$ and $\sigma = 5.09 \times 10^{-8}$, which implies the total polytrope mass $M = 1.97\times 10^{11}\,\Msun$. We see that in both cases, the best match is given by the non-relativistic polytropic configurations, with polytropic index close to $n = 3$.

%%%%%%%%%%%%%%%%%%%%%%%%%%%%%%%%%%%%%%%%%%%%%%%%%%%%%%%%%%%%%%%%%%%%%%%%%%%%%%%%%%%%%%%%%%%%%%%%%%%%%%%%%%%%%%%%%%%%%%%%%%%%%%%%%%%%%%%%%%%%
%%%%%%%%%%%%%%%%%%%%%%%%%%%%%%%%%%%%%%%%%%%%%%%%%%%%%%%%%%%%%%%%%%%%%%%%%%%%%%%%%%%%%%%%%%%%%%%%%%%%%%%%%%%%%%%%%%%%%%%%%%%%%%%%%%%%%%%%%%%%

\subsection{Cluster case}
For testing our polytropic spheres in modeling the halos of galaxy clusters, we use the previous study related to the rotational motion in galaxy clusters \citep{Man-Pli:2016:MNRAS:Gal}. In this work, the authors used for their validation test a virialized cluster with a mass of $4\times10^{14}\,\Msun$, radius $R_{\rm cl}$ = 1\,Mpc, and core radius $R_{\rm c}$ = 0.1\,Mpc. The density radial profile of the cluster is represented by the distribution proposed in \cite{King:1962:AJ:} that reads
    \begin{equation}
	   \rho(R) = \frac{\rho_{{\rm K};0}}{(1 +  (R/R_{\rm c})^2)^{3/2}}\, ,
    \end{equation}
where $\rho_{{\rm K};0}$ is the central density of the cluster, and its value is obtained by using the mass $M_{\rm cl}$, given by the relation
    \begin{equation}
        M_{\rm cl} = \frac{4}{3} \frac{\pi R^3 \rho_{{\rm K};0}}{\Big(1 +  (R/R_{\rm c})^2\Big)^{3/2}}\, .
    \end{equation}
By assuming the cluster is virialized, the rotational velocity profile $v_{\rm K}$ is given by
    \begin{equation}
        v_{\rm K}^2(R) = \frac{2}{3}\frac{G \pi \rho_{{\rm K};0} R^2}{\Big(1 +  (R/R_{\rm c})^2\Big)^{3/2}}\, .
    \end{equation}
We again compare the velocity curves related to the distant Newtonian observers.

\begin{figure}[htb]
	   \begin{center}
			\includegraphics[width=\linewidth]{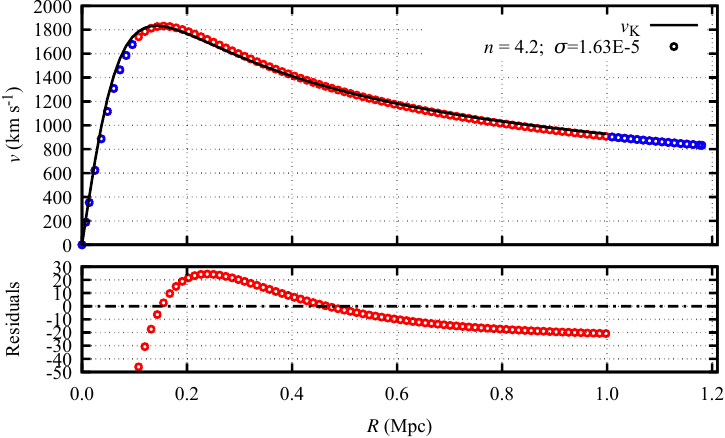}\\[3mm]
			\includegraphics[width=\linewidth]{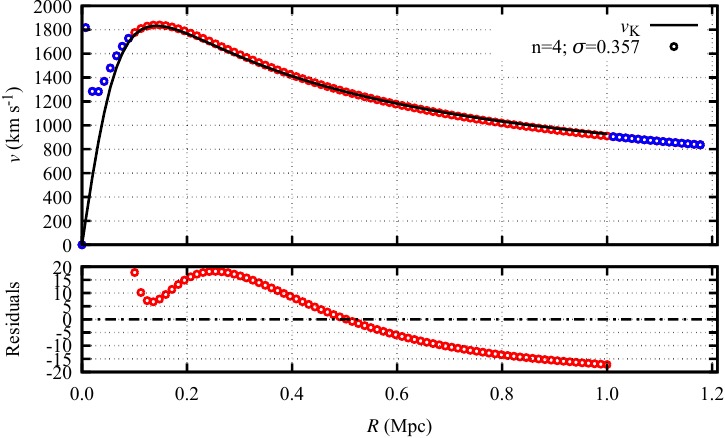}			
            \caption{\label{fig:vel_fit_cluster}Best match of the polytrope velocity profiles (circles) to the velocity profiles related to the King velocity profile $v_{\rm K}$ (line) is given for the non-relativistic polytropic sphere (top) and the relativistic polytrope of the spread kind (bottom). The red dots are the points included for the matching procedure, while the blue points are not included. We also show the residuals for monitoring of the process in dependence on the radius.}
	   \end{center}
    \end{figure}

The results of the matching procedure are presented in Fig.\,\ref{fig:vel_fit_cluster}. We have found that in this special case the matching is possible in two regimes: for the non-relativistic polytropes where the best match is obtained for the polytrope with $n = 4.2$ and the relativistic parameter $\sigma = 1.63 \times 10^{-5}$, and for the relativistic polytropes of the spread kind where the best match is obtained for the polytrope with $n = 4$ and the relativistic parameter $\sigma = 0.357$.

%%%%%%%%%%%%%%%%%%%%%%%%%%%%%%%%%%%%%%%%%%%%%%%%%%%%%%%%%%%%%%%%%%%%%%%%%%%%%%%%%%%%%%%%%%%%%%%%%%%%%%%%%%%%%%%%%%%%%%%%%%%%%%%%%%%%%%%%%%%%
%%%%%%%%%%%%%%%%%%%%%%%%%%%%%%%%%%%%%%%%%%%%%%%%%%%%%%%%%%%%%%%%%%%%%%%%%%%%%%%%%%%%%%%%%%%%%%%%%%%%%%%%%%%%%%%%%%%%%%%%%%%%%%%%%%%%%%%%%%%%

%%%%%%%%%%%%%%%%%%%%%%%%%%%%%%%%%%%%%%%%%%%%%%%%%%%%%%%%%%%%%%%%%%%%%%%%%%%%%%%%%%%%%%%%%%%%%%%%%%%%%%%%%%%%%%%%%%%%%%%%%%%%%%%%%%%%%%%%%%%%
\section{Discussion and Conclusions}\label{conclus}
We realized a detailed mapping of the extension and mass of the polytropic spheres in spacetimes with relict non-zero vacuum energy, corresponding to the observationally fixed cosmological constant, to the extension and mass of the DM halos in large galaxies or in clusters of galaxies. The static equilibrium polytropic spheres are determined by two coupled first-order nonlinear differential equations, governed by the polytropic index $n$, the relativistic parameter $\sigma$, central energy density $\rho_\mathrm{c}$, and the cosmological parameter $\lambda$ that is determined by the central density, if we consider a fixed value of the cosmological constant $\Lambda$. The exterior of the polytropic spheres is represented by the vacuum Schwarzschild-de Sitter spacetime. The extension of the static polytropic spheres cannot exceed the so-called static radius of the external spacetime where the gravitational attraction is just balanced by the cosmic repulsion, thus giving a natural limit on the extension of gravitationally bound systems \citep{Stu-Hle-Nov:2016:PHYSR4:}.

The mapping of the polytropic spheres has been done for the large galaxy ranges 100\,kpc $< R <$ 200\,kpc and $10^{12} < M/M_{\odot} < 5\times10^{12}$, and the galaxy-cluster ranges 1\,Mpc $< R <$ 5\,Mpc and $10^{15} < M/M_{\odot} < 5\times10^{15}$; extension to larger galaxies or their clusters is straightforward. Properties of the polytropic spheres from the regions of the parameter space allowing for the matching are summarized. For selected typical polytropic spheres, the rotation velocity curves determined by circular geodesics of the polytrope spacetime are given and compared to the observed rotation velocity curves. More precisely, the polytropic rotation velocity curves are compared to the velocity curves predicted by the standard models of the CDM halos applied in the explanation of the observed rotation velocity curves. Our general conclusions hold also for the values of $R$ and $M$ distributed between the chosen intervals and could thus be applied for the extremely large galaxies or smaller clusters.

We have found that the extension and mass parameters of the polytropic spheres could simultaneously correspond to the DM halo extension and mass related to the standard galaxies of the Milky Way type, large galaxies, or their clusters, having thus extension going up to $R \sim$~1\,Mpc and mass going up to $M \sim 10^{15}M_{\odot}$. The mapping of the correspondence can be realized in two regimes of construction of the polytropes, namely for the non-relativistic regime with $\sigma < 10^{-3}$ that is relevant for the whole range of the considered values of the polytropic index, $0 < n < 4.5$, representing CDM halos, and the relativistic regime with $\sigma \geq 0.1$ that can be applied only for $n > 3.3$ in vicinity of the critical values of the relativistic parameter $\sigma_{\mathrm f}$ (corresponding to unlimited spheres in the limit of vanishing vacuum energy) that can represent non-cold DM halos \citep{Stu-Hle-Nov:2016:PHYSR4:,Nil-Ugg:2000:ANNPH1:GRStarPoEqSt}. The allowed regions of the polytrope parameter space $\rho_{\rm c}$--$\sigma$ have the same character for all the values of the polytropic index in the non-relativistic regime, while they have a complex character being strongly dependent on the polytrope index in the relativistic regime.

For the non-relativistic polytropes, the matching with the CDM halos requires strongly non-relativistic fluid with $\sigma < 10^{-4}$ that have to be significantly diluted, with $\rho_\mathrm{c} < 10^{-22}\,\mathrm{g\, cm^{-3}}$. For the relativistic polytropes with index $ n >3.3$, and the relativistic parameter close to the critical values of the relativity parameter $\sigma_\mathrm{f}$, the simultaneous matching of the extension and mass of the DM halos is possible also for relatively high central densities that can be as large as $\rho_{\rm c} \sim 10^{-1}\,\mathrm{g\, cm^{-3}}$, but the central density is strongly case dependent and generally it increases with decreasing extension of the polytrope.

In our previous work we have demonstrated the applicability of non-relativistic polytropic spheres (in the sense of cold matter with $\sigma \ll 1$) in modeling of DM halos in dwarf galaxies \citep{Nov-Stu-Hla:AA:2021}. Here we have demonstrated that the non-relativistic polytropes corresponding to cold matter with $\sigma \ll 1$ can be relevant also for modeling of large galaxies where the general relativistic effects are substantial, but we have made substantial extensions to relativistic polytropic spheres ($\sigma \sim 1$) that open up an interesting new direction of investigations related to the so-called trapping polytropes \citep{Nov-Hla-Stu:2017:PRD:trapping}, where gravitational instability of the central trapping zone indicates the possibility of its collapse and creation of a supermassive central black hole of mass $M_{\rm BH} \sim 10^9\,\Msun$ inside the central region of an extremely extended halo of mass $M\sim 10^{12}\,\Msun$ \citep{Stu-etal:2017:JCAP:}. In the trapping polytropes, the very central region containing trapped null geodesics implying instability against gravitational perturbations contains around $10^{-3}M$ where $M$ denotes the total mass of the polytrope. Therefore, we can predict the existence of supermassive black holes having mass $\sim 10^{12}M_{\odot}$ in galaxy clusters with mass $M \sim 10^{15}M_{\odot}$, if governed by DM halos represented by trapping polytropes of such large mass. Notice that, as expected, the trapping polytropes are unstable against radial pulsations, as shown in Section 5. We could expect stabilization of the DM halos (the rest of the polytropes with collapsing center) due to rotation or some influences induced by the collapse of the central region. The possible creation of a central supermassive black hole due to the gravitational instability of the central region of the trapping polytropes can also be interesting in connection to the fact that a special family of the so-called spread polytropes (that are also trapping polytropes) can well fit the observed extension and mass of DM halos of large galaxies or their clusters, and can simultaneously also well fit the rotational curves of orbiting visible matter. Of course, such relevant coincidences require more detailed treatment of the behaviour of the relativistic polytropes and their coexistence with the central black hole in future studies.

The energy density, pressure, mass and metric coefficient radial profiles of the polytropic spheres differ significantly in the non-relativistic and relativistic regimes. In the non-relativistic regime, the profiles are relatively flat, but their concentration at the centre increases with increasing polytropic index $n$, along with the depth of the gravitational potential governed by the metric coefficient $g_{tt}$. In the allowed regions related to the relativistic regime of polytropes with $n>3.3$, we observe two qualitatively different kinds of the polytropic spheres. In the extremely concentrated ones, the energy density is almost singularly distributed in close vicinity of the centre, where essentially all the polytrope mass is concentrated; their gravitational potential monotonically decreases with an extremely large gradient near the centre, and its depth can be very large in the vicinity of $r = 0$. In the spread relativistic polytropes, the energy density decreases slowly, the polytrope mass increases correspondingly, and the gravitational potential can have a complex character demonstrating two local extrema.

The study of the radial profiles of the geodesic orbital velocity in the polytropic spheres demonstrates the possibility that for some parts of the allowed regions of the polytrope parameter space, the theoretical velocity profiles could correspond, at least partially, to the velocity curves observed in typical galaxies.

We have shown that for the non-relativistic polytropes with $n < 1$ the polytrope rotational velocity profile is strictly increasing, while for $1 < n < 2$ the profile has a maximum, but a nearly flat region behind the maximum. In such cases, the fitting to the observed rotational velocity profiles is possible, as for non-relativistic polytropes with index $n<2$ the outer regions of the geodetical polytrope velocity profile are nearly flat and could be thus related to the velocity curves observed in the external regions of galaxies. Notice that such non-relativistic polytropes could be relevant in dwarf galaxies as shown recently in \cite{Nov-Stu-Hla:AA:2021}. For polytropes with $n \geq 3$, the geodesic velocity profile is rapidly decreasing behind its maximum, demonstrating a nearly Keplerian character. Matching to the observational velocity curves is therefore not possible in such cases.

In the relativistic regime, the polytropes of the extremely concentrated kind give the geodesic rotational velocity profiles of nearly Keplerian character and cannot be applied for fitting to the observed velocity curves. On the other hand, the polytropes of the spread kind give geodesic rotational velocity profiles demonstrating a hump in the central region and a very flat profile at the outer regions of the polytrope, enabling thus the fitting to the observed velocity curves. Moreover, the velocity profiles demonstrating an extended hump could be applied in the case of a cluster of galaxies in order to explain some recent attempts to describe the rotational motion in the cluster structures.

In order to make comparison to the observational data obtained for the galaxy-like or cluster-like structures, we considered previous approximative studies of the fitting to the observational data.  We have demonstrated explicitly that our geodesic velocity profiles can be matched to the velocity profiles constructed due to the NFW (Burkert) CDM galaxy halo models and used in explanation of the rotation velocity curves in the galaxy M33. In such a case we can be approved to exclude from our consideration the influence of the galaxy disc (and bulge). We can see that the NFW (Burkert) velocity profile could be then well matched by our polytrope profiles, if the polytropic index is well tuned. In a similar way we are able to match the profiles introduced in the study of galaxy clusters by our relativistic polytropes of the spread kind, if the region of allowed parameters is properly chosen. Notice that for the galaxy matching the polytropic index $n \sim 3$, while for the galaxy cluster matching, $n \sim 4$; clearly, high polytropic indexes could occur in the early stages of the galaxy evolution what hot DM could be relevant.

We can conclude that because of our preliminary estimates, the polytropic spheres with the vacuum energy (relict cosmological constant) related to recent cosmological observations can be considered as a promising model of the DM halos in both the standard and large galaxies, and even for the galaxy clusters. Of special interest is the issue of the coexistence of supermassive black holes and DM halos. More detailed study is planned for direct matching of the velocity profiles related to both the non-relativistic and relativistic polytropes to the observational data in large galaxies, including the role of the galaxy disc and gas in the matching procedure.

\clearpage
%%%%%%%%%%%%%%%%%%%%%%%%%%%%%%%%%%%%%%%%%%%%%%%%%%%%%%%%%%%%%%%%%%%%%%%%%%%%%%%%%%%%%%%%%%%%%%%%%%%%%%%%%%%%%%%%%%%%%%%%%%%%%%%%%%%%%%%%%%%%
%%%%%%%%%%%%%%%%%%%%%%%%%%%%%%%%%%%%%%%%%%%%%%%%%%%%%%%%%%%%%%%%%%%%%%%%%%%%%%%%%%%%%%%%%%%%%%%%%%%%%%%%%%%%%%%%%%%%%%%%%%%%%%%%%%%%%%%%%%%%
\begin{appendix}

\section{\mbox{}\\ Profiles of the non-relativistic polytropes}\label{APPENDIX1}
\begin{figure*}[ht]
	   \begin{center}
			\includegraphics[width=.31\linewidth]{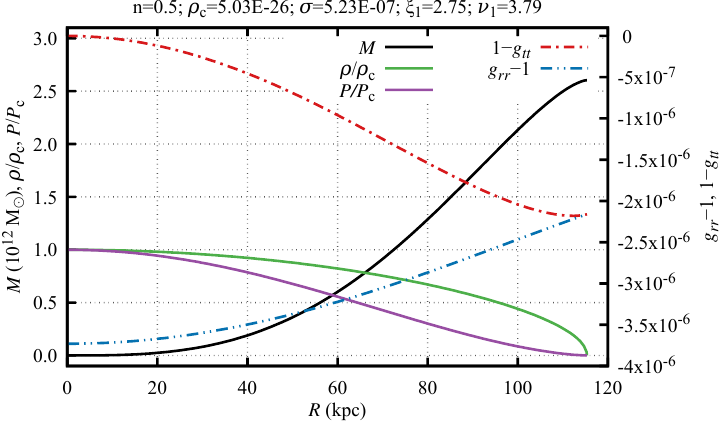}\hfill\includegraphics[width=.31\linewidth]{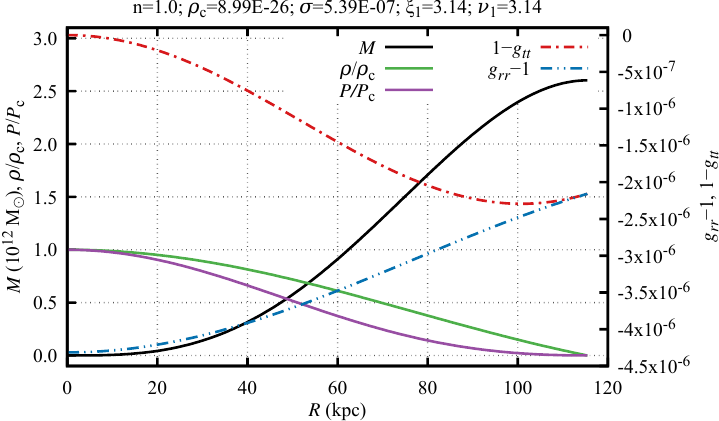}\hfill\includegraphics[width=.31\linewidth]{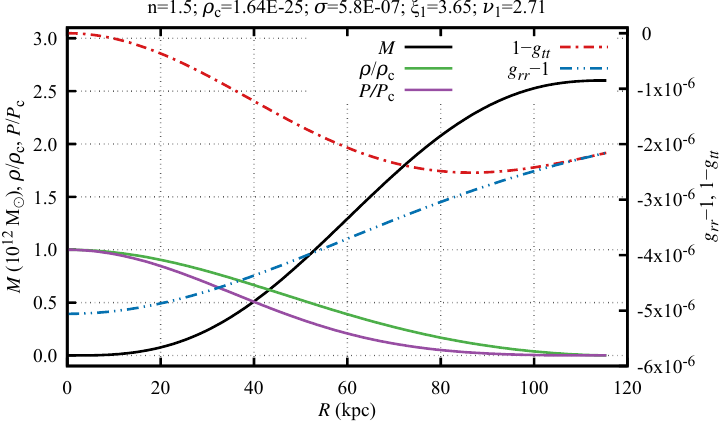}\\[2mm]
			\includegraphics[width=.31\linewidth]{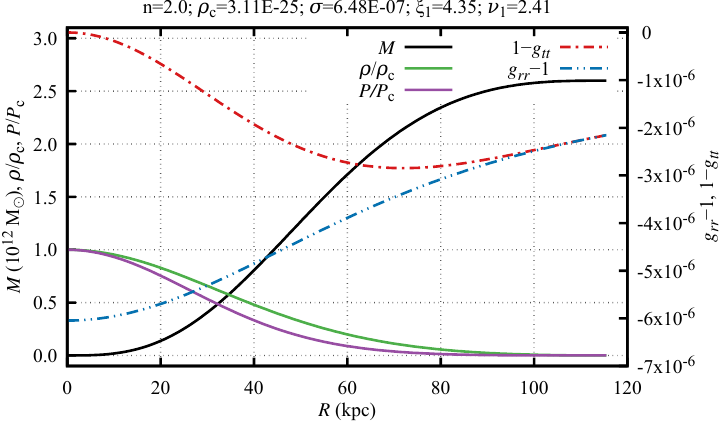}\hfill\includegraphics[width=.31\linewidth]{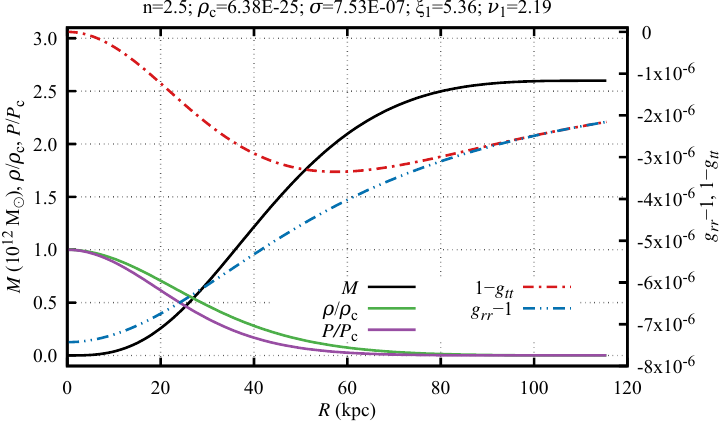}\hfill\includegraphics[width=.31\linewidth]{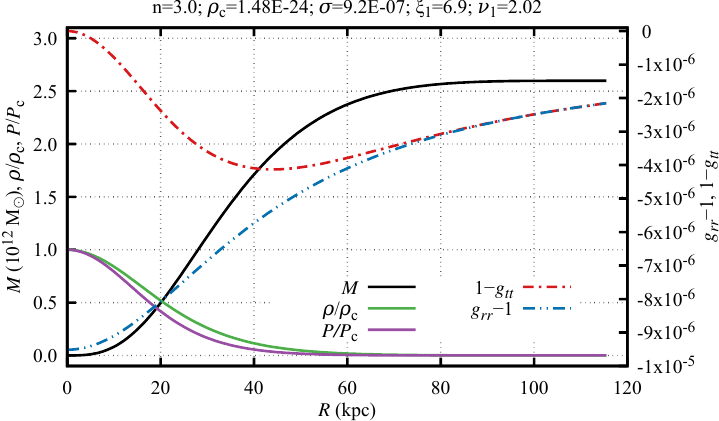}\\[2mm]
			\includegraphics[width=.31\linewidth]{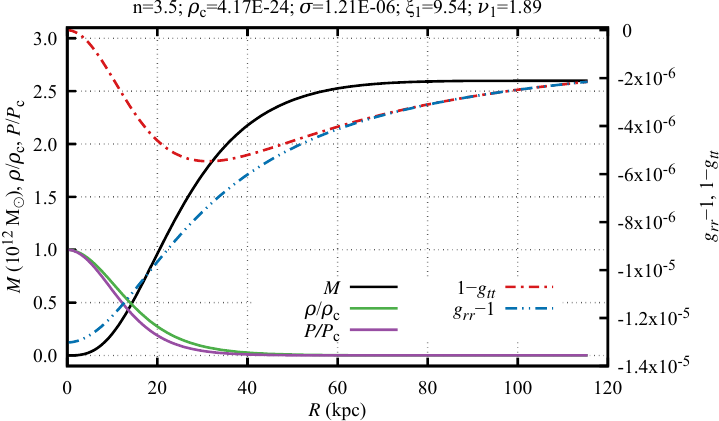}\hfill\includegraphics[width=.31\linewidth]{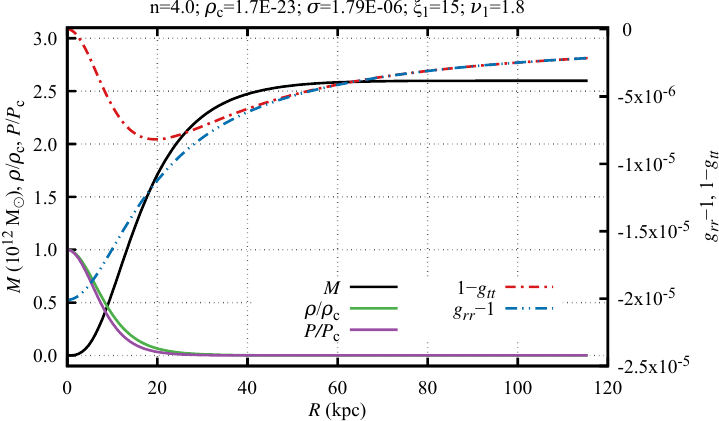}\hfill\includegraphics[width=.31\linewidth]{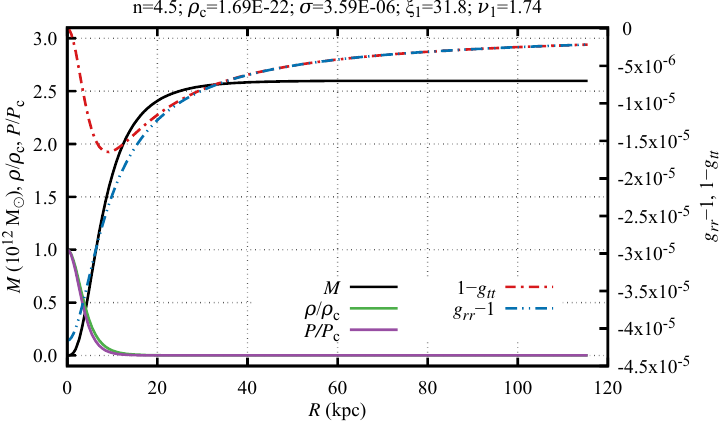}
            \caption{\label{fig:profiles_case1} Radial profiles of mass, energy density, pressure and metric coefficients for the non-relativistic regime polytropes, with $\sigma<10^{-4}$, related to typical galaxies ($R\sim 100$ kpc; $M \sim 10^{12}\,M_\odot$). For completeness we introduce for each chart the central energy density $\rho_{\mathrm{c}}$, $\sigma$, $\xi_1$ and $\nu_1$. Please notice that for the metric coefficients we have introduced a special scale on the right side of the chart.}
	   \end{center}
    \end{figure*}

\begin{figure*}[ht]
	   \begin{center}
			\includegraphics[width=.31\linewidth]{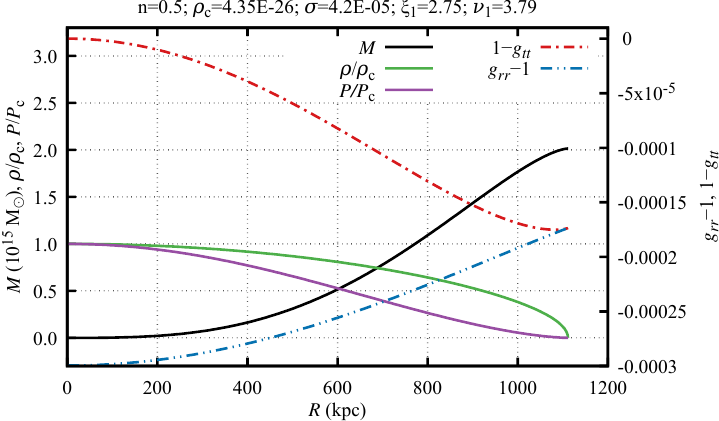}\hfill\includegraphics[width=.31\linewidth]{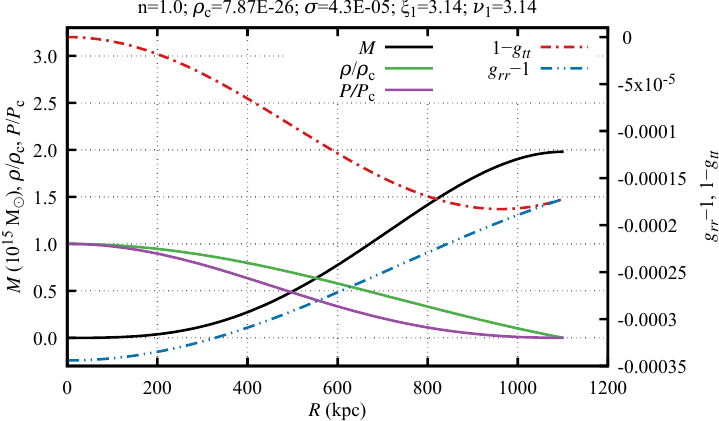}\hfill\includegraphics[width=.31\linewidth]{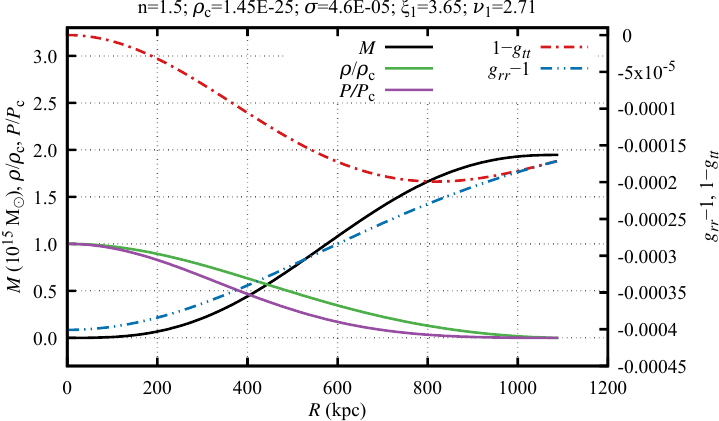}\\[2mm]
			\includegraphics[width=.31\linewidth]{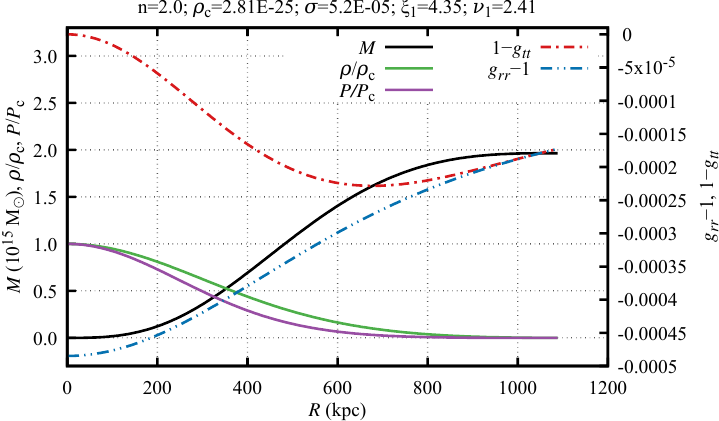}\hfill\includegraphics[width=.31\linewidth]{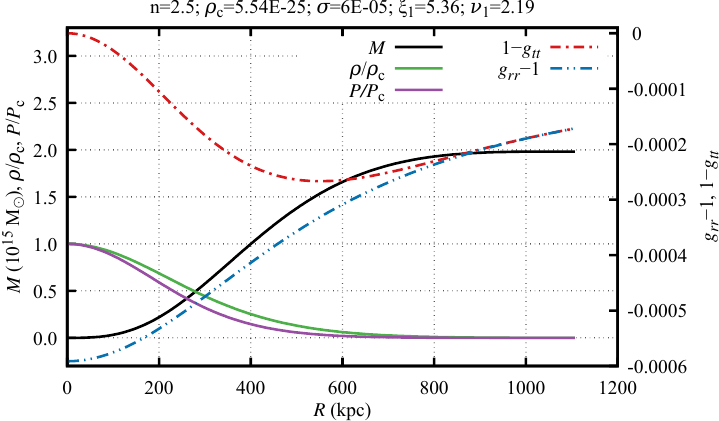}\hfill\includegraphics[width=.31\linewidth]{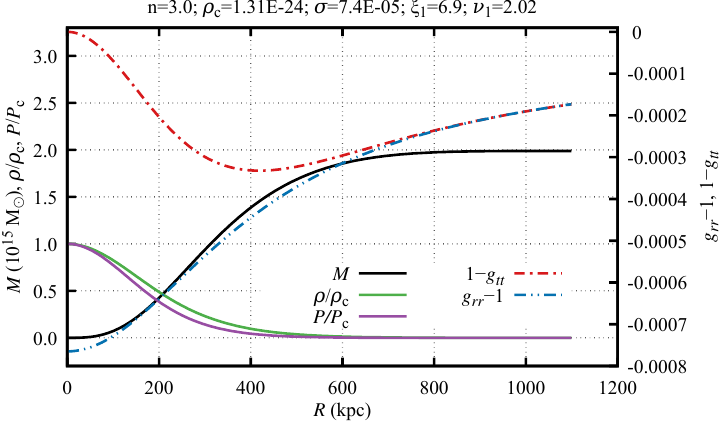}\\[2mm]
			\includegraphics[width=.31\linewidth]{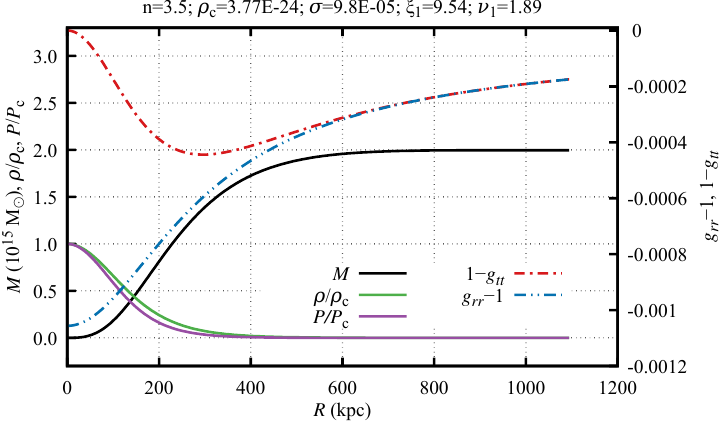}\hfill\includegraphics[width=.31\linewidth]{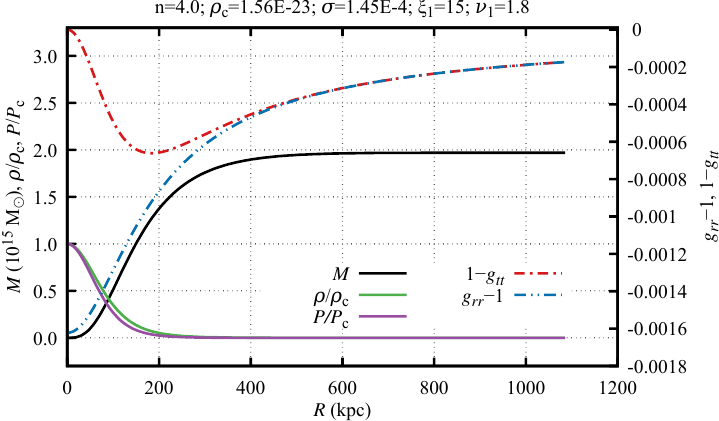}\hfill\includegraphics[width=.31\linewidth]{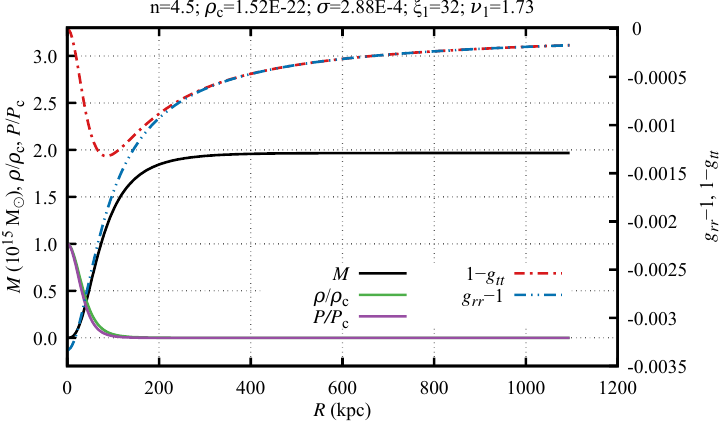}
            \caption{\label{fig:profiles_case2} Radial profiles of mass, energy density, pressure and metric coefficients for the non-relativistic regime polytropes, with $\sigma<10^{-3}$, related to typical galaxy clusters ($R\sim 1$ Mpc; $M \sim 10^{15}\,M_\odot$). Description is the same as in Fig.~\ref{fig:profiles_case1}.}
	   \end{center}
    \end{figure*}

\begin{figure*}[ht]
	   \begin{center}
            \includegraphics[width=.48\linewidth]{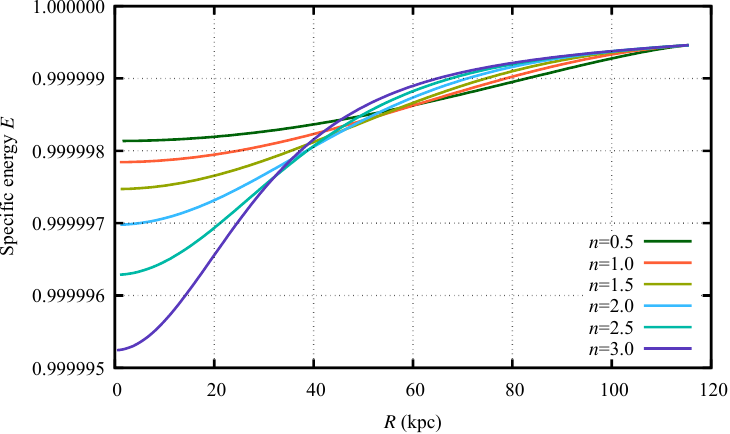}\hfill\includegraphics[width=.48\linewidth]{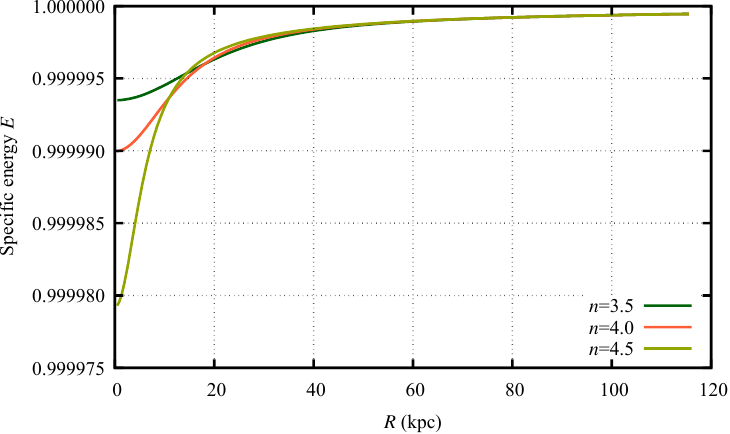}\\[3mm]
            \includegraphics[width=.48\linewidth]{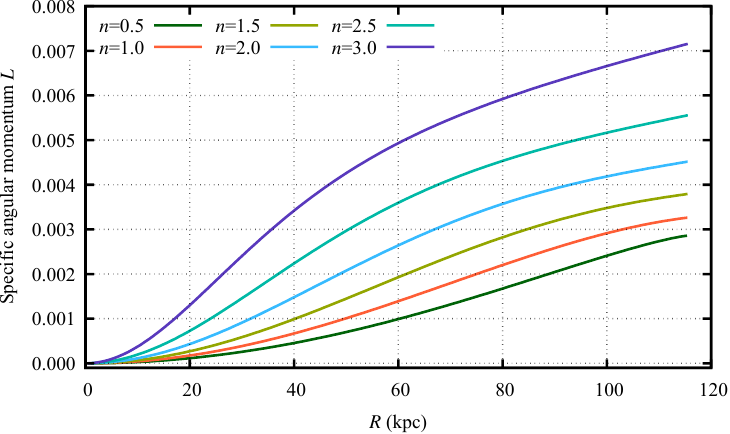}\hfill\includegraphics[width=.48\linewidth]{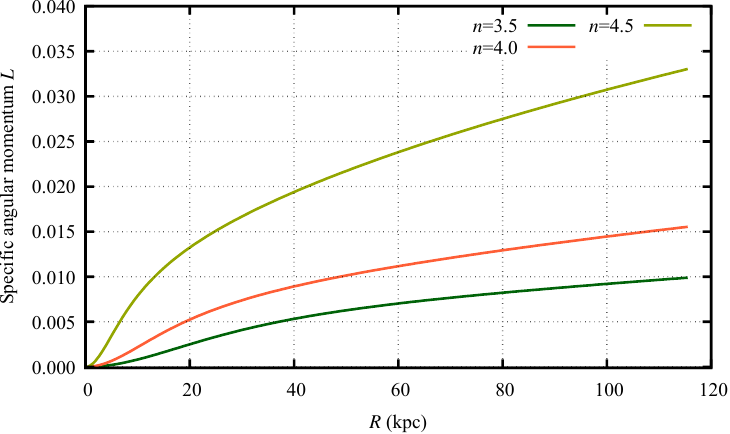}
            \caption{\label{fig:sp_case1} Specific energy $E$ and specific angular momentum $L$ for the galaxy-like profiles.}
	   \end{center}
    \end{figure*}

\begin{figure*}[ht]
	   \begin{center}
            \includegraphics[width=.48\linewidth]{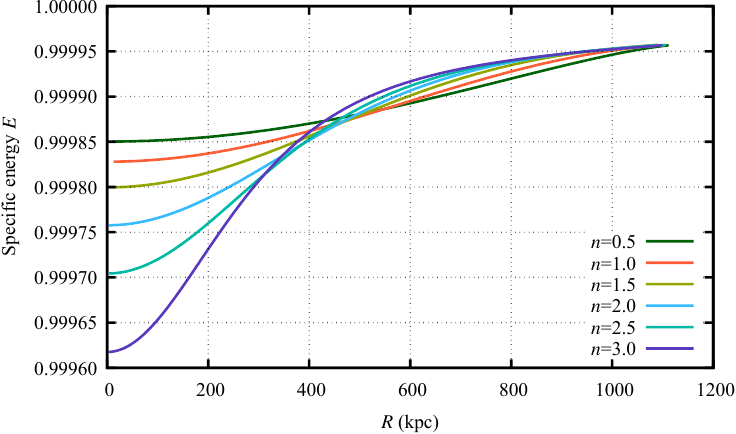}\hfill\includegraphics[width=.48\linewidth]{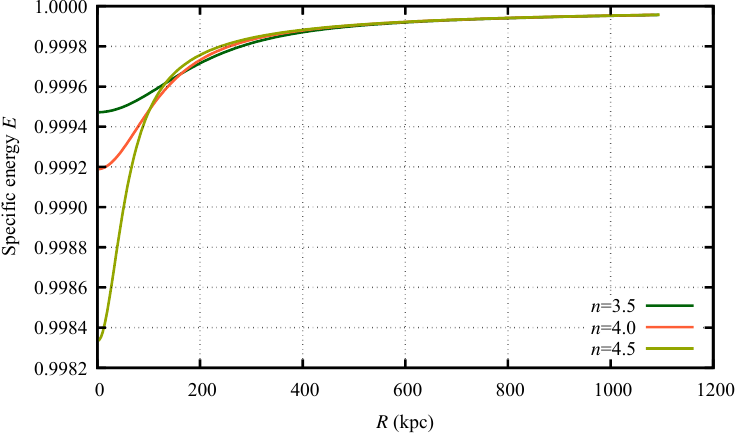}\\[3mm]
            \includegraphics[width=.48\linewidth]{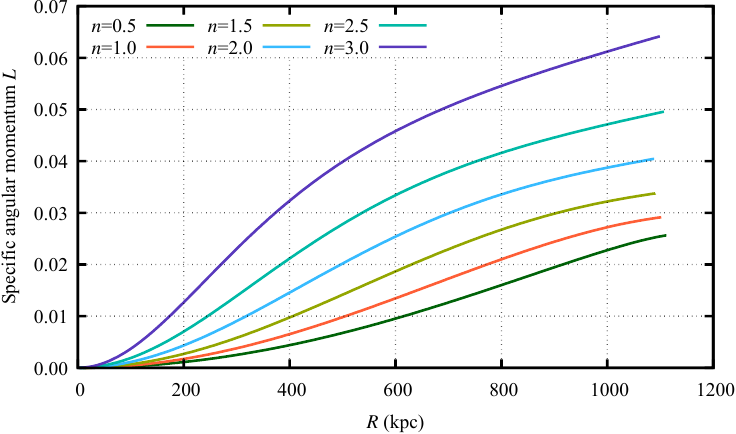}\hfill\includegraphics[width=.48\linewidth]{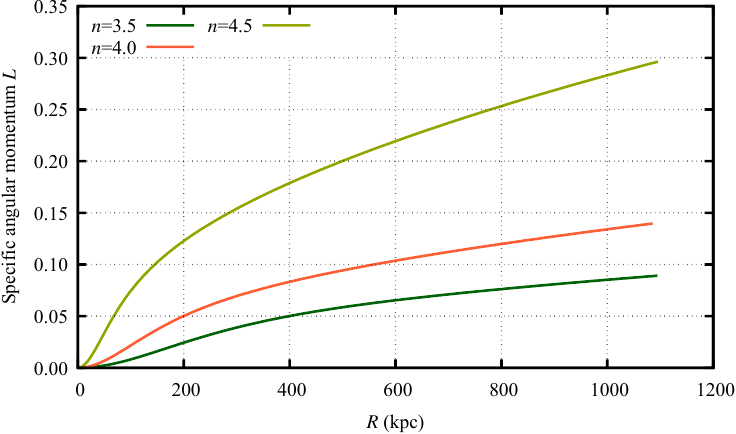}
			\caption{\label{fig:sp_case2} Specific energy $E$ and specific angular momentum $L$ for the cluster-like profiles.}
	   \end{center}
    \end{figure*}

In order to illustrate the character of the interior of the typical considered polytropic spheres, we give the radial profiles of their mass, energy density, pressure and metric coefficients (Figs.~\ref{fig:profiles_case1} and \ref{fig:profiles_case2}). The profiles of specific energy and specific angular momentum are given in Figs.~\ref{fig:sp_case1} \&~\ref{fig:sp_case2} for the case of non-relativistic polytropes (first regime of fitting procedure), covering thus both the galaxy-like polytropes and the cluster-like polytropes. Notice that for the non-relativistic polytropes the behaviour of the profiles is for given polytropic index $n$ of the same character for both the galaxy and cluster cases.

%%%%%%%%%%%%%%%%%%%%%%%%%%%%%%%%%%%%%%%%%%%%%%%%%%%%%%%%%%%%%%%%%%%%%%%%%%%%%%%%%%%%%%%%%%%%%%%%%%%%%%%%%%%%%%%%%%%%%%%%%%%%%%%%%%%%%%%%%%%%
%%%%%%%%%%%%%%%%%%%%%%%%%%%%%%%%%%%%%%%%%%%%%%%%%%%%%%%%%%%%%%%%%%%%%%%%%%%%%%%%%%%%%%%%%%%%%%%%%%%%%%%%%%%%%%%%%%%%%%%%%%%%%%%%%%%%%%%%%%%%

%%%%%%%%%%%%%%%%%%%%%%%%%%%%%%%%%%%%%%%%%%%%%%%%%%%%%%%%%%%%%%%%%%%%%%%%%%%%%%%%%%%%%%%%%%%%%%%%%%%%%%%%%%%%%%%%%%%%%%%%%%%%%%%%%%%%%%%%%%%%
%%%%%%%%%%%%%%%%%%%%%%%%%%%%%%%%%%%%%%%%%%%%%%%%%%%%%%%%%%%%%%%%%%%%%%%%%%%%%%%%%%%%%%%%%%%%%%%%%%%%%%%%%%%%%%%%%%%%%%%%%%%%%%%%%%%%%%%%%%%%

%%%%%%%%%%%%%%%%%%%%%%%%%%%%%%%%%%%%%%%%%%%%%%%%%%%%%%%%%%%%%%%%%%%%%%%%%%%%%%%%%%%%%%%%%%%%%%%%%%%%%%%%%%%%%%%%%%%%%%%%%%%%%%%%%%%%%%%%%%%%
%%%%%%%%%%%%%%%%%%%%%%%%%%%%%%%%%%%%%%%%%%%%%%%%%%%%%%%%%%%%%%%%%%%%%%%%%%%%%%%%%%%%%%%%%%%%%%%%%%%%%%%%%%%%%%%%%%%%%%%%%%%%%%%%%%%%%%%%%%%%

%%%%%%%%%%%%%%%%%%%%%%%%%%%%%%%%%%%%%%%%%%%%%%%%%%%%%%%%%%%%%%%%%%%%%%%%%%%%%%%%%%%%%%%%%%%%%%%%%%%%%%%%%%%%%%%%%%%%%%%%%%%%%%%%%%%%%%%%%%%%
%%%%%%%%%%%%%%%%%%%%%%%%%%%%%%%%%%%%%%%%%%%%%%%%%%%%%%%%%%%%%%%%%%%%%%%%%%%%%%%%%%%%%%%%%%%%%%%%%%%%%%%%%%%%%%%%%%%%%%%%%%%%%%%%%%%%%%%%%%%%
\clearpage

\section{\mbox{}\\ Profiles of the relativistic polytropes}\label{APPENDIX2}
\begin{figure*}[ht]
        \begin{center}
            \includegraphics[width=.495\linewidth]{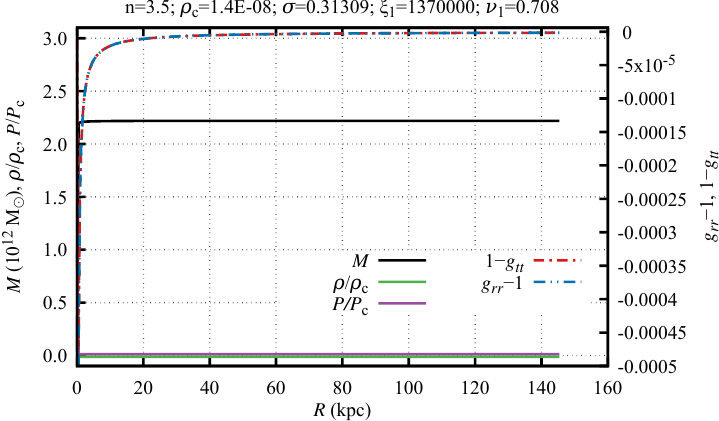}\hfill\includegraphics[width=.495\linewidth]{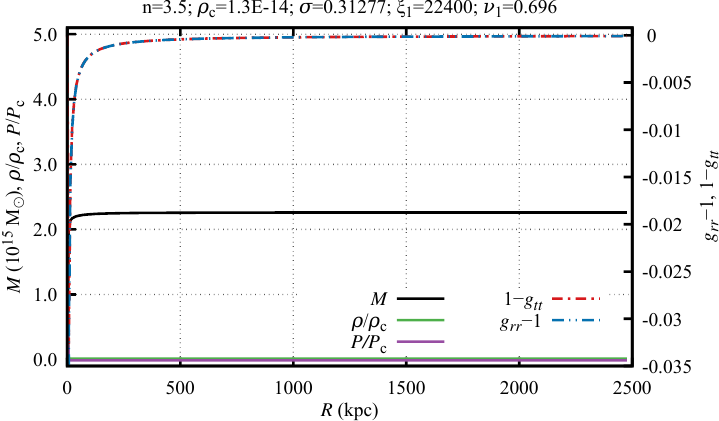}\\[3mm]
			\includegraphics[width=.495\linewidth]{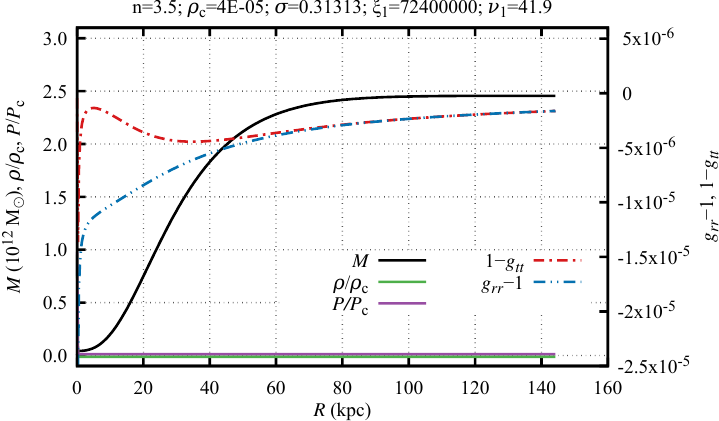}\hfill\includegraphics[width=.495\linewidth]{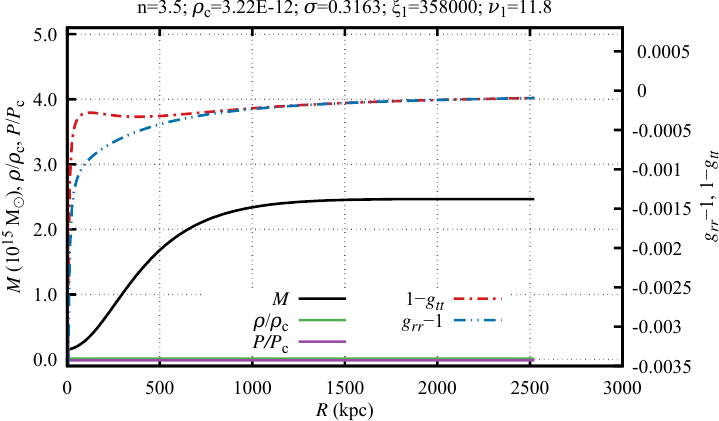}
			\caption{\label{fig:profiles_regime2_case2_n35} Profiles of energy density, pressure and metric coefficients for typical parameters of the different cases depicted by the Fig.~\ref{fig:profiles_regime2_case1}.}
        \end{center}
    \end{figure*}

\begin{figure*}[ht]
        \begin{center}
            \includegraphics[width=.49\linewidth]{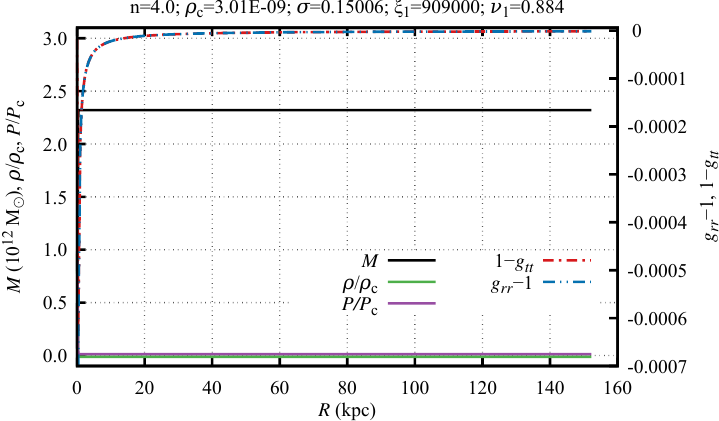}\hfill\includegraphics[width=.49\linewidth]{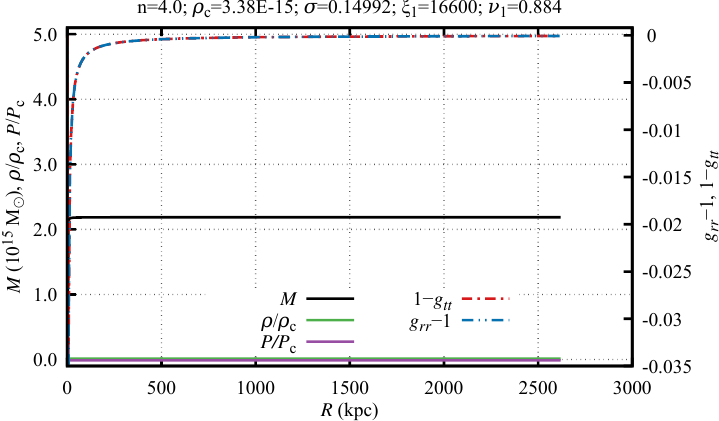}\\[3mm]
			\includegraphics[width=.49\linewidth]{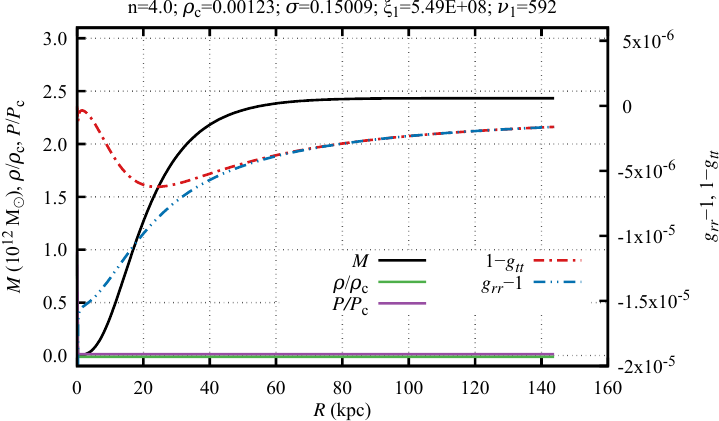}\hfill\includegraphics[width=.49\linewidth]{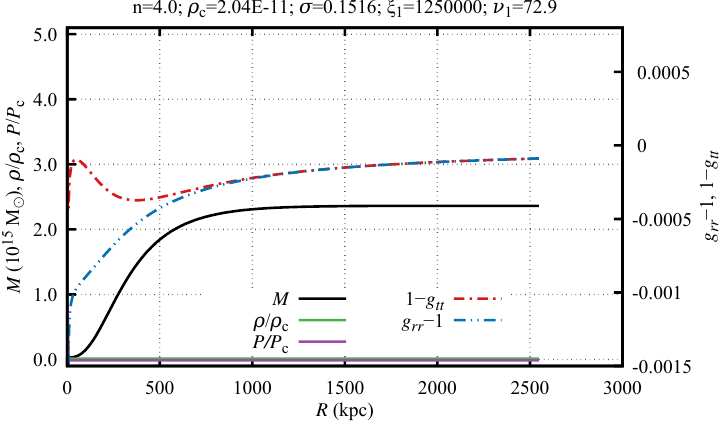}\\[3mm]
			\includegraphics[width=.49\linewidth]{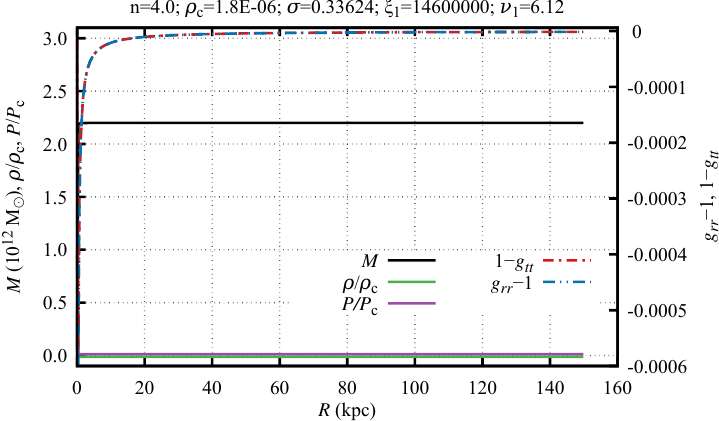}\hfill\includegraphics[width=.49\linewidth]{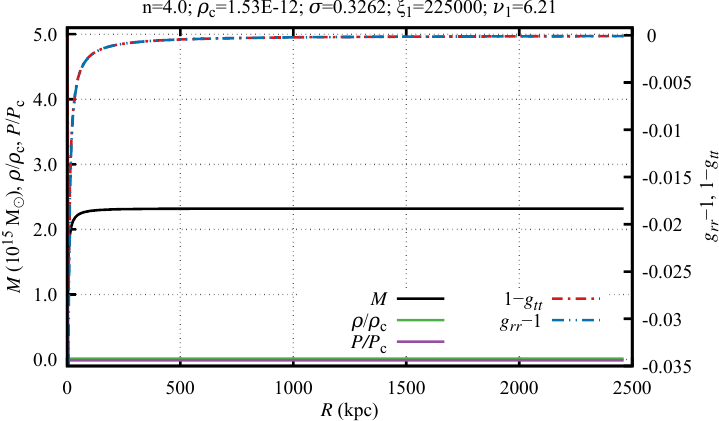}\\[3mm]
			\includegraphics[width=.49\linewidth]{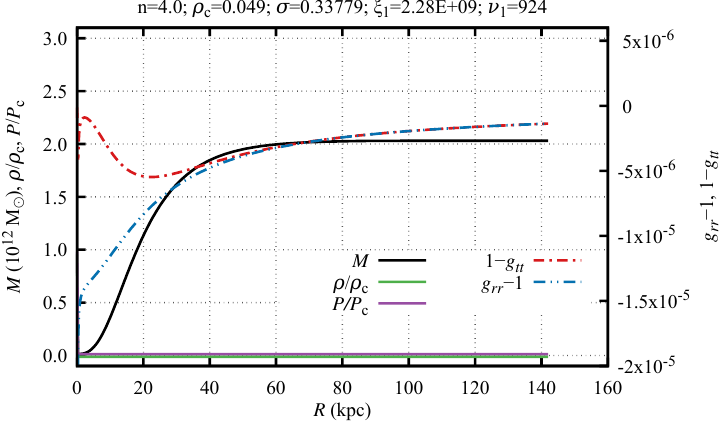}\hfill\includegraphics[width=.49\linewidth]{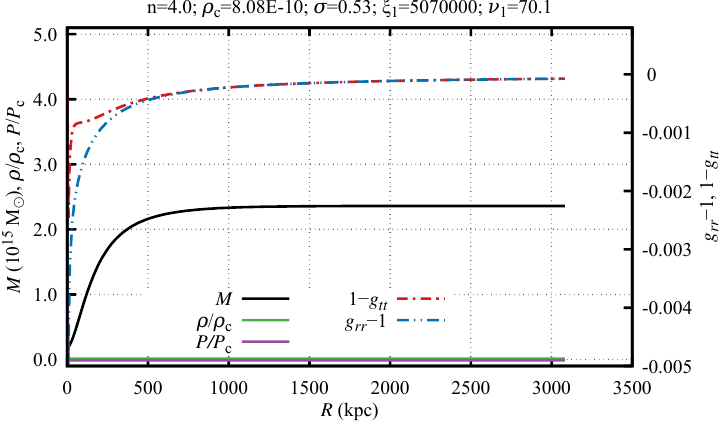}
			\caption{\label{fig:profiles_regime2_case2_n40} Profiles of energy density, pressure and metric coefficients for typical parameters of the different cases depicted by the Fig.~\ref{fig:profiles_regime2_case1_n40}.}
	   \end{center}
    \end{figure*}

\begin{figure*}[ht]
       \begin{center}
            \includegraphics[width=.48\linewidth]{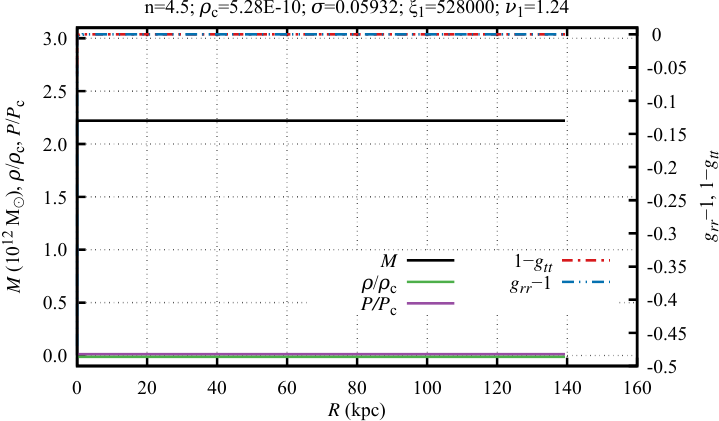}\hfill\includegraphics[width=.48\linewidth]{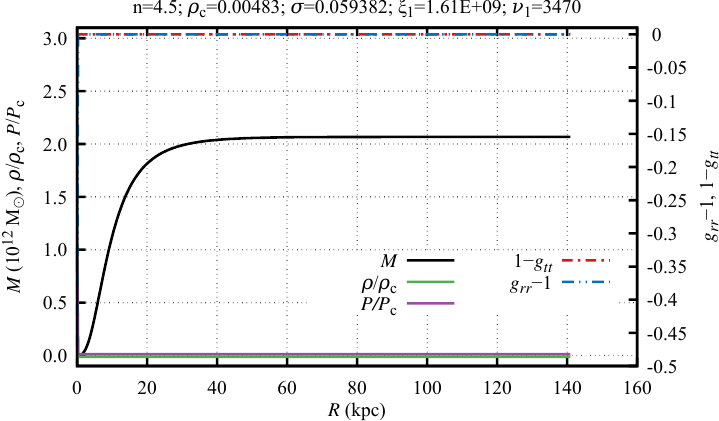}\\[3mm]
			\includegraphics[width=.48\linewidth]{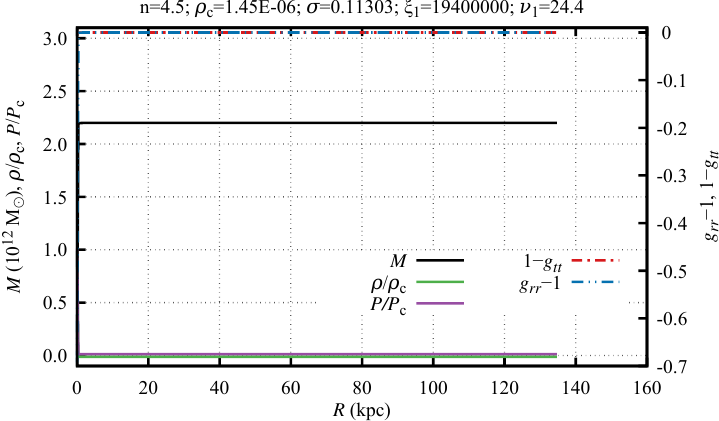}\hfill\includegraphics[width=.48\linewidth]{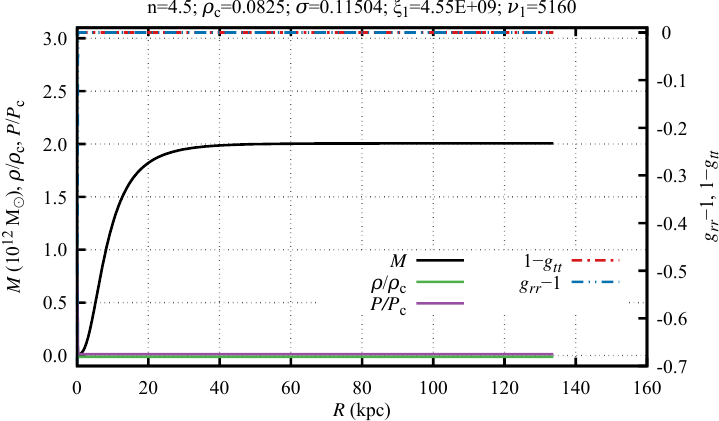}\\[3mm]
			\includegraphics[width=.48\linewidth]{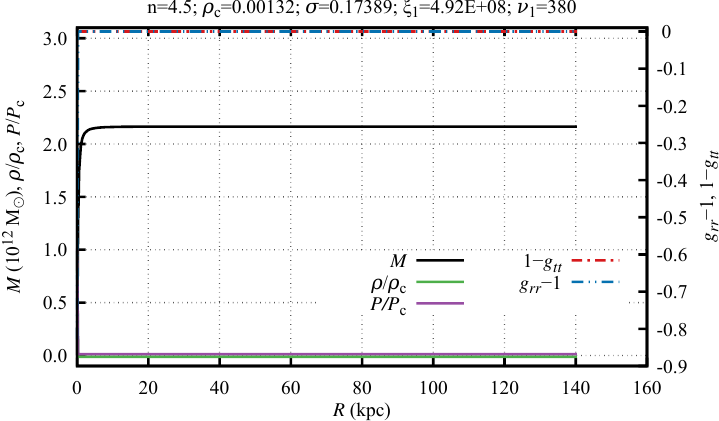}\hfill
            \caption{\label{fig:profiles_regime2_case2_n45} Profiles of energy density, pressure and metric coefficients for typical galaxy-like parameters of the different cases depicted by the Fig.~\ref{fig:profiles_regime2_case1_n45}.}
	   \end{center}
    \end{figure*}

\begin{figure*}[ht]
	   \begin{center}
            \includegraphics[width=.48\linewidth]{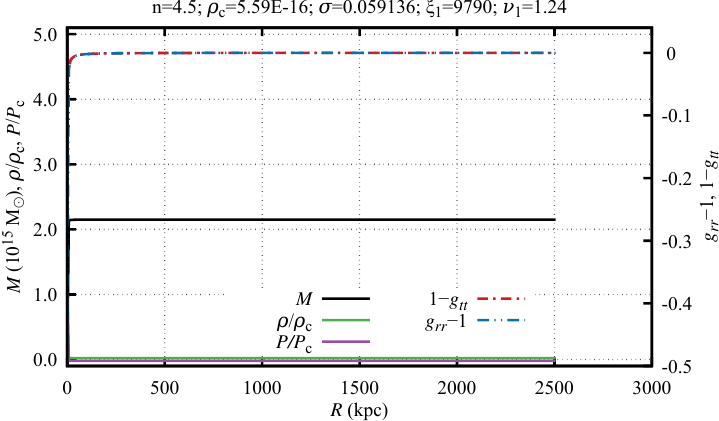}\hfill\includegraphics[width=.48\linewidth]{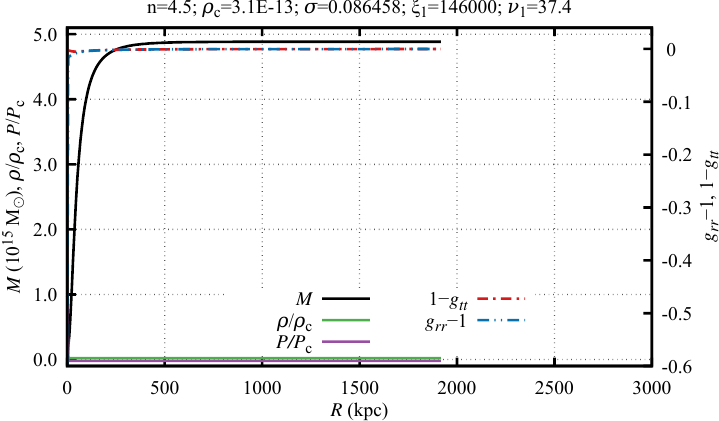}
			\caption{\label{fig:profiles_regime2_case2_n45_b} Profiles of energy density, pressure and metric coefficients for the typical cluster-like parameters of the different cases depicted by the Fig.~\ref{fig:profiles_regime2_case2_n45_a}.}
	   \end{center}
    \end{figure*}

\begin{figure*}[ht!]
	   \begin{center}
            \includegraphics[width=.33\linewidth]{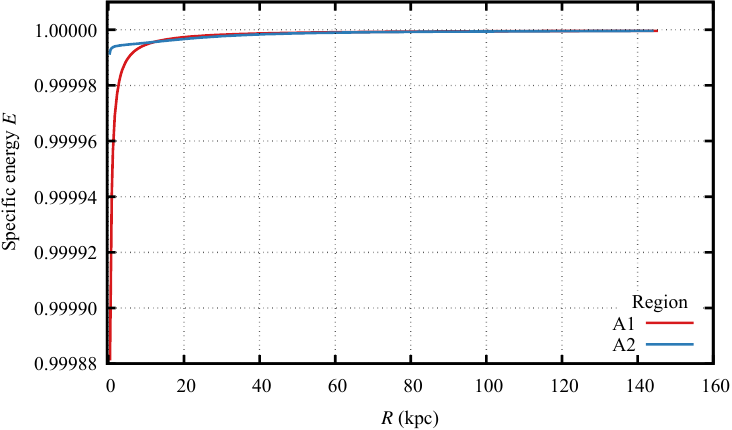}\hfill\includegraphics[width=.33\linewidth]{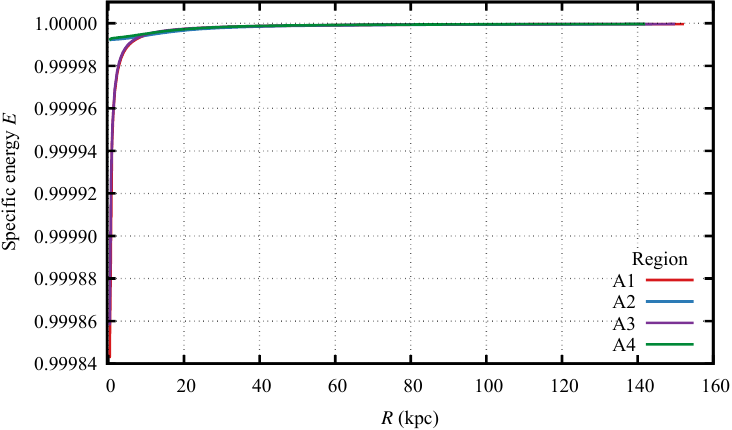}\hfill\includegraphics[width=.33\linewidth]{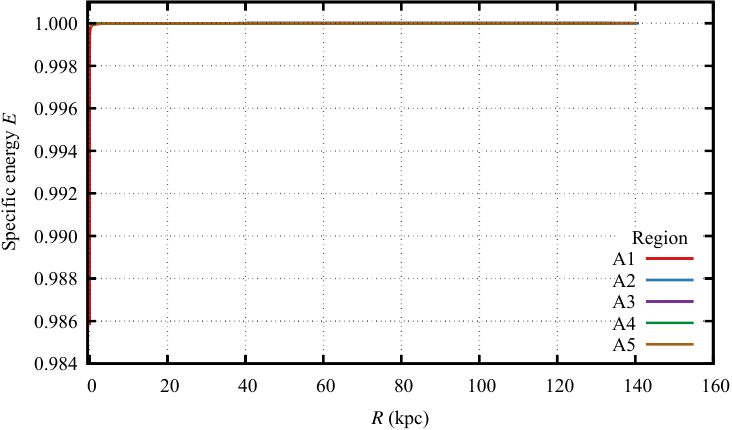}\\[3mm]
            \includegraphics[width=.33\linewidth]{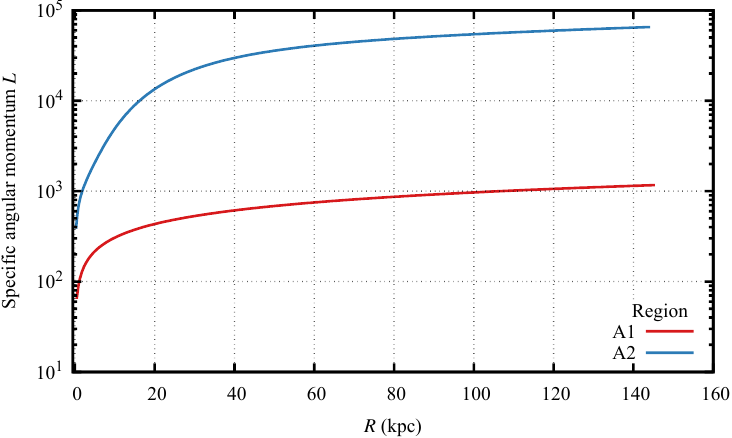}\hfill\includegraphics[width=.33\linewidth]{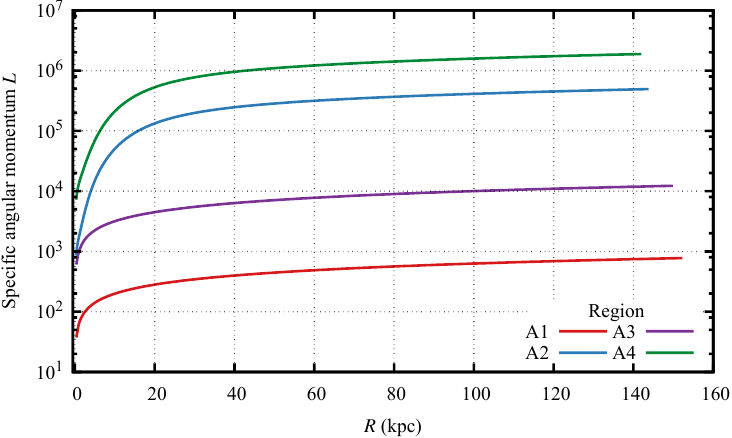}\hfill\includegraphics[width=.33\linewidth]{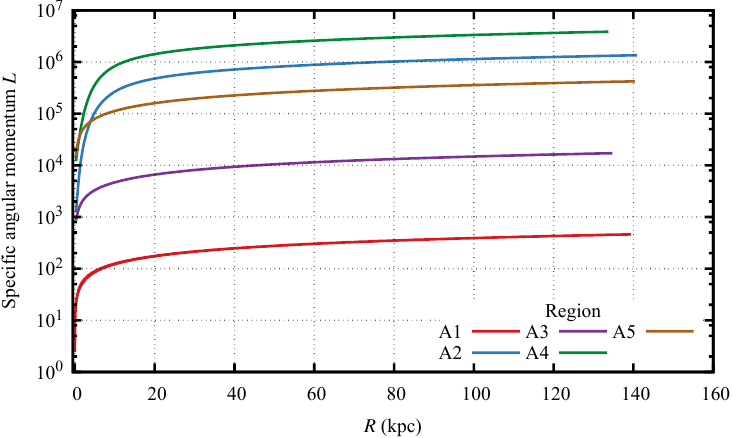}
            \caption{\label{fig:sp_case1_reg2} Specific energy $E$ and specific angular momentum $L$ for the galaxy-like profiles in the relativistic regime. The most left column corresponds to the polytropic index $n = 3.5$, the middle column corresponds to the index $n = 4.0$ and the last column to the index $n = 4.5$. For parameter $\sigma$ see Tables \ref{tab:centre_35}, \ref{tab:centre_40}, \ref{tab:centre_45}}.
	   \end{center}
    \end{figure*}

\begin{figure*}[ht]
	   \begin{center}
            \includegraphics[width=.33\linewidth]{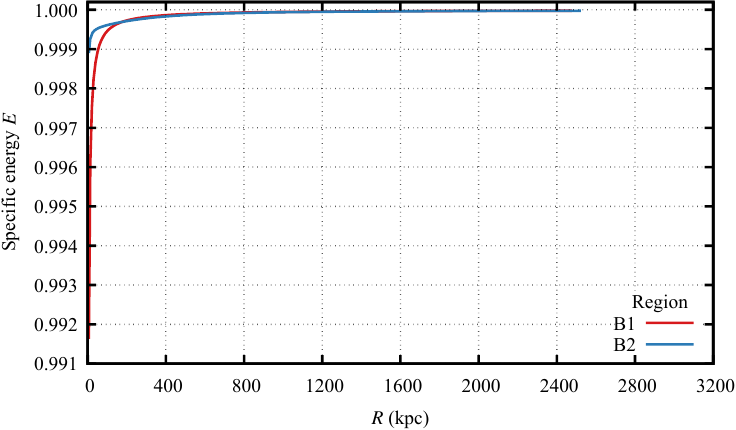}\hfill\includegraphics[width=.33\linewidth]{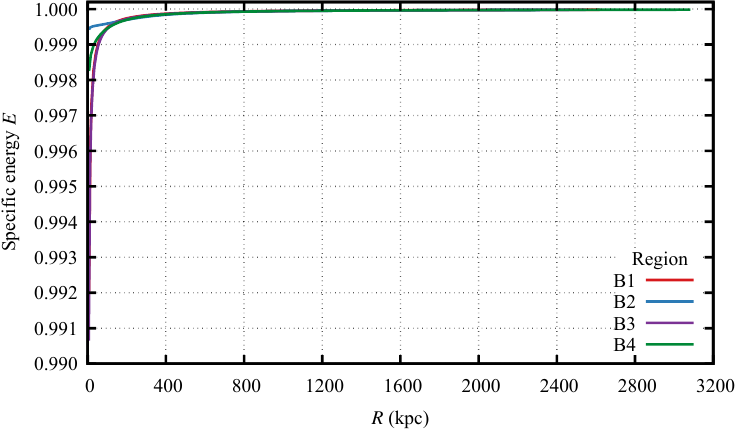}\hfill\includegraphics[width=.33\linewidth]{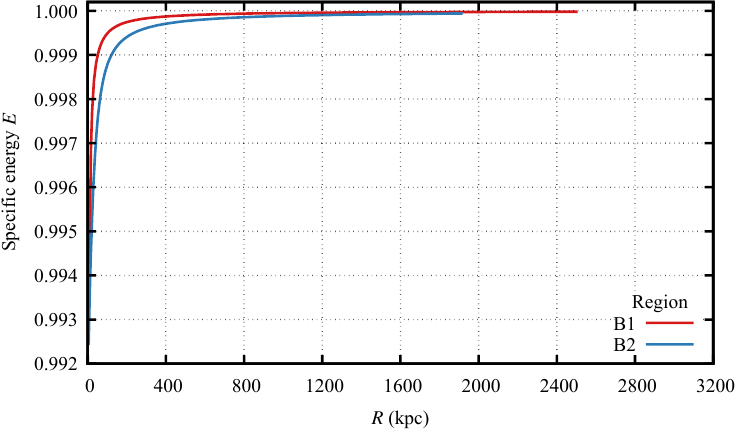}\\[3mm]
            \includegraphics[width=.33\linewidth]{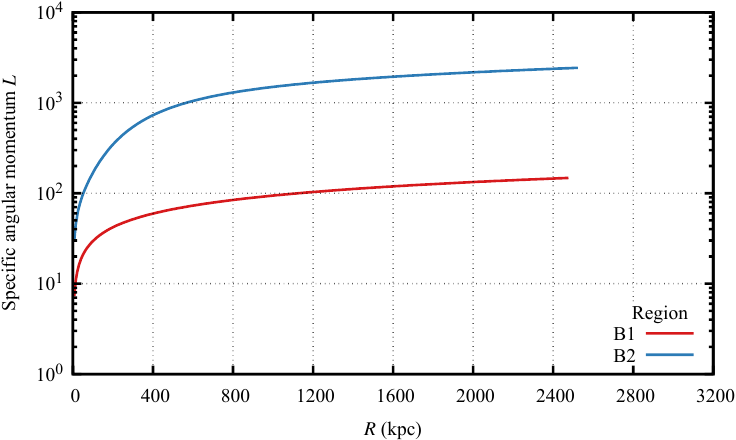}\hfill\includegraphics[width=.33\linewidth]{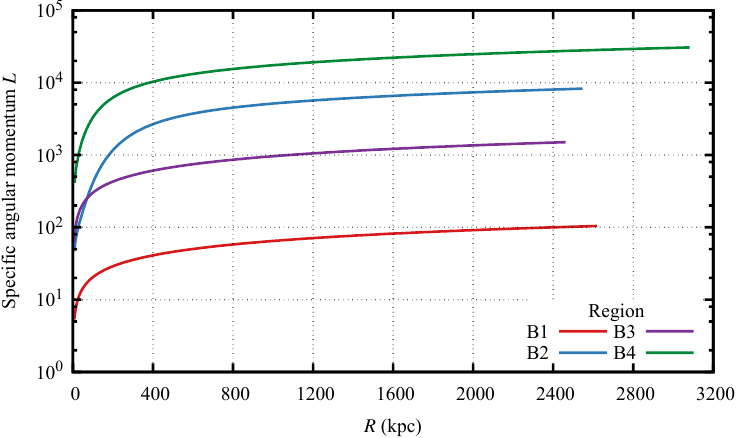}\hfill\includegraphics[width=.33\linewidth]{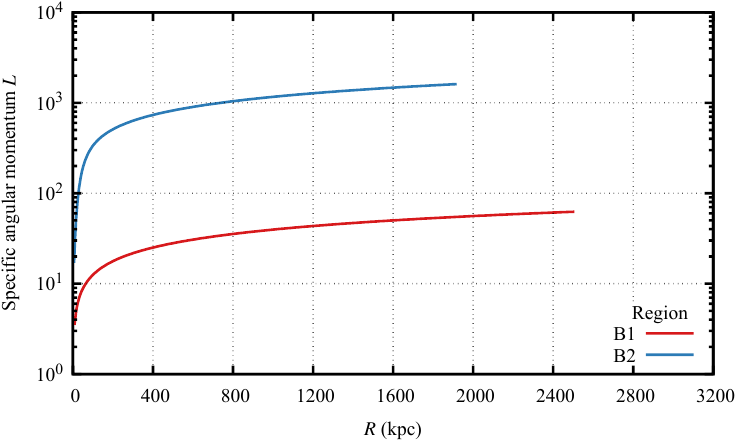}
            \caption{\label{fig:sp_case2_reg2} Specific energy $E$ and specific angular momentum $L$ for the cluster-like profiles in the relativistic regime. The most left column corresponds to the polytropic index $n = 3.5$, the middle column to the index $n = 4.0$ and the last column to the index $n = 4.5$. For parameter $\sigma$ see Tables \ref{tab:centre_35}, \ref{tab:centre_40}, \ref{tab:centre_45}}.
	   \end{center}
    \end{figure*}

In the case of relativistic (second regime) polytropes, we give for the typical values of the polytropic index $n$ = 3.5, 4.0, 4.5 the detailed shape of the allowed region in the $\sigma$--$\rho_\mathrm{c}$ parameter space for the galaxy-like ($A_i$) and cluster-like ($B_i$) polytropic spheres. For selected typical points (given by parameters $\sigma$ and $\rho_\mathrm{c}$) of the allowed regions, we give the profiles of energy density, pressure and metric coefficients in Figures~\ref{fig:profiles_regime2_case2_n35}, \ref{fig:profiles_regime2_case2_n40}, \ref{fig:profiles_regime2_case2_n45}, and \ref{fig:profiles_regime2_case2_n45_b}. Next, we are giving specific energy $E$ and specific angular momentum $L$ for the galaxy-like profiles (Fig.~\ref{fig:sp_case1_reg2}) and for the cluster-like profiles (Fig.~\ref{fig:sp_case2_reg2}) in the relativistic regime. In the case of the relativistic polytropes the situation is more complex than for the non-relativistic polytropes, and generally the profiles have different character for the galaxy-like and cluster-like polytropes. Moreover, the relativistic polytropes separate into several qualitatively different types.

%%%%%%%%%%%%%%%%%%%%%%%%%%%%%%%%%%%%%%%%%%%%%%%%%%%%%%%%%%%%%%%%%%%%%%%%%%%%%%%%%%%%%%%%%%%%%%%%%%%%%%%%%%%%%%%%%%%%%%%%%%%%%%%%%%%%%%%%%%%%
%%%%%%%%%%%%%%%%%%%%%%%%%%%%%%%%%%%%%%%%%%%%%%%%%%%%%%%%%%%%%%%%%%%%%%%%%%%%%%%%%%%%%%%%%%%%%%%%%%%%%%%%%%%%%%%%%%%%%%%%%%%%%%%%%%%%%%%%%%%%

%%%%%%%%%%%%%%%%%%%%%%%%%%%%%%%%%%%%%%%%%%%%%%%%%%%%%%%%%%%%%%%%%%%%%%%%%%%%%%%%%%%%%%%%%%%%%%%%%%%%%%%%%%%%%%%%%%%%%%%%%%%%%%%%%%%%%%%%%%%%
%%%%%%%%%%%%%%%%%%%%%%%%%%%%%%%%%%%%%%%%%%%%%%%%%%%%%%%%%%%%%%%%%%%%%%%%%%%%%%%%%%%%%%%%%%%%%%%%%%%%%%%%%%%%%%%%%%%%%%%%%%%%%%%%%%%%%%%%%%%%

%%%%%%%%%%%%%%%%%%%%%%%%%%%%%%%%%%%%%%%%%%%%%%%%%%%%%%%%%%%%%%%%%%%%%%%%%%%%%%%%%%%%%%%%%%%%%%%%%%%%%%%%%%%%%%%%%%%%%%%%%%%%%%%%%%%%%%%%%%%%
%%%%%%%%%%%%%%%%%%%%%%%%%%%%%%%%%%%%%%%%%%%%%%%%%%%%%%%%%%%%%%%%%%%%%%%%%%%%%%%%%%%%%%%%%%%%%%%%%%%%%%%%%%%%%%%%%%%%%%%%%%%%%%%%%%%%%%%%%%%%

%%%%%%%%%%%%%%%%%%%%%%%%%%%%%%%%%%%%%%%%%%%%%%%%%%%%%%%%%%%%%%%%%%%%%%%%%%%%%%%%%%%%%%%%%%%%%%%%%%%%%%%%%%%%%%%%%%%%%%%%%%%%%%%%%%%%%%%%%%%%
%%%%%%%%%%%%%%%%%%%%%%%%%%%%%%%%%%%%%%%%%%%%%%%%%%%%%%%%%%%%%%%%%%%%%%%%%%%%%%%%%%%%%%%%%%%%%%%%%%%%%%%%%%%%%%%%%%%%%%%%%%%%%%%%%%%%%%%%%%%%

%%%%%%%%%%%%%%%%%%%%%%%%%%%%%%%%%%%%%%%%%%%%%%%%%%%%%%%%%%%%%%%%%%%%%%%%%%%%%%%%%%%%%%%%%%%%%%%%%%%%%%%%%%%%%%%%%%%%%%%%%%%%%%%%%%%%%%%%%%%%
%%%%%%%%%%%%%%%%%%%%%%%%%%%%%%%%%%%%%%%%%%%%%%%%%%%%%%%%%%%%%%%%%%%%%%%%%%%%%%%%%%%%%%%%%%%%%%%%%%%%%%%%%%%%%%%%%%%%%%%%%%%%%%%%%%%%%%%%%%%%

%%%%%%%%%%%%%%%%%%%%%%%%%%%%%%%%%%%%%%%%%%%%%%%%%%%%%%%%%%%%%%%%%%%%%%%%%%%%%%%%%%%%%%%%%%%%%%%%%%%%%%%%%%%%%%%%%%%%%%%%%%%%%%%%%%%%%%%%%%%%
%%%%%%%%%%%%%%%%%%%%%%%%%%%%%%%%%%%%%%%%%%%%%%%%%%%%%%%%%%%%%%%%%%%%%%%%%%%%%%%%%%%%%%%%%%%%%%%%%%%%%%%%%%%%%%%%%%%%%%%%%%%%%%%%%%%%%%%%%%%%

\clearpage
\end{appendix}

%%%%%%%%%%%%%%%%%%%%%%%%%%%%%%%%%%%%%%%%%%%%%%%%%%%%%%%%%%%%%%%%%%%%%%%%%%%%%%%%%%%%%%%%%%%%%%%%%%%%%%%%%%%%%%%%%%%%%%%%%%%%%%%%%%%%%%%%%%%%
%%%%%%%%%%%%%%%%%%%%%%%%%%%%%%%%%%%%%%%%%%%%%%%%%%%%%%%%%%%%%%%%%%%%%%%%%%%%%%%%%%%%%%%%%%%%%%%%%%%%%%%%%%%%%%%%%%%%%%%%%%%%%%%%%%%%%%%%%%%%

\begin{acknowledgements}
    The authors were supported by the Research Centre for Theoretical Physics and Astrophysics, Institute of Physics, Silesian University in Opava.
\end{acknowledgements}

\bibliographystyle{aa}
\bibliography{bibliography}

\begin{thebibliography}{135}
\expandafter\ifx\csname natexlab\endcsname\relax\def\natexlab#1{#1}\fi

\bibitem[{Adhikari {et~al.}(2017)Adhikari, Agostini, Ky, Araki, Archidiacono,
  Bahr, Baur, Behrens, Bezrukov, Dev, Borah, Boyarsky, de~Gouvea, de~S.~Pires,
  de~Vega, Dias, Bari, Djurcic, Dolde, Dorrer, Durero, Dragoun, Drewes,
  Drexlin, Düllmann, Eberhardt, Eliseev, Enss, Evans, Faessler, Filianin,
  Fischer, Fleischmann, Formaggio, Franse, Fraenkle, Frenk, Fuller, Gastaldo,
  Garzilli, Giunti, Glück, Goodman, Gonzalez-Garcia, Gorbunov, Hamann, Hannen,
  Hannestad, Hansen, Hassel, Heeck, Hofmann, Houdy, Huber, Iakubovskyi, Ianni,
  Ibarra, Jacobsson, Jeltema, Jochum, Kempf, Kieck, Korzeczek, Kornoukhov,
  Lachenmaier, Laine, Langacker, Lasserre, Lesgourgues, Lhuillier, Li, Liao,
  Long, Maltoni, Mangano, Mavromatos, Menci, Merle, Mertens, Mirizzi, Monreal,
  Nozik, Neronov, Niro, Novikov, Oberauer, Otten, Palanque-Delabrouille,
  Pallavicini, Pantuev, Papastergis, Parke, Pascoli, Pastor, Patwardhan,
  Pilaftsis, Radford, Ranitzsch, Rest, Robinson, da~Silva, Ruchayskiy, Sanchez,
  Sasaki, Saviano, Schneider, Schneider, Schwetz, Schönert, Scholl, Shankar,
  Shrock, Steinbrink, Strigari, Suekane, Suerfu, Takahashi, Van, Tkachev,
  Totzauer, Tsai, Tully, Valerius, Valle, Venos, Viel, Vivier, Wang,
  Weinheimer, Wendt, Winslow, Wolf, Wurm, Xing, Zhou, \&
  Zuber}]{Adhi-etal:2017:JCAP}
Adhikari, R., Agostini, M., Ky, N.~A., {et~al.} 2017, Journal of Cosmology and
  Astroparticle Physics, 2017, 025

\bibitem[{Aliev(2007)}]{Ali:2007:PHYSR4:EMPropKadS}
Aliev, A.~N. 2007, Phys. Rev. D, 75, 084041

\bibitem[{{Alvarez-Castillo} {et~al.}(2017){Alvarez-Castillo}, {Blaschke}, \&
  {Typel}}]{Alv-Bla-Typ:2017:AN}
{Alvarez-Castillo}, D., {Blaschke}, D., \& {Typel}, S. 2017, Astronomische
  Nachrichten, 338, 1048

\bibitem[{{Anderhalden} {et~al.}(2012){Anderhalden}, {Diemand}, {Bertone},
  {Macci{\`o}}, \& {Schneider}}]{And-etal:2012:JCAP:}
{Anderhalden}, D., {Diemand}, J., {Bertone}, G., {Macci{\`o}}, A.~V., \&
  {Schneider}, A. 2012, Journal of Cosmology and Astroparticle Physics, 10, 047

\bibitem[{Armendariz-Picon {et~al.}(2000)Armendariz-Picon, Mukhanov, \&
  Steinhardt}]{ArP-Muk-Ste:2000:PHYRL:}
Armendariz-Picon, C., Mukhanov, V., \& Steinhardt, P.~J. 2000, Phys. Rev.
  Lett., 85, 4438

\bibitem[{Arraut(2013)}]{Ara:2013:MODPLA:}
Arraut, I. 2013, Modern Phys. Lett. A, 28

\bibitem[{Arraut(2014{\natexlab{a}})}]{Ara:2014:arXiv:}
Arraut, I. 2014{\natexlab{a}}, Internat. J. Modern Phys. D, 23

\bibitem[{Arraut(2014{\natexlab{b}})}]{Arra:2014:PHYSR4:}
Arraut, I. 2014{\natexlab{b}}, Phys. Rev. D, 90, 124082

\bibitem[{Arraut(2015)}]{Ara:2013:arXiv:}
Arraut, I. 2015, Internat. J. Modern Phys. D, 24, 1550022

\bibitem[{Bahcall {et~al.}(1999)Bahcall, Ostriker, Perlmutter, \&
  Steinhardt}]{Bah-etal:1999:SCIEN:}
Bahcall, N., Ostriker, J.~P., Perlmutter, S., \& Steinhardt, P.~J. 1999,
  Science, 284, 1481

\bibitem[{Barreira {et~al.}(2015)Barreira, Li, Jennings, Merten, King, Baugh,
  \& Pascoli}]{Bar:etal:2015:MNRAS:GalLensing:}
Barreira, A., Li, B., Jennings, E., {et~al.} 2015, Monthly Notices Roy.
  Astronom. Soc., 454, 4085

\bibitem[{Baur {et~al.}(2016)Baur, Palanque-Delabrouille, Y{\`{e}}che,
  Magneville, \& Viel}]{Bau-etal:2016:JCAP:WDM}
Baur, J., Palanque-Delabrouille, N., Y{\`{e}}che, C., Magneville, C., \& Viel,
  M. 2016, Journal of Cosmology and Astroparticle Physics, 2016, 012

\bibitem[{Binney \& Tremaine(1988)}]{Bin-Tre:1988:GalacDynam:}
Binney, J. \& Tremaine, S. 1988, {Galactic Dynamics}, {Princeton Series in
  Astrophysics} (Princeton: Princeton University Press), 755

\bibitem[{{Blanchard} {et~al.}(2024){Blanchard}, {H{\'e}loret}, {Ili{\'c}},
  {Lamine}, \& {Tutusaus}}]{Blanchard:2024:OJAp}
{Blanchard}, A., {H{\'e}loret}, J.-Y., {Ili{\'c}}, S., {Lamine}, B., \&
  {Tutusaus}, I. 2024, The Open Journal of Astrophysics, 7, 32

\bibitem[{Bode {et~al.}(2009)Bode, Ostriker, \&
  Vikhlinin}]{Bod-Jer-Vik:2009:Apj}
Bode, P., Ostriker, J.~P., \& Vikhlinin, A. 2009, The Astrophysical Journal,
  700, 989

\bibitem[{B{\"o}hmer(2004)}]{Boh:2004:GENRG2:}
B{\"o}hmer, C.~G. 2004, Gen. Relativity Gravitation, 36, 1039

\bibitem[{B{\" o}hmer \& Fodor(2008)}]{Boe-Fod:2008:PHYSR4:}
B{\" o}hmer, C.~G. \& Fodor, G. 2008, Phys. Rev. D, 77, 064008

\bibitem[{{B{\"o}hmer} \& {Harko}(2005)}]{Boh:2005:PRD:}
{B{\"o}hmer}, C.~G. \& {Harko}, T. 2005, \prd, 71, 084026

\bibitem[{B{\"o}rner(1993)}]{Bor:1993:EarlyUniv:}
B{\"o}rner, G. 1993, {The Early Universe} (Berlin--Heidelberg--New York:
  Springer-Verlag)

\bibitem[{Bosma(1981)}]{Bos:1981:ASTRJ1:21cmSpiGal}
Bosma, A. 1981, Astronom. J., 86, 1791

\bibitem[{Boyda {et~al.}(2003)Boyda, Ganguli, Ho{\v{r}}ava, \&
  Varadarajan}]{Boy-etal:2003:PHYSR4:HoloProtChron}
Boyda, E.~K., Ganguli, S., Ho{\v{r}}ava, P., \& Varadarajan, U. 2003, Phys.
  Rev. D, 67, 106003

\bibitem[{{Brout} {et~al.}(2022){Brout}, {Scolnic}, {Popovic}, {Riess}, {Carr},
  {Zuntz}, {Kessler}, {Davis}, {Hinton}, {Jones}, {Kenworthy}, {Peterson},
  {Said}, {Taylor}, {Ali}, {Armstrong}, {Charvu}, {Dwomoh}, {Meldorf},
  {Palmese}, {Qu}, {Rose}, {Sanchez}, {Stubbs}, {Vincenzi}, {Wood}, {Brown},
  {Chen}, {Chambers}, {Coulter}, {Dai}, {Dimitriadis}, {Filippenko}, {Foley},
  {Jha}, {Kelsey}, {Kirshner}, {M{\"o}ller}, {Muir}, {Nadathur}, {Pan}, {Rest},
  {Rojas-Bravo}, {Sako}, {Siebert}, {Smith}, {Stahl}, \&
  {Wiseman}}]{Brout:2022:ApJ}
{Brout}, D., {Scolnic}, D., {Popovic}, B., {et~al.} 2022, \apj, 938, 110

\bibitem[{Burkert(1995)}]{Bur:1995:AAL:}
Burkert, A. 1995, The Astrophysical Journal Letters, 447, L25

\bibitem[{{Cabral-Rosetti} {et~al.}(2004){Cabral-Rosetti}, {Matos},
  {N{\'u}{\~n}ez}, {Sussman}, \& {Zavala}}]{Cab-etal:2004:GRG:NFWpoly}
{Cabral-Rosetti}, L.~G., {Matos}, T., {N{\'u}{\~n}ez}, D., {Sussman}, R.~A., \&
  {Zavala}, J. 2004, ArXiv Astrophysics e-prints [\eprint{astro-ph/0405242}]

\bibitem[{{Carroll}(2001)}]{Carroll:2001:LRR}
{Carroll}, S.~M. 2001, Living Reviews in Relativity, 4, 1

\bibitem[{Chandrasekhar(1964)}]{Cha:1964:ASTRJ2:}
Chandrasekhar, S. 1964, Astrophys. J., 140, 417

\bibitem[{Chen \& Wang(2008)}]{Che:2008:CHINPB:DkEnGeoMorSchw}
Chen, J.-H. \& Wang, Y.-J. 2008, Chinese Physics~B, 17, 1184

\bibitem[{{Corbelli, Edvige} {et~al.}(2014){Corbelli, Edvige}, {Thilker,
  David}, {Zibetti, Stefano}, {Giovanardi, Carlo}, \& {Salucci,
  Paolo}}]{Cor-etal:2014:AA:Dyn:}
{Corbelli, Edvige}, {Thilker, David}, {Zibetti, Stefano}, {Giovanardi, Carlo},
  \& {Salucci, Paolo}. 2014, Astronomy and Astrophysics, 572, A23

\bibitem[{Cremaschini \& Stuchl{\'i}k(2013)}]{Cre-Stu:2013:IJMPD:}
Cremaschini, C. \& Stuchl{\'i}k, Z. 2013, Internat. J. Modern Phys. D, 22,
  50077

\bibitem[{Cruz {et~al.}(2005)Cruz, Olivares, \&
  Villanueva}]{Cru-Oli-Vil:2005:CLAQG:GeoSdSBH}
Cruz, N., Olivares, M., \& Villanueva, J.~R. 2005, Classical Quantum Gravity,
  22, 1167

\bibitem[{da~Silva {et~al.}(2013)da~Silva, Fontanini, \&
  Guariento}]{Sil-Fon-Gua:2013:PHYSR4:}
da~Silva, A.~M., Fontanini, M., \& Guariento, D.~C. 2013, Phys. Rev. D, 87,
  064030

\bibitem[{de~Felice(1974)}]{deFel:1974:ASTRA:}
de~Felice, F. 1974, Astronomy and Astrophysics, 34, 15

\bibitem[{de~Felice(1978)}]{deFel:1978:NATURE:InstabNS}
de~Felice, F. 1978, Nature, 273, 429

\bibitem[{Faraoni(2016)}]{Far:2016:PDU:}
Faraoni, V. 2016, Physics of the Dark Universe, 11, 11

\bibitem[{Faraoni {et~al.}(2015)Faraoni, Lapierre-L{\'{e}}onard, \&
  Prain}]{Far-Lap-Pra:2015:JCAP:}
Faraoni, V., Lapierre-L{\'{e}}onard, M., \& Prain, A. 2015, Journal of
  Cosmology and Astroparticle Physics, 2015, 013

\bibitem[{Faraoni {et~al.}(2014)Faraoni, Moreno, \&
  Prain}]{Far-Mor-Pra:2014:PHYSR4:}
Faraoni, V., Moreno, A. F.~Z., \& Prain, A. 2014, Phys. Rev. D, 89, 103514

\bibitem[{Fleury {et~al.}(2013)Fleury, Dupuy, \&
  Uzan}]{Fle-Dup-Uza:2013:PHYSR4:}
Fleury, P., Dupuy, H., \& Uzan, J.-P. 2013, Phys. Rev. D, 87, 123526

\bibitem[{Gariazzo {et~al.}(2017)Gariazzo, Escudero, Diamanti, \&
  Mena}]{Gar-etal:2017:ARx}
Gariazzo, S., Escudero, M., Diamanti, R., \& Mena, O. 2017, Physical Review D,
  96

\bibitem[{Gimon \& Ho{\v{r}}ava(2004)}]{Gim-Hor:2004:hep-th0405019:GodHolo}
Gimon, E.~G. \& Ho{\v{r}}ava, P. 2004, {Over-Rotating Black Holes, G{\"o}del
  Holography and the Hypertube}

\bibitem[{Gimon \& Ho{\v{r}}ava(2009)}]{Gim-Hor:2009:PHYLB:AstVioSignStr}
Gimon, E.~G. \& Ho{\v{r}}ava, P. 2009, Phys. Lett. B, 672, 299

\bibitem[{Grenon \& Lake(2010)}]{Gre-Lak:2010:PHYSR4:}
Grenon, C. \& Lake, K. 2010, Phys. Rev. D, 81, 023501

\bibitem[{Gu \& Cheng(2007)}]{Gu-Cheng:2007:GENRG2:CircLoopKdS}
Gu, Z. \& Cheng, H. 2007, Gen. Relativity Gravitation, 39, 1

\bibitem[{{Guth}(1981)}]{Guth:1981:PRD:}
{Guth}, A.~H. 1981, \prd, 23, 347

\bibitem[{{Guth} {et~al.}(2014){Guth}, {Kaiser}, \& {Nomura}}]{Guth:2014:PhLB}
{Guth}, A.~H., {Kaiser}, D.~I., \& {Nomura}, Y. 2014, Physics Letters B, 733,
  112

\bibitem[{Hackmann {et~al.}(2010)Hackmann, Hartmann, L{\" a}mmerzahl, \&
  Sirimachan}]{Hac-etal:2010:PHYSR4:KerrBHCoStr:}
Hackmann, E., Hartmann, B., L{\" a}mmerzahl, C., \& Sirimachan, P. 2010, Phys.
  Rev. D, 82, 044024

\bibitem[{Hendi \& Momeni(2011)}]{Hen-Mom:2011:EPJC:}
Hendi, S.~H. \& Momeni, D. 2011, The European Physical Journal C, 71, 1823

\bibitem[{Hendi {et~al.}(2012)Hendi, Panah, \&
  Mousavi}]{Hen-Pan-Mou:2012:GENRG:}
Hendi, S.~H., Panah, B.~E., \& Mousavi, S.~M. 2012, General Relativity and
  Gravitation, 44, 835

\bibitem[{Hioki \& Maeda(2009)}]{Hio-Mae:2009:PHYSR4:KerrSpinMeas}
Hioki, K. \& Maeda, K.-i. 2009, Phys. Rev. D, 80, 024042

\bibitem[{{Hlad{\'\i}k} {et~al.}(2020){Hlad{\'\i}k}, {Posada}, \&
  {Stuchl{\'\i}k}}]{Hla-Pos-Stu:2020:Modern}
{Hlad{\'\i}k}, J., {Posada}, C., \& {Stuchl{\'\i}k}, Z. 2020, International
  Journal of Modern Physics D, 29, 2050030

\bibitem[{Hlad{\'i}k \& Stuchl{\'i}k(2011)}]{Hla-Stu:2011:JCAP:}
Hlad{\'i}k, J. \& Stuchl{\'i}k, Z. 2011, Journal of Cosmology and Astroparticle
  Physics, 2011, 012

\bibitem[{Hui {et~al.}(2017)Hui, Ostriker, Tremaine, \&
  Witten}]{Hui-etal:2017:PRD:Ultra}
Hui, L., Ostriker, J.~P., Tremaine, S., \& Witten, E. 2017, Phys. Rev. D, 95,
  043541

\bibitem[{{Huterer} \& {Shafer}(2018)}]{Huterer:2018:RPPh}
{Huterer}, D. \& {Shafer}, D.~L. 2018, Reports on Progress in Physics, 81,
  016901

\bibitem[{Iorio(2009)}]{Ior:2009:NEWASTR:CCDGPGrav}
Iorio, L. 2009, New Astronomy, 14, 196

\bibitem[{Iorio(2010)}]{Ior:2010:MONNR:GalOrbMoDarkMat}
Iorio, L. 2010, Monthly Notices Roy. Astronom. Soc., 401, 2012

\bibitem[{Kagramanova {et~al.}(2006)Kagramanova, Kunz, \&
  Lammerzahl}]{Kag-Kun-Lam:2006:PHYLB:SolarSdS}
Kagramanova, V., Kunz, J., \& Lammerzahl, C. 2006, Phys. Lett. B, 634, 465

\bibitem[{Kaloper {et~al.}(2010)Kaloper, Kleban, \&
  Martin}]{Kal-Kle-Mar:2010:PHYSR4:}
Kaloper, N., Kleban, M., \& Martin, D. 2010, Phys. Rev. D, 81, 104044

\bibitem[{{King}(1962)}]{King:1962:AJ:}
{King}, I. 1962, Astrophys. J., 67, 471

\bibitem[{Kolb \& Turner(1990)}]{Kol-Tur:1990:EarUni:}
Kolb, E.~W. \& Turner, M.~S. 1990, {The Early Universe} (Redwood City,
  California: Addison-Wesley)

\bibitem[{Kolo{\v{s}} \&
  Stuchl{\'i}k(2010)}]{Kol-Stu:2010:PHYSR4:CurCarStrLoops}
Kolo{\v{s}}, M. \& Stuchl{\'i}k, Z. 2010, Phys. Rev. D, 82, 125012

\bibitem[{Kraniotis(2004)}]{Kra:2004:CLAQG:}
Kraniotis, G.~V. 2004, Classical Quantum Gravity, 21, 4743

\bibitem[{Kraniotis(2005)}]{Kra:2005:DARK:CCPerPrec}
Kraniotis, G.~V. 2005, in {Dark matter in astro- and particle physics.
  Proceedings of the International Conference DARK 2004, College Station,
  Texas, USA, 3--9 October, 2004}, ed. H.~V. Klapdor-Kleingrothaus \&
  R.~Arnowitt (Berlin: Springer), 469--479

\bibitem[{Kraniotis(2007)}]{Kra:2007:CLAQG:Periapsis}
Kraniotis, G.~V. 2007, Classical Quantum Gravity, 24, 1775

\bibitem[{Krauss(1998)}]{Kra:1998:ASTRJ2:}
Krauss, L.~M. 1998, Astrophys. J., 501, 461

\bibitem[{Krauss \& Turner(1995)}]{Kra-Tur:1995:GENRG2:}
Krauss, L.~M. \& Turner, M.~S. 1995, Gen. Relativity Gravitation, 27, 1137

\bibitem[{Lake(2002)}]{Lak:2002:PHYSR4:BendLiCC:}
Lake, K. 2002, Phys. Rev. D, 65

\bibitem[{Lake \& Abdelqader(2011)}]{Lak-Abd:2011:PHYSR4:}
Lake, K. \& Abdelqader, M. 2011, Phys. Rev. D, 84, 044045

\bibitem[{{Lapi} \& {Danese}(2015)}]{Lap-Dan:2015:JCAP:}
{Lapi}, A. \& {Danese}, L. 2015, Journal of Cosmology and Astroparticle
  Physics, 9, 003

\bibitem[{Lattimer \& Prakash(2001)}]{Lat-Pra:2001:ASTRJ2:NS}
Lattimer, J.~M. \& Prakash, M. 2001, Astrophys. J., 550, 426

\bibitem[{Linde(1990)}]{Lin:1990:InfCos:}
Linde, A.~D. 1990, Particle Physics and Inflationary Cosmology (New York:
  Gordon and Breach)

\bibitem[{Manolopoulou \& Plionis(2016)}]{Man-Pli:2016:MNRAS:Gal}
Manolopoulou, M. \& Plionis, M. 2016, Monthly Notices Roy. Astronom. Soc., 465,
  2616

\bibitem[{Marsh \& Silk(2013)}]{Mar-Sil:2014:MONNR:}
Marsh, D. J.~E. \& Silk, J. 2013, Monthly Notices of the Royal Astronomical
  Society, 437, 2652

\bibitem[{McVittie(1933)}]{McV:1933:MONRAS:}
McVittie, G.~C. 1933, Monthly Notices Roy. Astronom. Soc., 93, 325

\bibitem[{{Merafina} \& {Ruffini}(1989)}]{Merafina:1989}
{Merafina}, M. \& {Ruffini}, R. 1989, Astron. Astrophys., 221, 4

\bibitem[{Misner {et~al.}(1973)Misner, Thorne, \&
  Wheeler}]{Mis-Tho-Whe:1973:Gra:}
Misner, C.~W., Thorne, K.~S., \& Wheeler, J.~A. 1973, Gravitation (New York,
  San Francisco: W. H. Freeman and Co)

\bibitem[{{Mukhopadhyay} {et~al.}(2008){Mukhopadhyay}, {Ray}, \&
  {Choudhury}}]{Mukhopadhyay:2008:IJMPD}
{Mukhopadhyay}, U., {Ray}, S., \& {Choudhury}, S.~B.~D. 2008, International
  Journal of Modern Physics D, 17, 301

\bibitem[{M{\"u}ller(2008)}]{Mul:2008:GENRG2:FallSchBH}
M{\"u}ller, T. 2008, Gen. Relativity Gravitation, 56

\bibitem[{Murgia {et~al.}(2017)Murgia, Merle, Viel, Totzauer, \&
  Schneider}]{Mur-etal:2017:Arxiv}
Murgia, R., Merle, A., Viel, M., Totzauer, M., \& Schneider, A. 2017, Journal
  of Cosmology and Astroparticle Physics, 2017, 046–046

\bibitem[{Nandra {et~al.}(2012)Nandra, Lasenby, \&
  Hobson}]{Nan-Las-Hob:2012:MONRAS:}
Nandra, R., Lasenby, A.~N., \& Hobson, M.~P. 2012, Monthly Notices Roy.
  Astronom. Soc., 422, 2945

\bibitem[{Navarro {et~al.}(1997)Navarro, Frenk, \&
  White}]{Nav-Fre-Whi:1997:ASTRJ2:UniDeProHiCl}
Navarro, J.~F., Frenk, C.~S., \& White, S.~D.~M. 1997, Astrophys. J., 490, 493

\bibitem[{Nilsson \& Uggla(2000{\natexlab{a}})}]{Nilsson2000}
Nilsson, U.~S. \& Uggla, C. 2000{\natexlab{a}}, Ann. Physics, 286, 278

\bibitem[{Nilsson \&
  Uggla(2000{\natexlab{b}})}]{Nil-Ugg:2000:ANNPH1:GRStarPoEqSt}
Nilsson, U.~S. \& Uggla, C. 2000{\natexlab{b}}, Ann. Physics, 286, 292

\bibitem[{Nolan(1998)}]{Nol:1998:PHYSR4:}
Nolan, B.~C. 1998, Phys. Rev. D, 58, 064006

\bibitem[{Nolan(1999)}]{Nol:1999:CLAQG:}
Nolan, B.~C. 1999, Classical Quantum Gravity, 16, 1227

\bibitem[{Nolan(2014)}]{Nol:2014:CLAQG:}
Nolan, B.~C. 2014, Classical Quantum Gravity, 31, 235008

\bibitem[{{Novotn{\'y}} {et~al.}(2017){Novotn{\'y}}, {Hlad{\'{\i}}k}, \&
  {Stuchl{\'{\i}}k}}]{Nov-Hla-Stu:2017:PRD:trapping}
{Novotn{\'y}}, J., {Hlad{\'{\i}}k}, J., \& {Stuchl{\'{\i}}k}, Z. 2017, Phys.
  Rev. D, 95, 043009

\bibitem[{{Novotn{\'y}} {et~al.}(2021){Novotn{\'y}}, {Stuchl{\'\i}k}, \&
  {Hlad{\'\i}k}}]{Nov-Stu-Hla:AA:2021}
{Novotn{\'y}}, J., {Stuchl{\'\i}k}, Z., \& {Hlad{\'\i}k}, J. 2021, \aap, 647,
  A29

\bibitem[{Ostriker \& Steinhardt(1995)}]{Ost-Ste:1995:NATURE:}
Ostriker, J.~P. \& Steinhardt, P.~J. 1995, Nature, 377, 600

\bibitem[{{{\"O}zel} \& {Psaltis}(2009)}]{Oze-Psa:2009:PRD:}
{{\"O}zel}, F. \& {Psaltis}, D. 2009, Phys.~Rev.~D, 80, 103003

\bibitem[{{Peebles} \& {Ratra}(2003)}]{Peebles:2003:RvMP}
{Peebles}, P.~J. \& {Ratra}, B. 2003, Reviews of Modern Physics, 75, 559

\bibitem[{{Planck Collaboration} {et~al.}(2020{\natexlab{a}}){Planck
  Collaboration}, {Aghanim}, {Akrami}, {Ashdown}, {Aumont}, {Baccigalupi},
  {Ballardini}, {Banday}, {Barreiro}, {Bartolo}, {Basak}, {Battye}, {Benabed},
  {Bernard}, {Bersanelli}, {Bielewicz}, {Bock}, {Bond}, {Borrill}, {Bouchet},
  {Boulanger}, {Bucher}, {Burigana}, {Butler}, {Calabrese}, {Cardoso},
  {Carron}, {Challinor}, {Chiang}, {Chluba}, {Colombo}, {Combet}, {Contreras},
  {Crill}, {Cuttaia}, {de Bernardis}, {de Zotti}, {Delabrouille}, {Delouis},
  {Di Valentino}, {Diego}, {Dor{\'e}}, {Douspis}, {Ducout}, {Dupac}, {Dusini},
  {Efstathiou}, {Elsner}, {En{\ss}lin}, {Eriksen}, {Fantaye}, {Farhang},
  {Fergusson}, {Fernandez-Cobos}, {Finelli}, {Forastieri}, {Frailis},
  {Fraisse}, {Franceschi}, {Frolov}, {Galeotta}, {Galli}, {Ganga},
  {G{\'e}nova-Santos}, {Gerbino}, {Ghosh}, {Gonz{\'a}lez-Nuevo}, {G{\'o}rski},
  {Gratton}, {Gruppuso}, {Gudmundsson}, {Hamann}, {Handley}, {Hansen},
  {Herranz}, {Hildebrandt}, {Hivon}, {Huang}, {Jaffe}, {Jones}, {Karakci},
  {Keih{\"a}nen}, {Keskitalo}, {Kiiveri}, {Kim}, {Kisner}, {Knox},
  {Krachmalnicoff}, {Kunz}, {Kurki-Suonio}, {Lagache}, {Lamarre}, {Lasenby},
  {Lattanzi}, {Lawrence}, {Le Jeune}, {Lemos}, {Lesgourgues}, {Levrier},
  {Lewis}, {Liguori}, {Lilje}, {Lilley}, {Lindholm}, {L{\'o}pez-Caniego},
  {Lubin}, {Ma}, {Mac{\'\i}as-P{\'e}rez}, {Maggio}, {Maino}, {Mandolesi},
  {Mangilli}, {Marcos-Caballero}, {Maris}, {Martin}, {Martinelli},
  {Mart{\'\i}nez-Gonz{\'a}lez}, {Matarrese}, {Mauri}, {McEwen}, {Meinhold},
  {Melchiorri}, {Mennella}, {Migliaccio}, {Millea}, {Mitra},
  {Miville-Desch{\^e}nes}, {Molinari}, {Montier}, {Morgante}, {Moss}, {Natoli},
  {N{\o}rgaard-Nielsen}, {Pagano}, {Paoletti}, {Partridge}, {Patanchon},
  {Peiris}, {Perrotta}, {Pettorino}, {Piacentini}, {Polastri}, {Polenta},
  {Puget}, {Rachen}, {Reinecke}, {Remazeilles}, {Renzi}, {Rocha}, {Rosset},
  {Roudier}, {Rubi{\~n}o-Mart{\'\i}n}, {Ruiz-Granados}, {Salvati}, {Sandri},
  {Savelainen}, {Scott}, {Shellard}, {Sirignano}, {Sirri}, {Spencer},
  {Sunyaev}, {Suur-Uski}, {Tauber}, {Tavagnacco}, {Tenti}, {Toffolatti},
  {Tomasi}, {Trombetti}, {Valenziano}, {Valiviita}, {Van Tent}, {Vibert},
  {Vielva}, {Villa}, {Vittorio}, {Wandelt}, {Wehus}, {White}, {White},
  {Zacchei}, \& {Zonca}}]{PlanckB:2020:AaA}
{Planck Collaboration}, {Aghanim}, N., {Akrami}, Y., {et~al.}
  2020{\natexlab{a}}, \aap, 641, A6

\bibitem[{{Planck Collaboration} {et~al.}(2020{\natexlab{b}}){Planck
  Collaboration}, {Akrami}, {Arroja}, {Ashdown}, {Aumont}, {Baccigalupi},
  {Ballardini}, {Banday}, {Barreiro}, {Bartolo}, {Basak}, {Benabed}, {Bernard},
  {Bersanelli}, {Bielewicz}, {Bock}, {Bond}, {Borrill}, {Bouchet}, {Boulanger},
  {Bucher}, {Burigana}, {Butler}, {Calabrese}, {Cardoso}, {Carron},
  {Challinor}, {Chiang}, {Colombo}, {Combet}, {Contreras}, {Crill}, {Cuttaia},
  {de Bernardis}, {de Zotti}, {Delabrouille}, {Delouis}, {Di Valentino},
  {Diego}, {Donzelli}, {Dor{\'e}}, {Douspis}, {Ducout}, {Dupac}, {Dusini},
  {Efstathiou}, {Elsner}, {En{\ss}lin}, {Eriksen}, {Fantaye}, {Fergusson},
  {Fernandez-Cobos}, {Finelli}, {Forastieri}, {Frailis}, {Franceschi},
  {Frolov}, {Galeotta}, {Galli}, {Ganga}, {Gauthier}, {G{\'e}nova-Santos},
  {Gerbino}, {Ghosh}, {Gonz{\'a}lez-Nuevo}, {G{\'o}rski}, {Gratton},
  {Gruppuso}, {Gudmundsson}, {Hamann}, {Handley}, {Hansen}, {Herranz}, {Hivon},
  {Hooper}, {Huang}, {Jaffe}, {Jones}, {Keih{\"a}nen}, {Keskitalo}, {Kiiveri},
  {Kim}, {Kisner}, {Krachmalnicoff}, {Kunz}, {Kurki-Suonio}, {Lagache},
  {Lamarre}, {Lasenby}, {Lattanzi}, {Lawrence}, {Le Jeune}, {Lesgourgues},
  {Levrier}, {Lewis}, {Liguori}, {Lilje}, {Lindholm}, {L{\'o}pez-Caniego},
  {Lubin}, {Ma}, {Mac{\'\i}as-P{\'e}rez}, {Maggio}, {Maino}, {Mandolesi},
  {Mangilli}, {Marcos-Caballero}, {Maris}, {Martin},
  {Mart{\'\i}nez-Gonz{\'a}lez}, {Matarrese}, {Mauri}, {McEwen}, {Meerburg},
  {Meinhold}, {Melchiorri}, {Mennella}, {Migliaccio}, {Mitra},
  {Miville-Desch{\^e}nes}, {Molinari}, {Moneti}, {Montier}, {Morgante}, {Moss},
  {M{\"u}nchmeyer}, {Natoli}, {N{\o}rgaard-Nielsen}, {Pagano}, {Paoletti},
  {Partridge}, {Patanchon}, {Peiris}, {Perrotta}, {Pettorino}, {Piacentini},
  {Polastri}, {Polenta}, {Puget}, {Rachen}, {Reinecke}, {Remazeilles}, {Renzi},
  {Rocha}, {Rosset}, {Roudier}, {Rubi{\~n}o-Mart{\'\i}n}, {Ruiz-Granados},
  {Salvati}, {Sandri}, {Savelainen}, {Scott}, {Shellard}, {Shiraishi},
  {Sirignano}, {Sirri}, {Spencer}, {Sunyaev}, {Suur-Uski}, {Tauber},
  {Tavagnacco}, {Tenti}, {Toffolatti}, {Tomasi}, {Trombetti}, {Valiviita}, {Van
  Tent}, {Vielva}, {Villa}, {Vittorio}, {Wandelt}, {Wehus}, {White}, {Zacchei},
  {Zibin}, \& {Zonca}}]{Planck:2020:AaA}
{Planck Collaboration}, {Akrami}, Y., {Arroja}, F., {et~al.}
  2020{\natexlab{b}}, \aap, 641, A10

\bibitem[{{Posada} {et~al.}(2020){Posada}, {Hlad{\'\i}k}, \&
  {Stuchl{\'\i}k}}]{Pos-Hla-Stu:2020:PHYS4:DynStability:}
{Posada}, C., {Hlad{\'\i}k}, J., \& {Stuchl{\'\i}k}, Z. 2020, \prd, 102, 024056

\bibitem[{Pugliese \& Stuchl{\'{i}}k(2016)}]{Pug-Stu:2016:ApJS:}
Pugliese, D. \& Stuchl{\'{i}}k, Z. 2016, Astrophys. J. Suppl., 223, 27

\bibitem[{{Pugliese} \& {Stuchl{\'\i}k}(2024)}]{Pug-Stu:2024:EPJC:}
{Pugliese}, D. \& {Stuchl{\'\i}k}, Z. 2024, European Physical Journal C, 84,
  486

\bibitem[{Raccanelli {et~al.}(2016)Raccanelli, Kovetz, Dai, \&
  Kamionkowski}]{Rac-etal:PRD:2016:}
Raccanelli, A., Kovetz, E., Dai, L., \& Kamionkowski, M. 2016, Phys. Rev. D,
  93, 083512

\bibitem[{Rezzolla {et~al.}(2003)Rezzolla, Zanotti, \&
  Font}]{Rez-Zan-Fon:2003:ASTRA:}
Rezzolla, L., Zanotti, O., \& Font, J.~A. 2003, Astronomy and Astrophysics,
  412, 603

\bibitem[{Riess {et~al.}(2004)}]{Rie-etal:2004:ASTRJ2:}
Riess, A.~G. {et~al.} 2004, Astrophys. J., 123, 145

\bibitem[{Rubin(1982)}]{Rub:1982:HIContNorGal:SaSbScGal}
Rubin, V.~C. 1982, in {Comparative HI Content of Normal Galaxies, Proceedings
  of the Workshop}, ed. Knudsen, 42, 1982chcn.conf...42R

\bibitem[{Sartoris {et~al.}(2014)Sartoris, Biviano, Rosati, Borgani, Umetsu,
  Bartelmann, Girardi, Grillo, Lemze, Zitrin, Balestra, Mercurio, Nonino,
  Postman, Czakon, Bradley, Broadhurst, Coe, Medezinski, Melchior, Meneghetti,
  Merten, Annunziatella, Benitez, Czoske, Donahue, Ettori, Ford, Fritz, Kelson,
  Koekemoer, Kuchner, Lombardi, Maier, Moustakas, Munari, Presotto, Scodeggio,
  Seitz, Tozzi, Zheng, \& Ziegler}]{Sar-etal:2014:APJ:DMtoEOS:}
Sartoris, B., Biviano, A., Rosati, P., {et~al.} 2014, The Astrophysical Journal
  Letters, 783, L11

\bibitem[{Sch{\"{u}}cker \& Zaimen(2008)}]{Sch-Zai:2008:0801.3776:CCTimeDelay}
Sch{\"{u}}cker, T. \& Zaimen, N. 2008, Astronomy and Astrophysics, 484, 103

\bibitem[{Sereno(2008)}]{Ser:2008:PHYSR4:CCLens}
Sereno, M. 2008, Phys. Rev. D, 77, 043004

\bibitem[{Shapiro \& Teukolsky(1983)}]{Sha-Teu:1983:CompStar:}
Shapiro, S.~L. \& Teukolsky, S.~A. 1983, {Black Holes, White Dwarfs and Neutron
  Stars: The Physics of Compact Objects} (New York: John Wiley \&{} Sons), 672

\bibitem[{Slan{\'y} \& Stuchl{\'i}k(2005)}]{Sla-Stu:2005:CLAQG:}
Slan{\'y}, P. \& Stuchl{\'i}k, Z. 2005, Classical Quantum Gravity, 22, 3623

\bibitem[{Spergel {et~al.}(2007)Spergel, Bean, Dore, Nolta, Bennett, Dunkley,
  Hinshaw, Jarosik, Komatsu, Page, Peiris, Verde, Halpern, Hill, Kogut, Limon,
  Meyer, Odegard, Tucker, Weiland, Wollack, \&
  Wright}]{Spe-etal:2007:ASTJS:3yrWMAP}
Spergel, D.~N., Bean, R., Dore, O., {et~al.} 2007, Astrophys. J. Suppl., 170,
  377

\bibitem[{Stuchl{\'i}k(1980)}]{Stu:1980:BULAI:}
Stuchl{\'i}k, Z. 1980, Bull. Astronom. Inst. Czechoslovakia, 31, 129

\bibitem[{Stuchl{\'i}k(1983)}]{Stu:1983:BULAI:}
Stuchl{\'i}k, Z. 1983, Bull. Astronom. Inst. Czechoslovakia, 34, 129

\bibitem[{Stuchl{\'i}k(1984)}]{Stu:1984:BULAI:}
Stuchl{\'i}k, Z. 1984, Bull. Astronom. Inst. Czechoslovakia, 35, 205

\bibitem[{Stuchl{\'i}k(2000)}]{Stu:2000:ACTPS2:}
Stuchl{\'i}k, Z. 2000, Acta Phys. Slovaca, 50, 219

\bibitem[{Stuchl{\'i}k(2005)}]{Stu:2005:MODPLA:}
Stuchl{\'i}k, Z. 2005, Modern Phys. Lett. A, 20, 561

\bibitem[{Stuchl{\'i}k \& Calvani(1991)}]{Stu-Cal:1991:GENRG2:}
Stuchl{\'i}k, Z. \& Calvani, M. 1991, Gen. Relativity Gravitation, 23, 507

\bibitem[{{Stuchl{\'\i}k} \& {Charbul{\'a}k}(2024)}]{Stu-Char:2024:PhRvD}
{Stuchl{\'\i}k}, Z. \& {Charbul{\'a}k}, D. 2024, \prd, 109, 064008

\bibitem[{{Stuchl{\'\i}k} {et~al.}(2018){Stuchl{\'\i}k}, {Charbul{\'a}k}, \&
  {Schee}}]{Stu-Char-Sche:2018:EPJC}
{Stuchl{\'\i}k}, Z., {Charbul{\'a}k}, D., \& {Schee}, J. 2018, European
  Physical Journal C, 78, 180

\bibitem[{Stuchl{\'i}k \& Hled{\'i}k(1999)}]{Stu-Hle:1999:PHYSR4:}
Stuchl{\'i}k, Z. \& Hled{\'i}k, S. 1999, Phys. Rev. D, 60, 044006 (15~pages)

\bibitem[{Stuchl{\'i}k \& Hled{\'i}k(2002)}]{Stu-Hle:2002:ACTPS2:}
Stuchl{\'i}k, Z. \& Hled{\'i}k, S. 2002, Acta Phys. Slovaca, 52, 363

\bibitem[{{Stuchl{\'\i}k} \& {Hled{\'\i}k}(2005)}]{Stu-Hle:2005:RAGtime:}
{Stuchl{\'\i}k}, Z. \& {Hled{\'\i}k}, S. 2005, in RAGtime 6/7: Workshops on
  black holes and neutron stars, ed. S.~{Hled{\'\i}k} \& Z.~{Stuchl{\'\i}k},
  209--222

\bibitem[{Stuchl\'{\i}k {et~al.}(2016)Stuchl\'{\i}k, Hled\'{\i}k, \&
  Novotn\'y}]{Stu-Hle-Nov:2016:PHYSR4:}
Stuchl\'{\i}k, Z., Hled\'{\i}k, S., \& Novotn\'y, J. 2016, Phys. Rev. D, 94,
  103513

\bibitem[{Stuchl{\'i}k {et~al.}(2011)Stuchl{\'i}k, Hled{\'i}k, \&
  Truparov\'{a}}]{Stu-Hle-Tru:2011:CLAQG:}
Stuchl{\'i}k, Z., Hled{\'i}k, S., \& Truparov\'{a}, K. 2011, Classical Quantum
  Gravity, 28

\bibitem[{{Stuchl{\'\i}k} {et~al.}(2020){Stuchl{\'\i}k}, {Kolo{\v{s}}},
  {Kov{\'a}{\v{r}}}, {Slan{\'y}}, \& {Tursunov}}]{Stu-etal:2020:Universe}
{Stuchl{\'\i}k}, Z., {Kolo{\v{s}}}, M., {Kov{\'a}{\v{r}}}, J., {Slan{\'y}}, P.,
  \& {Tursunov}, A. 2020, Universe, 6, 26

\bibitem[{Stuchl{\'i}k \&
  Kov{\'a}{\v{r}}(2008)}]{Stu-Kov:2008:INTJMD:PsNewtSdS}
Stuchl{\'i}k, Z. \& Kov{\'a}{\v{r}}, J. 2008, INTJMD, 17, 2089

\bibitem[{Stuchl{\'i}k \& Schee(2011)}]{Stu-Sch:2011:JCAP:CCMagOnCloud}
Stuchl{\'i}k, Z. \& Schee, J. 2011, Journal of Cosmology and Astroparticle
  Physics, 9, 018

\bibitem[{Stuchl{\'i}k \& Schee(2012)}]{Stu-Sch:2012:CLAQG:}
Stuchl{\'i}k, Z. \& Schee, J. 2012, Classical Quantum Gravity, 29, 025008

\bibitem[{Stuchl{\'i}k \& Schee(2013)}]{Stu-Sch:2013:CLAQG:UHEKerrGeo}
Stuchl{\'i}k, Z. \& Schee, J. 2013, Classical Quantum Gravity, 30, 075012

\bibitem[{Stuchl\'{\i}k {et~al.}(2017)Stuchl\'{\i}k, Schee, Toshmatov,
  Hlad\'{\i}k, \& Novotn\'{y}}]{Stu-etal:2017:JCAP:}
Stuchl\'{\i}k, Z., Schee, J., Toshmatov, B., Hlad\'{\i}k, J., \& Novotn\'{y},
  J. 2017, Journal of Cosmology and Astroparticle Physics, 2017, 056

\bibitem[{Stuchl{\'i}k \& Slan{\'y}(2004)}]{Stu-Sla:2004:PHYSR4:}
Stuchl{\'i}k, Z. \& Slan{\'y}, P. 2004, Phys. Rev. D, 69, 064001

\bibitem[{Stuchl{\'i}k {et~al.}(2000)Stuchl{\'i}k, Slan{\'y}, \&
  Hled{\'i}k}]{Stu-Sla-Hle:2000:ASTRA:}
Stuchl{\'i}k, Z., Slan{\'y}, P., \& Hled{\'i}k, S. 2000, Astronomy and
  Astrophysics, 363, 425

\bibitem[{Stuchl{\'i}k {et~al.}(2009)Stuchl{\'i}k, Slan{\'y}, \&
  Kov{\'a}{\v{r}}}]{Stu-Sla-Kov:2009:CLAQG:}
Stuchl{\'i}k, Z., Slan{\'y}, P., \& Kov{\'a}{\v{r}}, J. 2009, Classical Quantum
  Gravity, 26, 215013 (34~pp)

\bibitem[{Tooper(1964)}]{Too:1964:ASTRJ2:}
Tooper, R.~F. 1964, Astrophys. J., 140, 434

\bibitem[{Tooper(1965)}]{Too:1964:ApJ:}
Tooper, R.~F. 1965, Astrophys. J., 142, 1541

\bibitem[{{Tristram} {et~al.}(2024){Tristram}, {Banday}, {Douspis}, {Garrido},
  {G{\'o}rski}, {Henrot-Versill{\'e}}, {Hergt}, {Ili{\'c}}, {Keskitalo},
  {Lagache}, {Lawrence}, {Partridge}, \& {Scott}}]{Tristram:2024:AaA}
{Tristram}, M., {Banday}, A.~J., {Douspis}, M., {et~al.} 2024, \aap, 682, A37

\bibitem[{Uzan {et~al.}(2011)Uzan, Ellis, \&
  Larena}]{Uza-Ell-Lar:2011:GENRG2:2MassExp:}
Uzan, J.-P., Ellis, G.~F.~R., \& Larena, J. 2011, Gen. Relativity Gravitation,
  43, 191

\bibitem[{Villanueva {et~al.}(2012)Villanueva, Saavedra, Olivares, \&
  Cruz}]{Vil-etal:2013:ASTSS1:PhMoChgAdS:}
Villanueva, J.~R., Saavedra, J., Olivares, M., \& Cruz, N. 2012, Astrophys. and
  Space Sci., 344, 437

\bibitem[{Wang {et~al.}(2000)Wang, Caldwell, Ostriker, \&
  Steinhardt}]{Wan-etal:2000:ASTRJ2:}
Wang, L., Caldwell, R.~R., Ostriker, J.~P., \& Steinhardt, P.~J. 2000,
  Astrophys. J., 530, 17

\bibitem[{Wang \& Cheng(2012)}]{Wan-Che:2012:PHYLB:CirLoopPerTens}
Wang, L. \& Cheng, H. 2012, Phys. Lett. B, 713, 59

\bibitem[{Ziolkowski(2005)}]{Zio:2005:NUOC2:GalCollObj}
Ziolkowski, J. 2005, Nuovo Cimento della Societa Italiana di fisica B~--
  General physics relativity astronomy and mathematical physics and methods,
  120, 757, {Vulcano Workshop 2004 on Frontier Objects in Astrophysics,
  Vulcano, Italy, May 24--29, 2004}

\bibitem[{Ziolkowski(2008)}]{Zio:2008:CHIAA:MassBHU}
Ziolkowski, J. 2008, Chinese Astronom. Astrophys., 8, 273, {7th International
  Workshop on Multifrequency Behaviour of High Energy Cosmic Sources, Vulcano,
  Italy, May~28--Jun~02, 2007}

\end{thebibliography}

\end{document}